\documentclass[twocolumn]{aastex6}
\usepackage{natbib}
\usepackage{amsmath,amsthm,amssymb,amsfonts}
\extrafloats{200}
\newcommand{\braket}[1] {\left< #1 \right>}
 
\shorttitle{New Constraints on the Surface Densities of Protoplanetary Disks}

\begin{document}

\title{New Constraints From Dust Lines on the Surface Densities of Protoplanetary Disks}
\author{Diana Powell\altaffilmark{1,2}, Ruth Murray-Clay\altaffilmark{1}, Laura M. P\'erez\altaffilmark{3}, Hilke E. Schlichting\altaffilmark{4,5}, Mickey Rosenthal\altaffilmark{1}}

\altaffiltext{1}{Department of Astronomy and Astrophysics, University of California, Santa Cruz, CA 95064}
\altaffiltext{2}{\href{mailto:dkpowell@ucsc.edu}{dkpowell@ucsc.edu} }
\altaffiltext{3}{Departamento de Astronom\'ia, Universidad de Chile, Camino El Observatorio 1515, Las Condes, Santiago, Chile}
\altaffiltext{4}{Department of Earth, Planetary, and Space Sciences, University of California, Los Angeles, CA 90095}
\altaffiltext{5}{Department of Earth, Atmospheric and Planetary Sciences, Massachusetts Institute of Technology, 77 Massachusetts Avenue, Cambridge, MA 02139}

\begin{abstract}
We present new determinations of disk surface density, independent of an assumed dust opacity, for a sample of 7 bright, diverse protoplanetary disks using measurements of disk dust lines. We develop a robust method for determining the location of dust lines by modeling disk interferometric visibilities at multiple wavelengths. The disks in our sample have newly derived masses that are 9-27\% of their host stellar mass, substantially larger than the minimum mass solar nebula. All are stable to gravitational collapse except for one which approaches the limit of Toomre-Q stability. Our mass estimates are 2-15 times larger than estimates from integrated optically thin dust emission. We derive depleted dust-to-gas ratios with typical values of $\sim$$10^{-3}$ in the outer disk. Using coagulation models we derive dust surface density profiles that are consistent with millimeter dust observations. In these models, the disks formed with an initial dust mass that is a factor of $\sim$10 greater than is presently observed. Of the three disks in our sample with resolved CO line emission, the masses of HD 163296, AS 209, and TW Hya are roughly 3, 115, and 40 times more massive than estimates from CO respectively. This range indicates that CO depletion is not uniform across different disks and that dust is a more robust tracer of total disk mass. Our method of determining surface density using dust lines is robust even if particles form as aggregates and is useful even in the presence of dust substructure caused by pressure traps. The low Toomre-Q values observed in this sample indicate that at least some disks do not accrete efficiently.
\end{abstract}


\section{Introduction}\label{intro}
Protoplanetary disks are the likely initial conditions of planet formation. One of the most fundamental parameters in planet formation theory is the disk surface density -- or the total disk mass inventory. The most common method of observationally determining disk surface densities is to infer the total mass through the use of a mass tracer that emits more readily than the main mass constituent -- molecular hydrogen. The two most commonly used tracers of mass are dust and rotational lines of carbon monoxide gas, since both emit substantially in the millimeter. From dust observations, the solid surface density is inferred from the dust's assumed optically thin thermal emission and is converted to a total surface density via an assumed dust-to-gas ratio. From observations of rotational lines of CO, the gaseous surface density is inferred from observations of one or more CO isotopologues that are thought to be optically thin and is converted to a total surface density via an assumed CO-to-H$_2$ ratio. Mass estimates derived from these methods, however are often inconsistent and can vary by orders of magnitude \citep[e.g.,][]{2013Natur.493..644B}. There are several reasons to question the accuracy of these methods. 

When inferring solid surface densities, a dust grain opacity must be assumed. However, the opacity of dust grains in disks is highly uncertain \citep[e.g.,][]{1987ApJ...320..818W,2000prpl.conf..533B,1991ApJ...381..250B,2005ApJ...631.1134A,2018ApJ...869L..45B} and dust continuum observations lose sensitivity to solids that are much larger than the observing wavelength \citep[e.g.,][]{2011ARAA..49...67W}. It is therefore possible that measurements from dust observations are missing a reservoir of mass. Furthermore, the dust-to-gas ratio in disks should differ from the ISM value of 10$^{-2}$ due to processes such as grain growth and drift \citep[e.g.,][]{2012MNRAS.423..389H}. A differing dust-to-gas ratio will change the inferred total gas surface density even if it does not change the inferred solid surface density. These effects work to complicate extrapolations of disk mass from continuum dust observations as the total mass in dust is uncertain and the ratio used to convert to a total gaseous surface density may be incorrect by orders of magnitude \citep[e.g.,][]{2012A&A...539A.148B}. 

Recent observational work has also indicated that the typically assumed CO-to-H$_2$ ratio of 10$^{-4}$ (derived primarily from studies of the interstellar medium (ISM)) is likely over simplified. This is true not only in the context of protoplanetary disks but also in the context of star-forming molecular clouds where CO isotopologues have been shown to be weaker tracers of mass than dust and to be depleted in regions near B-stars as well as at the location of protostellar cores \citep[e.g.,][]{2009ApJ...692...91G,2015ApJ...803...38I}. In particular, there are several observational lines of evidence pointing to a depletion or lack of CO, and potentially all gas phase carbon compounds, in disks. Observations of HD gas, the most direct observational probe of disk mass as it is a hydrogen molecule line with a well defined ratio with respect to H$_2$, derive a disk mass for TW Hya that is significantly higher than observations of CO alone by $\sim$2 orders of magnitude \citep{2013Natur.493..644B,2013ApJ...776L..38F,2016ApJ...823...91S,2016A&A...592A..83K}. The HD derived masses for two other disks (GM Aur and DM Tau) are also significantly larger than those derived from CO isotopologues \citep{2016ApJ...831..167M}. Unfortunately, while HD gas is the most direct available tracer of total disk mass, observations were only made for a few disks by the Herschel Space Observatory before its decommissioning. 

Furthermore, a recent survey of disks in the Lupus star-forming region by \citet{2016ApJ...828...46A} found that assuming an ISM CO-to-H$_2$ ratio leads to anomalously small derived disk masses (often less than 1 M$_\text{jup}$). These disk masses seem to be inconsistent with observations of accretion onto these stars which indicate the presence of abundant gas, indicating that CO is a poor tracer of the total mass in these systems. Indeed, for the same sample of disks the derived dust masses are correlated with the measured accretion rates as predicted by viscous accretion theory, while the gas mass derived from CO observations has no correlation with measured accretion rates, suggesting that dust is a better tracer of disk mass \citep{2016A&A...591L...3M,2017A&A...599A.113M}. An analogous CO survey of disks in the Chameleon star forming region also derives implausibly low gas masses for the objects with detected emission \citep{2017ApJ...844...99L}. A separate large survey of disks done by \citet{2016A&A...588A.108K} using carbon lines thought to be less affected by the photodissociation of CO also finds that many systems are either carbon-depleted or gas-poor disks. 

Chemical modeling of observed disks around more massive stars suggest that the gaseous carbon abundance is depleted, suggesting low dust-to-gas ratios \citep{2008A&A...488..565C,2012A&A...541A..91B}. Similar modeling of recent observations of DCO$^+$ in HD 169142 also require a CO depletion of a factor of 5 relative to the fiducial literature model to reproduce the observed DCO$^+$ radial intensity profile \citep{2018A&A...614A.106C}.

CO has been historically used as a tracer of total gas mass because it is believed to have stable chemistry and to remain in the gas phase for temperatures $>20$ K in disks around sun-like stars \citep{2011ApJ...743L..16O,2013Sci...341..630Q}. However, the disk-averaged CO abundance can be much lower than the canonical ISM value due to freezeout and CO photodissociation \citep[e.g.,][]{2001ApJ...561.1074T,2003A&A...402.1003D,2010A&A...520A..61C}. Newer theoretical studies have further found that CO chemistry in the disk environment is more complicated than previously assumed. \citet{2016ApJ...822...53Y} use a chemical model that includes detailed photochemistry to propose that the CO abundance varies with distance in the planet forming regions of disks and that the CO-to-H$_2$ ratio drops to an order of magnitude below the interstellar value inside the CO freeze-out radius and is also a function of time due in part to the formation of complex organic molecules. Recent work using this modeling technique further shows that CO depletion in the outer disk driven by ionization is a robust result for realistic T-Tauri star ionization rates \citep{2018ApJ...868L..37D}. These factors may cause disk masses measured from standard CO observations to be under-predicted due to CO being chemically depleted in the outer disk where emission is optically thin \citep{2017ApJ...841...39Y}. 

The observational and theoretical evidence presented thus far points towards disks potentially having more mass, or a broader range in mass, than standard observational and theoretical assumptions derive. There are several other reasons to expect that disks may be more massive, or exhibit a broader range in mass, than typically assumed. For example, recent observations in the millimeter have uncovered a new class of disks with spiral arms. These disks have morphologies that potentially indicate that they are massive, gravitationally unstable objects \citep{2016Sci...353.1519P,2018arXiv181204193H}. Furthermore, planet formation models for our solar system often require around an order of magnitude enhancement in density from the minimum mass solar nebula (MMSN) to form Jupiter and the other giant planets within a disk lifetime \citep[e.g.,][]{1996Icar..124...62P,2005Icar..179..415H,2008Sci...321..814T,2009Icar..199..338L,2009ApJ...691.1764M,2010Icar..207..491D,2014Icar..241..298D}. \citet{2014ApJ...795L..15S} show that if close-in Earth-to-Neptune-sized planets formed \textit{in situ} as isolation masses, then the disk in which they formed would be gravitationally unstable assuming standard dust-to-gas ratios. If these planets instead formed at smaller isolation masses and then grew to their present size by giant impacts, then the surface density of the disks in which they formed is at least a factor of 20 larger than the MMSN when giant impacts are considered, close to the limit of gravitational stability. A derivation of a standard minimum mass extrasolar nebula (MMEN) by \citet{2013MNRAS.431.3444C} further derives an average minimum disk mass that is a factor of 5 larger than the MMSN, while other work has indicated that there is no universal MMEN and extrasolar disks must have a variety of different properties \citep{2014MNRAS.440L..11R}. In addition, the properties of a protoplanetary disk are set by the initial properties of the star forming cloud core, which vary from cloud to cloud \citep[e.g.,][]{2010ApJ...708.1585K,2011ARAA..49...67W}. Recent simulations of embedded disks derive masses that are greater than those typically inferred from observations of dust by at least a factor of 2-3 and that exceed the MMSN for objects with stellar masses as low as 0.05-0.1 M$_\sun$ \citep{2011ApJ...729..146V}.

\citet{2017ApJ...840...93P} suggest an alternative method for determining disk mass that does not rely on an assumed tracer-to-hydrogen mass ratio. They demonstrate that it may be possible to use dust to trace the total disk mass through a consideration of the aerodynamic properties of the grains. This can be achieved empirically through the consideration of spatially resolved multiwavelength observations of disks in the millimeter. These aerodynamic grain properties are thought to cause particle drift radially inward towards the star. Recent multiwavelength observations of disks appear to show signatures of particle drift as the radial extent of several disks becomes smaller at longer wavelengths \citep[e.g.,][]{2010ApJ...714.1746I,2017ApJ...845...44T,2011AA...529A.105G,2011A&A...525A..12B,2012ApJ...760L..17P,2015ApJ...813...41P,2016AA...588A..53T}. The radial extent of a disk at a particular wavelength is known as a disk dust line \citep{2017ApJ...840...93P} as, in the millimeter, we can assume that emission at the observed wavelength is dominated by particles with a size comparable to that wavelength. As these particles are all in the Epstein drag regime, the surface density of a given disk can be readily determined given the maximum radius where particles of a given size are present. This model was successfully applied to the disk TW Hya, yielding a large total disk mass, consistent with measurements of HD gas and far in excess of measurements based on CO emission \citep{2017ApJ...840...93P}. In this work we further develop and test this model through applying it to six new disks.

The two input parameters of this model are the wavelength of observation and the radial extent of the disk. This model is thus independent of an assumed tracer-to-H$_2$ ratio or dust opacity model. The observational studies that find a decrease in disk radial extent as a function of wavelength also tend to find that the continuum emission at each wavelength exhibits a markedly sharp decrease over a very narrow radial range such that $\Delta r/r \lesssim 0.1$ \citep{2012ApJ...744..162A,2013A&A...557A.133D}. This is encouraging, as models of radial drift predict such a cut-off. However, accurately determining the outer edge of disk emission empirically is not trivial \citep{2017ApJ...845...44T}. 

In this work, after summarizing the \citet{2017ApJ...840...93P} model in Section \ref{modelit}, we introduce several model updates. In Section \ref{fitroutine}, we adapt the method derived in \citet{2017ApJ...845...44T} to accurately determine the outer edge of disk emission through modeling the interferometric visibilities. We describe the archival data used in this modeling work in Section \ref{data}. In Section \ref{modeleddisks}, we apply this method to multiwavelength observations of six new disks plus TW Hya. We compare our estimates of disk surface density and disk mass to previous observations and to limits from gravitational stability. We provide a validation of our analytic model using the semi-analytic model from \citet{2012A&A...539A.148B}. In Section \ref{discuss}, we comment on how to recognize whether the extent of dust emission at a given wavelength is set by drift or pressure bumps and provide a discussion of the effects of particle porosity. We provide a summary and conclusion of our results in Section \ref{summarize}.

\section{Disk Surface Density Derivation}\label{modelit}
To determine the disk surface density without assuming a tracer-to-H$_2$ ratio we use recent resolved images of disks in the millimeter to infer the maximum radial location of different particle sizes in the disk. We then use reasonable assumptions about the aerodynamic properties of the grains to determine the total gaseous disk surface density profile. Through a consideration of particle growth, we further calculate the surface density profile in dust which provides a consistency check with observations of total integrated dust emission. 

The location of the protoplanetary disk outer edge at a given millimeter wavelength is meaningful because it indicates that the particles that primarily contribute to the emission do not extend to larger radii. We refer to the empirically measured disk outer radius at a given millimeter wavelength as a disk ``dust line" \citep{2017ApJ...840...93P}. A dust line could be set by particle trapping in a ring or be set by the inward radial drift rate of solid particles. Particle trapping in rings likely occurs (see Section \ref{atrap}), but for many disks the fact that the dust lines are at different locations at different wavelengths suggests a differentiation of particle size with radial extent. As a significant particle trap should be efficient for particles across a range of sizes, this indicates that the dust line is not set by a strong particle trap for disks with an outer edge that varies with wavelength (see Section \ref{atrap} for a more detailed discussion, including the effects of an inefficient particle trap in the outer disk). We therefore assume that in these cases the dust lines are set by particle drift. 

There are several theoretical reasons to think that the disk radial extent is governed by particle drift. In evolved disks, particle growth is typically limited by fragmentation in the inner disk and drift in the outer disk \citep{2012A&A...539A.148B}. Particles in the outer disk will therefore grow until they reach a size such that their motion is sufficiently decoupled from the motion of the gas and they begin to experience a significant headwind. This headwind will rob the particle of angular momentum and it will begin to drift radially inwards \citep{1977MNRAS.180...57W}. In the outer disk, large particles will drift more quickly than smaller particles and will not be present at larger radii as they drift faster than they can be replenished due to particle growth \citep[e.g.,][]{2012A&A...539A.148B}. This is known as the drift-limited regime because the local particle size is limited by drift. Observationally we would expect disks in this regime to look smaller at wavelengths that probe larger particle sizes and for there to be a sharp decrease in flux exterior to the disk dust line. There are several disks in the literature that show evidence of particle drift \citep[e.g.,][]{2012ApJ...760L..17P,2015ApJ...813...41P,2012ApJ...744..162A,2013A&A...557A.133D}. Disks that demonstrate this behavior are good candidates for this new method of determining disk surface density. 

Following the method described in \citet{2017ApJ...840...93P}, the disk surface density can be derived using dust lines that are set by radial drift. Assuming that we are in the drift-limited regime, we expect that the drift timescale of the maximally sized particle at a given dust line is equal to the age of the system, $t_\text{drift} = t_\text{disk}$. We further assume that the timescale at a dust line can be determined using the current disk surface density profile. This assumption is reasonable because, for particle sizes of interest in the outer regions of a disk, drift is faster at larger separations. The time that it takes for a particle to drift to its observed location is thus dominated by the local drift timescale. Furthermore, when the overall surface density was higher, which was likely true at earlier times, the overall drift rate was slower. Using the current surface density profile to determine the disk surface density is therefore a conservative assumption. Under these assumptions, the disk surface density can be determined by

\begin{equation}\label{surf_dens}
\Sigma_g(r) \approx \frac{2.5t_\text{disk}v_0 \rho_s s}{r}
\end{equation}

\noindent where $\Sigma_g$ is the disk surface density which varies with semi-major axis, $t_\text{disk}$ is the current age of the system, $v_0$ approximately corresponds to the maximum drift velocity and is defined as $v_0 \equiv c_s^2/2v_k$ where $v_k$ is the Keplerian velocity, $\rho_s$ is the internal particle density, $s$ is the particle size, and $r$ is the maximum radius in which particles of size $s$ are present in the disk. By defining $v_0$ in this way, we implicitly set the power-law index of the gas pressure profile to unity as it is not known \textit{a priori}. 

This assumption about the power-law index is not entirely self-consistent. However, one can self-consistently determine the surface density, including this factor, through iteratively fitting a surface density profile to the derived surface densities at the dust line locations. We have done this iteration for the disks in our sample (not shown). Given the small number of current observational data points, we fix the inner disk index and vary the critical radius and total surface density profile normalization (see Equation \ref{similarity}). Unsurprisingly, doing this iteration with current data results in an excellent fit because the number of data points is comparable to the number of fitting parameters. More importantly, this fitting procedure changes our derived critical radius and total disk mass by 20-30 \%, well within the anticipated error of an order of magnitude model. As such, we move forward with the more simplified modeling described below with the note that when more data-points are available it may be appropriate to determine a surface density profile through iterative fitting of the data alone, without reference to previously inferred profiles.

To derive Equation (\ref{surf_dens}) we use $t_\text{stop} = m\Delta v/F_\text{drag}$, where $F_\text{drag} = 4/3 \pi \rho_g \Delta v \bar{v}_\text{th} s^2$, the volumetric gas density $\rho_g = \Sigma_g/2H$, $H$ is the scale height of the gas, $\Delta v$ is the relative velocity between a particle of mass $m$ and the gas, and $\bar{v}_\text{th} =(8/\pi)^{1/2}c_s$ is the mean thermal velocity of the gas (assuming a Maxwellian velocity distribution). In this derivation we have assumed that particles are in the Epstein drag regime in the outer disk which we find to be true for all currently modeled disks.

Using Equation (\ref{surf_dens}), the disk's total surface density can be derived as a function of radius given empirically determined dust lines. If the wavelength directly corresponds to the size of the emitting particle, as is typically assumed, then we can associate each observed dust line with a particle size. The optical depth of dust grains $\tau \propto n(s) \sigma(s)$ where $n$ is the number density, $\sigma$ is the interaction cross section between particles and light, and $s$ is the particle radius. For particles larger than the observed wavelength, $\lambda_\text{obs}$, $\sigma$ is the geometric cross section ($\pi s^2$). Particles somewhat smaller than $\lambda_\text{obs}$ are in the Mie scattering regime such that $\sigma = \pi s^2 (2\pi s/\lambda_\text{obs})$. Size distributions are typically expressed as $dN/ds \propto s^{-q}$ which gives $n(s) \propto s^{-q+1}$, so that

\begin{equation}\label{opacity}
\begin{array}{@{} r @{} c @{} l @{} }
&\tau &{}\propto \displaystyle
\begin{cases}
s^{3-q} & \lambda_\text{obs}< 2\pi s,\\
s^{4-q} &  \lambda_\text{obs}> 2\pi s.
\end{cases}
\end{array}
\end{equation}

\noindent We note that for values of $q$ less than 3, the largest particles in the disk should dominate the emission at all wavelengths and for $q$ values greater than 4 the smallest particles dominate the emission at all wavelengths. In either case, we would see the same disk dust line at all wavelengths, inconsistent with the observations. For example, we would expect relatively smaller disk sizes for $q<3$ with the outer edge tracing large particles that have drifted inwards, and larger disk sizes for $q>4$ with the outer edge tracing small particles that are present throughout the disk. We therefore assume that $3<q<4$, which implies that the particles dominating the observed emission at the disk outer edge have size $s = \lambda_\text{obs}/2\pi$.

For the commonly invoked Dohnanyi size distribution, $q=3.5$, which is in our preferred range \citep{1969JGR....74.2531D}. There is no \textit{a priori} reason that dust in the drift-limited regime will have a Dohnanyi size distribution because a collisional cascade is not expected. However, if $q$ instead had a value of 2.5 as has been suggested to explain observations of objects where grain growth may be significant \citep[e.g.,][]{2001ApJ...553..321D, 2004ASPC..323..279N, 2010AA...512A..15R, 2010A&A...512A..15R}, the largest grains would always dominate the emission and the dust line should be at the same location across all wavelengths. The observational fact that some disks have dust lines at different locations thus suggests that $q$ is indeed between 3 and 4.

Since, emission at an observed wavelength $\lambda_{\rm obs}$ is dominated by particles of size $s = \lambda_{\rm obs}/2\pi$, the dust line (maximum disk radius) observed at $\lambda_{\rm obs}$ gives the maximum radial extent of particles of this size. The radius $r$ of the disk dust line can therefore be used to determine the disk surface density at the dust line location following Equation (\ref{surf_dens}). If the dust emission is optically thin, it is straightforward to associate the dust line with a drop off in dust density. If the dust emission is instead optically thick, one might worry that the dust line would be measured exterior to the location at which the density falls off. However, because the observed decrease in emission at a given wavelength is sharp \citep[][see schematic in \citet{2017ApJ...840...93P}]{2012ApJ...744..162A,2013A&A...557A.133D,2014ApJ...780..153B}, which we do not expect to result from a transition in optical depth in a disk with smoothly declining density, dust lines are likely associated with steep decreases in dust density even in the case of optically thick emission.

In our modeling we assume, and later verify, that the disks are dominated thermodynamically by passive stellar irradiation at the radii of interest. This assumption is valid for all but the innermost radii for disks with average accretion rates of $\sim 10^{-8}$ M$_\sun$ yr$^{-1}$ or less \citep{2007A&A...473..457D}. We therefore parameterize the disk midplane temperature following \citet{1997ApJ...490..368C}, where the canonical temperature profile in the disk midplane is

\begin{equation}\label{disk_temperature}
T(r) = T_0\times\left(\frac{r}{r_0}\right)^{-3/7}
\end{equation}

\noindent where the temperature $T_0$, defined at $r_0 = 1$ au, is

\begin{equation}
T_0 = L_\star^{2/7}\left(\frac{1}{4\sigma_\text{SB}\pi}\right)^{2/7}\left(\frac{2}{7}\right)^{1/4}\left(\frac{k}{\mu GM_\star}\right)^{1/7}r^{-3/7}
\end{equation}

\noindent where $L_\star$ is the stellar luminosity, $\sigma_\text{SB}$ is the Stefan-Boltzmann constant, $k$ is the Boltzmann constant, $\mu$ is the reduced mass taken to be 2.3$\text{m}_\text{H}$ assuming a hydrogen/helium disk composition, $G$ is Newton's gravitational constant, $M_\star$ is the stellar mass, and $r$ is the disk semi-major axis. 

While CO may be depleted in disks, the shape of the surface density profile derived from resolved observations may still roughly correspond to the distribution of the underlying hydrogen and helium gas mass. Therefore, if this method is valid in determining surface density, we might expect that the derived surface density points will follow the shape of the CO emission although the normalization of the surface density profile is expected to differ. Alternatively, the surface density profiles derived from simultaneously modeling multiwavelength millimeter observations of dust might approximate the shape of the surface density profile. While it is not obvious that either profiles should necessarily match the distribution of the underlying gas disk completely, this comparison provides a useful initial method check.

\subsection{Dust Surface Density}\label{dustsd}

As described in the above model, knowing the dust surface density is not necessary to determine the gas surface density. We can, however, derive the dust surface density profile from our gas surface density profile using a drift and coagulation model without the need for an assumed dust opacity model. Comparing our derived dust surface density with the observed profile provides a consistency check for our model of total gaseous surface density.

In the drift-limited regime, the maximum particle size at a given radius is the particle whose growth timescale is equal to its drift timescale, as larger particles with higher drift velocities will be removed by drift before they are replenished by growth \citep{2012A&A...539A.148B,2014ApJ...780..153B}. We can therefore expand our assumptions regarding the drift-limited regime to include the growth timescale such that $t_\text{drift} = t_\text{grow} = t_\text{disk}$. This differs from the method described in \citet{2014A&A...572A.107L} as we do not prescribe a dust-to-gas ratio or dust surface density profile \textit{a priori} and we further consider a constantly evolving disk at the outer edge instead of a disk in steady state. We assume that these disks are formed with an ISM dust-to-gas ratio; if our growth model finds a lower dust-to-gas ratio this implies that the additional solids inherited from the ISM have drifted into the interior of the disk.

Before particles reach the regime of drift-limited growth they must grow from very small, submicron grains to roughly millimeter size grains that are affected by gas drag. This initial stage of growth can potentially be significant. This timescale is approximately given by:

\begin{equation}\label{slowdatgrow}
t_\text{early growth} \approx \frac{0.033\alpha^{-0.63}}{\Omega f_d}\ln\left(\frac{a_\text{max}}{a_0}\right)
\end{equation}

\noindent where $f_d = \Sigma_d/\Sigma_g$ is the dust-to-gas ratio, $\Sigma_d$ is the surface density in dust, $\Omega$ is the local Keplerian orbital angular velocity, a$_\text{max}$ is the maximum particle size at a given location which in this case is set by particle drift, and $a_0$ is the initial particle size inherited from the ISM which we assume to be $\sim 0.1\;\mu$m.  The dimensionless value $\alpha$ is the standard Shakura-Sunyayev parameter describing disk viscosity \citep{1973A&A....24..337S}, which in this work we use only to parameterize the local eddy diffusivity of the gas and which may be a function of location in the disk. Equation (\ref{slowdatgrow}) is based on the approximation for this timescale derived in \citet{2012A&A...539A.148B} where particle growth is collisional and particles are assumed to grow by collisions with similarly sized grains. We modify the \citet{2012A&A...539A.148B} expression, however, to account for the slower growth of very small particles that is affected by the amount of turbulence in the disk (see Appendix \ref{coefficient}). Our modification increases the growth timescale by a factor of two for $\alpha = 10^{-3}$ and by larger factors for smaller values of $\alpha$. Given an initial dust-to-gas ratio of 10$^{-2}$, particles will have grown to the drift-regulated stage of growth in disks with ages roughly $\gtrsim$1 Myr as long as $\alpha \gtrsim 10^{-7}$. We include this early growth phase in our models, but it does not affect our results.

Once particles have undergone a phase of early growth, we model particle growth in more detail. Our growth timescale is derived by first considering the collisional growth rate of particles following:

\begin{equation}\label{grow_rate}
\dot{m} = \rho_d \sigma \Delta v
\end{equation}

\noindent where $\rho_d$ is the volumetric density of particles in the disk (not the particle internal density, $\rho_s$) and $\sigma=\pi s^2$ is the particle cross section where $s$ is the size of the largest particles at a given radius. We can convert this growth rate to a growth timescale such that:

\begin{equation}
\tau_\text{grow} = \frac{m}{\dot{m}} \sim \frac{8s\rho_sH_d}{3\Sigma_d \Delta v f} \sim \frac{8s\rho_sH_d}{3f_d \Sigma_g \Delta v f}
\end{equation}\label{growy}

\noindent where $H_d$ is the particle scale height and is given by $H_d = H \sqrt{\alpha/(\alpha+St})$ \citep{2012ApJ...747..115O} and $\Delta v$ is the relative particle velocity. Particle relative velocities loosely fall into three regimes as discussed and derived in an order of magnitude scheme in Appendix \ref{appgrow} to \ref{appgrow1}. As this model can derive a range of disk masses, we do not prescribe a relative velocity \textit{a priori}. We instead use the full expression from \citet{2007A&A...466..413O} (see their Equation 16) that encapsulates the three different regimes of particle growth. For ease of comparison we also introduce a coagulation efficiency parameter, $f$, to calibrate our coagulation estimates with detailed numerical simulations. We adopt a value of $f = 0.55$ following \citet{2012A&A...539A.148B} which produces results in agreement with numerical models. 

We assume a Dohnanyi particle size distribution, which has a value of $q=3.5$ lying between 3 and 4 (see Section \ref{modelit}) and which is commonly used in disk modeling. The Dohnanyi size distribution is dominated in mass by the largest sized particles (mass $\propto s^{0.5}$). In our calculations, this choice is roughly consistent with choosing any size distribution that is also dominated in mass by the largest particles such as the drift-limited size distribution defined in \citet{2015ApJ...813L..14B}. For size distributions with this attribute, the growth of the large particles can be modeled through collisions with similarly sized grains. This is valid because the largest grains dominate both the density and cross section terms in Equation (\ref{grow_rate}). For particle sizes probed by millimeter observations, the intermediate relative velocity regime is typically appropriate. In this regime, the relative velocity is roughly independent of the small body size. Thus, the growth rate is dominated by the largest particles.

Assuming that particles have taken time $t_\text{early growth}$ to grow to a size such that their growth timescale is given by $\tau_\text{growth}$, the growth timescale is given by $\tau_\text{growth} = t_\text{disk}-t_\text{early growth}$. With this known formulation for the particle growth timescale we are able to solve for the dust-to-gas ratio for the maximally sized particles at a given dust line such that

\begin{equation}\label{dust-to-gas}
f_d \sim \frac{8s\rho_sH_d}{3\tau_\text{grow} \Sigma_g \Delta v f}.
\end{equation}

This empirically derived dust-to-gas ratio, calculated at each dust line given the above assumptions, can then be used to convert the total surface density profile to a dust surface density profile following:

\begin{equation}
\Sigma_d(r) = \Sigma_g(r) f_d
\end{equation}

\noindent where $\Sigma_d$ is the dust surface density and $\Sigma_g$ is the total surface density which is dominated by the gas mass. In a Dohnanyi size distribution, roughly 70 $\%$ of the mass is in grains whose radii are within an order of magnitude of the maximally sized particles. After deriving the dust surface density for the maximally sized grains we therefore add the additional surface density in smaller grains such that $\Sigma_\text{d,tot}(r) \approx 1.3\Sigma_d$.

\subsection{Main Sources of Uncertainty}\label{uncertain}

In this modeling, there are several sources of uncertainty that may not be well constrained such as the age of the system and the distance to the system. The disk age is usually assumed to be the same as the stellar age. However, for young stars, stellar ages are subject to significant observational uncertainties. For example, literature age estimates for several disks in our sample span many millions of years. In this modeling it may therefore be appropriate to tune the disk age. The inferred disk surface density and therefore the disk mass is linearly proportional to disk age (see Equation \ref{surf_dens}). In this work, however, a single inferred age is used for each disk in this study as described in Section \ref{modeleddisks}.

The distances to particular disks is another likely source of uncertainty in this work. For example, the disk HD 163296 is located at 100 pc \citep[Gaia DR2,][]{2018A&A...616A...1G} as opposed to the previously determined location of 122 pc \citep{1998A&A...330..145V}. This amount of uncertainty in distance introduces an uncertainty of $\sim 10 \%$ when determining the dust line location. In this work we use the previously derived distance of 122 pc so as to easily compare with previous observations that use this distance. We do, however, calculate the dust line locations for this disk given the updated distance and provide these values in Section \ref{modeleddisks}.

Our modeling is not a complete model of all growth in protoplanetary disks as we model the small particles whose aerodynamic properties give rise to substantial particle drift. Larger particles may form following different processes or may form at earlier times in disk evolution and persist to later stages in the disk lifetime. As there are little observational constraints on larger planetesimal sized particles they are not included in our modeling. We note, however, that our model does not preclude their presence. 

\section{Determining the Disk Outer Edge and the Dust Line Locations}\label{fitroutine}

We increase the accuracy of this model by using a detailed method of empirically deriving the location of the disk dust lines. The most accurate determination of disk radial scale at a given wavelength can be derived through modeling of the interferometric visibilities using a Monte Carlo Markov Chain (MCMC) method. We model multi-wavelength millimeter observations following the fitting routine described below and derive radial distances of disk dust lines with characterized errors. In particular, the quantity of interest is the radius at which the flux falls off steeply. We expect such a radius to exist as previous work has indicated that the continuum emission at each wavelength exhibits a markedly sharp decrease over a very narrow radial range such that $\Delta r/r \leq 0.1$ \citep{2012ApJ...744..162A,2013A&A...557A.133D}. 

Previous work typically models disk continuum emission using either a power-law brightness profile with a sharp cut-off  \citep[e.g.,][]{2008ApJ...678L.133A,2012ApJ...744..162A,2016A&A...586A..99H} or a similarity solution brightness profile that follows from models of viscous accretion disks \citep[e.g.,][]{2010AJ....140..887H,2010ApJ...714.1746I,2011ApJ...742L...5A}. \citet{2017ApJ...845...44T} find that different disks are better described by one of these two options and therefore invokes a flexible surface brightness profile.

As we are most interested in the disk radial scale in which the flux drops off steeply, and not necessarily the shape of the disk brightness profile, we use two different models for the disk surface brightness profile and test their accuracy in finding the disk outer edge. In particular we test both the computationally less intensive power-law profile with a sharp cut-off and the more flexible Nuker surface brightness profile \citep{1995AJ....110.2622L} using a method adapted from \citet{2017ApJ...845...44T}. From a given surface brightness profile we compute visibilities from the model's Fourier transform \citep[for further details see][]{2017ApJ...839...99P}. The model visibilities are sampled at the same spatial frequencies as the simulated or real data. The modeled visibilities are also transformed to account for the disk position angle and inclination which we constrain from the literature (see Table \ref{obs}). 

 \begin{figure}[tbp]
\epsscale{1.17}
\plotone{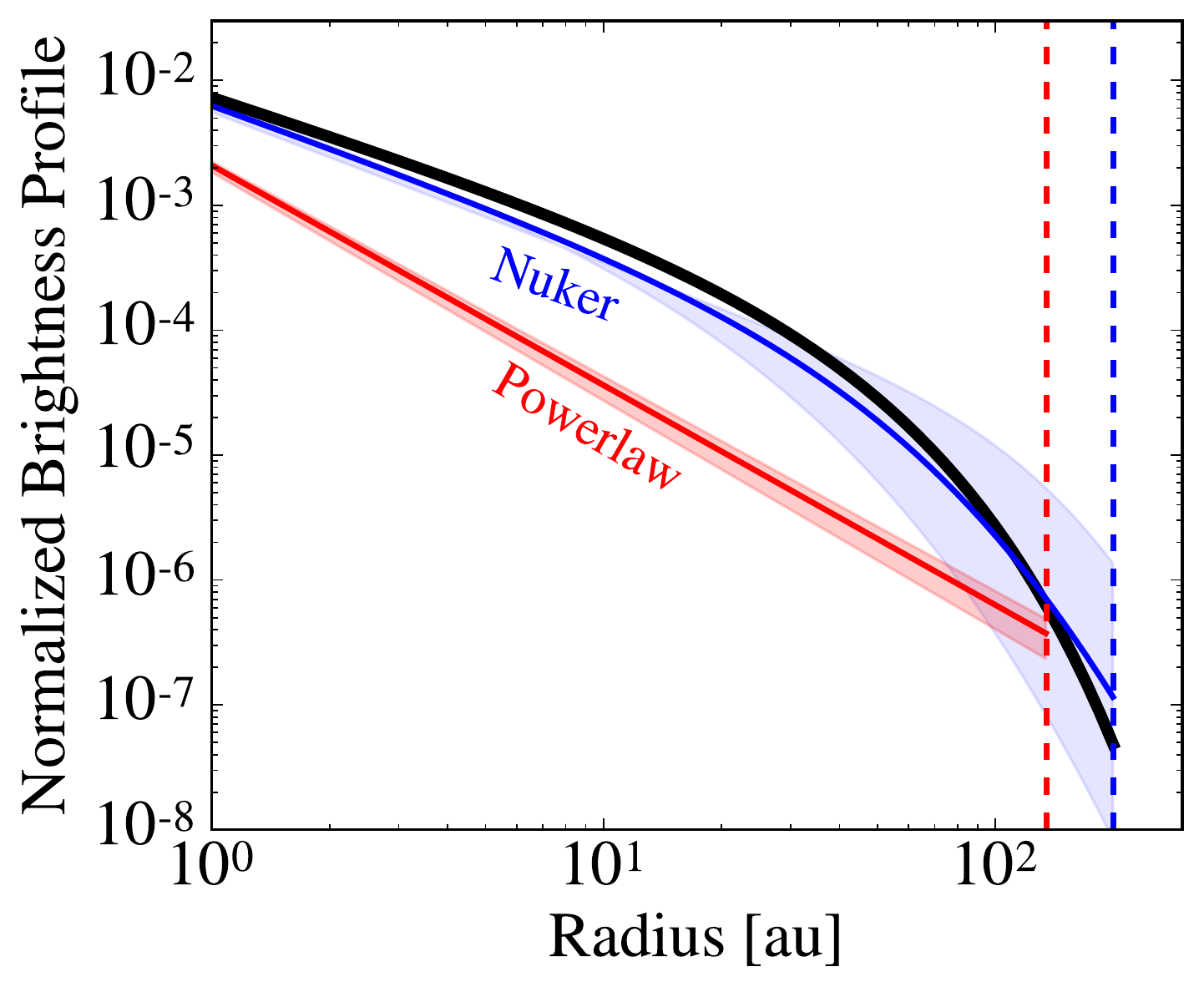}
\caption{The Nuker brightness profile fit (blue, solid line) matches the model brightness profile used to generate simulated ALMA data (black line) well and also finds the correct disk outer edge (blue, dashed line). The power-law brightness profile (red, solid line) does not find the disk outer edge as accurately (red, dashed line). The shaded regions correspond to the one sigma errors for the different profile fitting parameters.}
\end{figure}\label{brightness_prof_fits}

 \begin{figure*}[tbp]
\epsscale{1.}
\plottwo{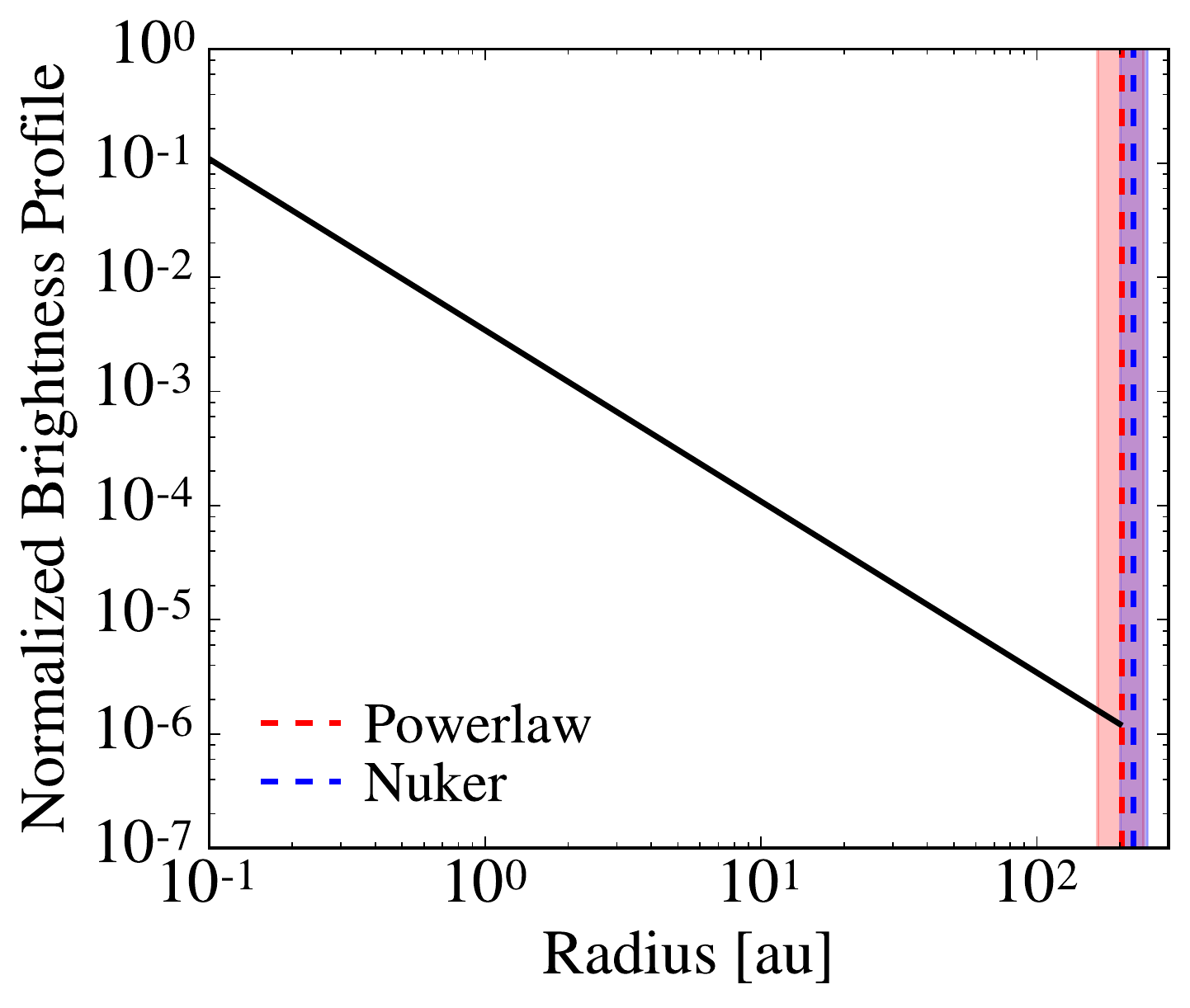}{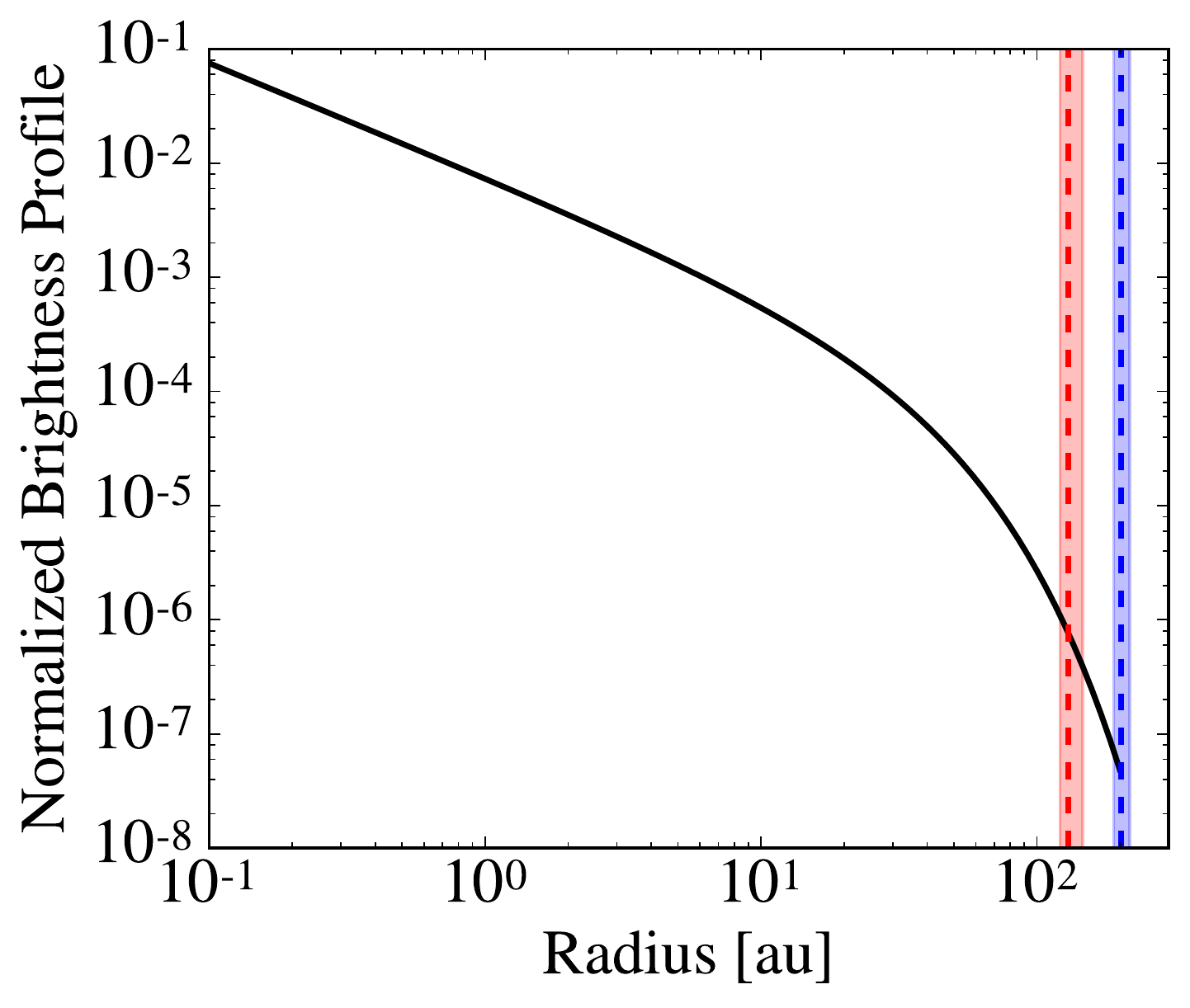}
\plottwo{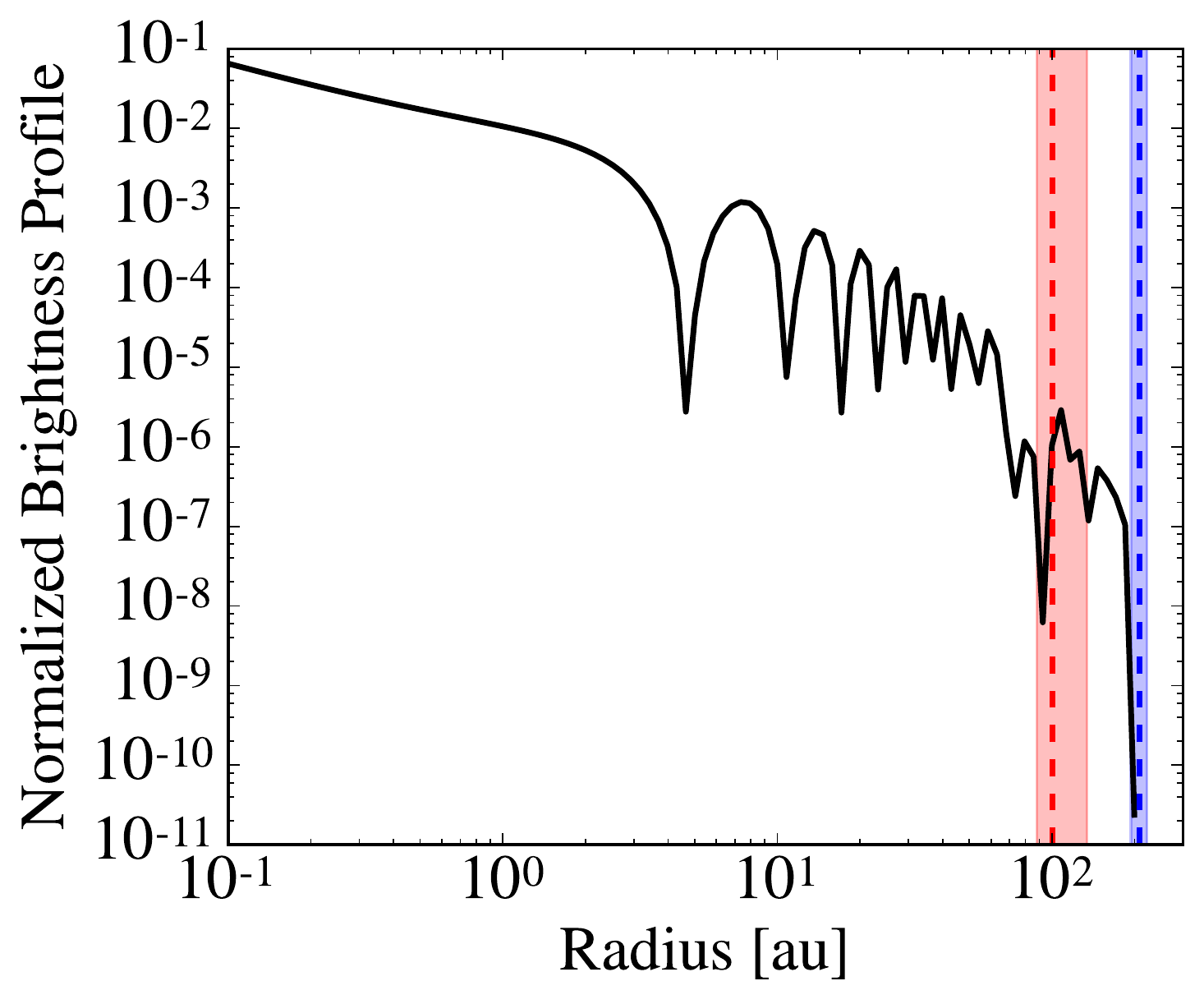}{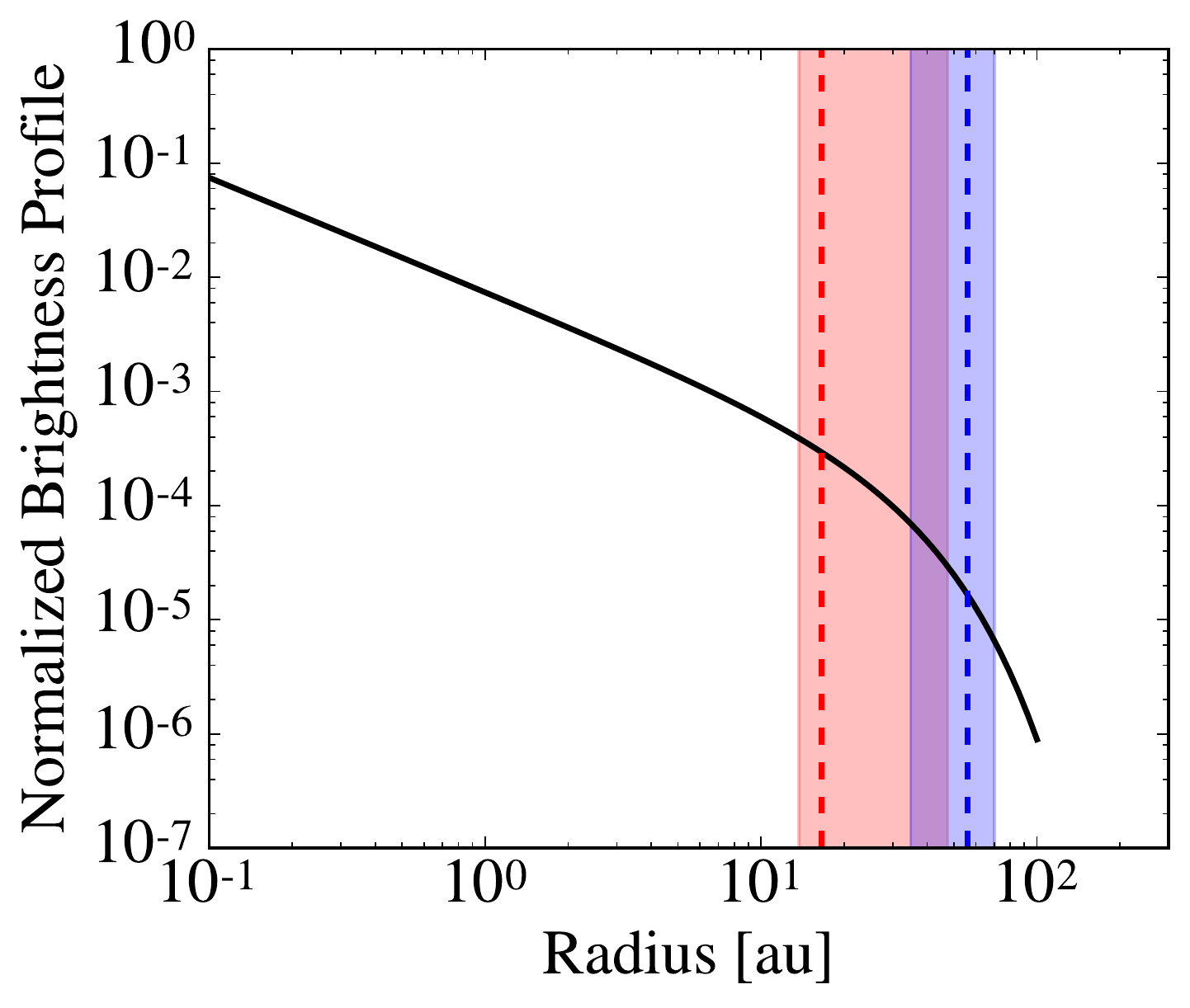}
\caption{The Nuker profile finds the outer radius well (blue, dashed) for several different simulated surface brightness profiles (black) even when there are confounding dips in brightness (bottom left) or a significant taper in brightness (bottom right). The best fit for the outer radius using the power-law brightness profile only finds the outer edge of the disk (red, dashed) more accurately for a simple power law disk brightness profile (top left). The shaded regions corresponding to the colored dashed lines show the one sigma errors for the disk outer edge.}
\end{figure*}\label{diff_tested_disks}

For our initial tests of the method we simulate ALMA data and use an MCMC method to minimize the free parameters in the surface brightness profile which we fit to the data in visibility space. The advantages of fitting to data in visibility space are well described in the literature \citep[e.g.,][]{2013ApJ...762L..21M,2015ApJ...801...59M,2015ApJ...809...47M}. We employ the ensemble sampler proposed in \citet{2010CAMCS...5...65G} and implemented it as described in \citet{2013PASP..125..306F}. To simulate ALMA observations we first create model FITS images from a known surface brightness profile and then derive simulated noise-free interferometric visibilities using the CASA software package \citep{2007ASPC..376..127M}\footnote{In the mock observations, we chose a configuration with baselines ranging from 50 m to 4 km. When fitting, we binned the visibilities in 40 k$\lambda$ sized bins, a common bin size for observations at 1.3 mm \citep[e.g.,][]{2015ApJ...813...41P}. With inflated error bars we derive similar fits to the simulated data.}. 

The power-law brightness profile with a sharp cut-off (see Figure \ref{brightness_prof_fits}) is the least computationally intensive of the two profiles as it only has three free parameters. The power-law visibility profile has the following form:

\begin{equation}
I_\nu(\varrho) \propto \left(\frac{\varrho}{\varrho_0}\right)^{-\gamma}, 
\end{equation}

\noindent where $\varrho$ is the radial coordinate projected on the sky, $\gamma$ is the disk index and $\varrho_0$ is a reference radial location which we set to 10 au. The power-law profile has three free parameters assuming that the position angle and inclination are well-constrained in the literature: the total flux $F_\text{tot}$, $\gamma$, and $R_\text{out}$. All three parameters have uniform priors in linear space such that: $p(F_\nu) = \mathcal{U}$(0, 10 Jy), $p(\gamma) = \mathcal{U}$(-3, 3), and $p(R_\text{out}) = \mathcal{U}$(0, 300 au).  

The Nuker profile introduces several free parameters and is well suited for approximating the behavior of both a power-law disk with a sharp cut-off and a disk with an exponential fall-off (see Figure \ref{brightness_prof_fits}). We therefore additionally choose this model to test as it has a comparable number of free parameters as a similarity solution brightness profile but is more flexible and well-suited for modeling multiple disks with a range in profile shapes. This profile takes the following form:

\begin{equation}
I_\nu(\varrho) \propto \left(\frac{\varrho}{\varrho_t}\right)^{-\gamma}\left[1+\left(\frac{\varrho}{\varrho_t}\right)^\alpha\right]^{(\gamma-\beta)/\alpha}, 
\end{equation}

\noindent where $\varrho_t$ is the transition radius, $\gamma$ is the inner disk index, $\beta$ is the outer disk index, and $\alpha$ is the transition index. The Nuker profile is a flexible brightness profile such that when $\varrho \ll \varrho_t$ or $\varrho \gg \varrho_t$ the brightness profile scales as $\varrho^{-\gamma}$ or $\varrho^{-\beta}$ respectively. The index $\alpha$ controls the asymptotic behavior and when $\alpha$ is small the profile behaves like a similarity solution brightness profile. The behavior of this profile is discussed in more detail in \citet{2017ApJ...845...44T}.

\begin{figure}[tbp]
\epsscale{1.1}
\plotone{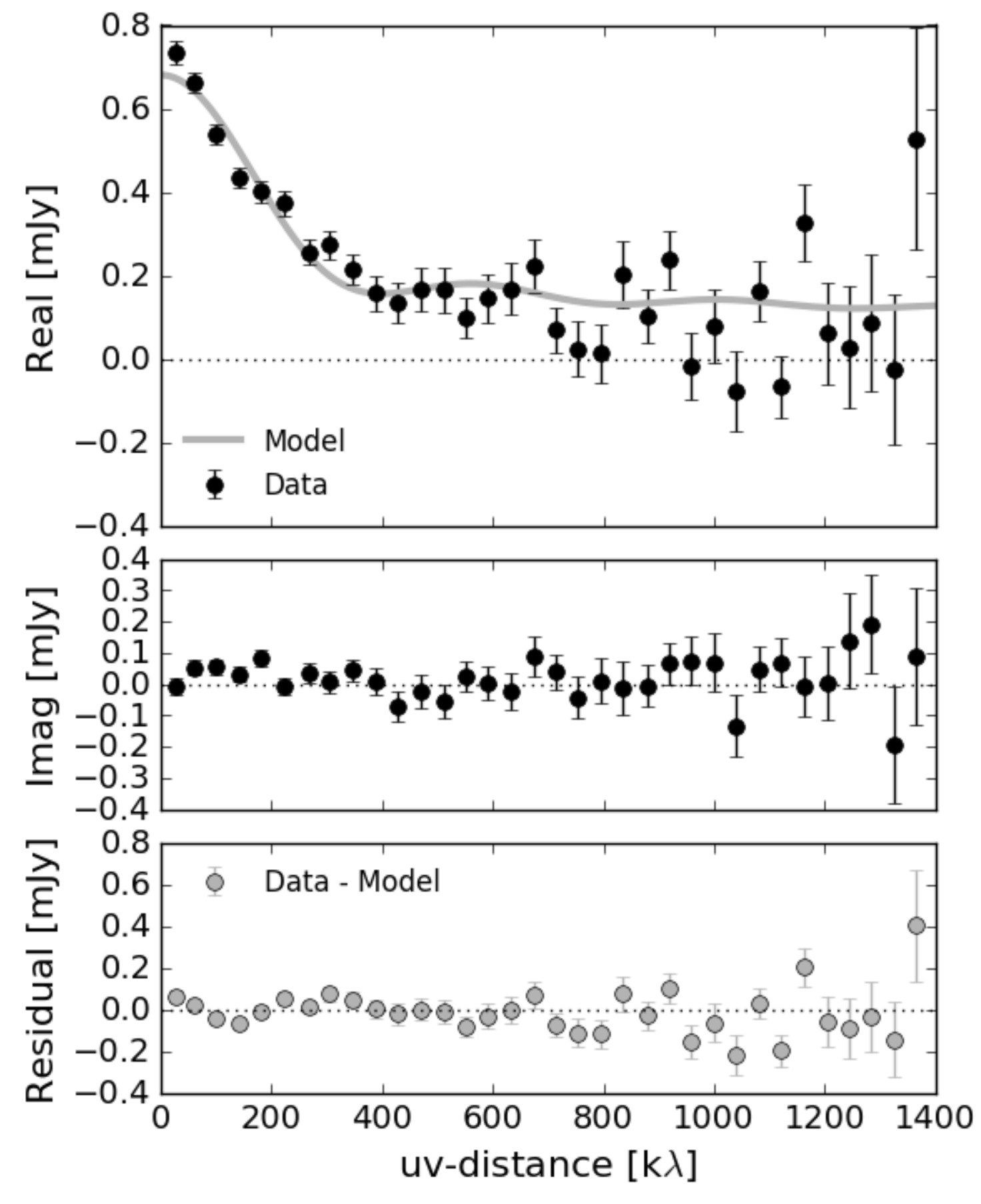}
\caption{Modeled interferometric visibility data for the disk AS 209. The data was taken using the VLA at 10 mm and was originally presented in \citet{2012ApJ...760L..17P}. The top panel shows the modeled real component of the visibilities, the middle panel shows the imaginary component of the visibilities which the model assumes to be zero, and the residuals of the real visibilities and the model are shown in the bottom panel.}
\end{figure}\label{example_vis}

The Nuker profile has 6 free parameters: $F_\text{tot}$, $\gamma$, $\beta$, $\alpha$, $\varrho_t$, and $R_\text{out}$. The priors on these parameters are uniform in linear space except for $\alpha$ which has a log-uniform prior as most of the variation occurs in the first decade of prior space. The priors on these values are: $p(F_\nu) = \mathcal{U}$(0, 10 Jy), $p(\varrho_t) = \mathcal{U}$(0, 300 au), $p(R_\text{out}) = \mathcal{U}$(0, 300 au), $p(\gamma) = \mathcal{U}$(-3, 3), $p(\beta) = \mathcal{U}$(2, 10), $p(\log_{10}\alpha) = \mathcal{U}$(0, 2). These priors are set following \citet{2017ApJ...845...44T}, however, we neglect for the time being the logistic tapers in the prior on $\gamma$ which we find to have a negligible effect on fitting the disks in our current sample. For both $\varrho_t$ and $R_\text{out}$ we convert distances in au to the projected radial location on the sky.

Because we are interested in the outer edge of the disk where we assume that the emission falls of steeply, we prescribe an outer radius as a free parameter in our model brightness profiles that is defined as the location where disk emission is roughly zero. We therefore do not define an effective disk size metric as defined in \citet{2017ApJ...845...44T} where they prescribe an effective disk size that encompasses 68 \% of the total flux. As an assumed effective radius depends strongly on the strength of the inner disk index \citep{2017ApJ...845...44T} this metric will not likely determine the dust line location. For example, a disk that is very bright within a few AU of the star as compared to the outer disk will have a 95 \% or 98 \% flux threshold outer radius that is not the true outer disk radius of interest. By instead assuming that there is a true outer radius we introduce comparatively larger errors primarily because the observations have the poorest sensitivity at larger radial coordinates. A comparison between R$_\text{out}$ derived in this work and R$_\text{eff}$ described in \citet{2017ApJ...845...44T} is discussed in Section \ref{modeleddisks}. 

A comparison between the Nuker and power-law brightness profiles is shown for a simulated ALMA observation in Figure \ref{brightness_prof_fits}. The Nuker profile does a better job of constraining the shape of the surface brightness profile as well as the location of the disk outer edge. Furthermore, the Nuker brightness profile more consistently finds the outer edge of disk emission accurately even when there are confounding dips in brightness or when there is a width/taper to the emission cutoff. This is shown for several simulated disk brightness profiles in Figure \ref{diff_tested_disks} where the outer edge derived using the Nuker and power-law brightness profiles is shown. In each case the fit using the Nuker brightness profile accurately finds the location of the disk outer edge when the surface brightness falls off quickly and approximates the disk outer edge relatively well when there is a significant taper to the disk emission profile. This is encouraging as we expect the decrease in millimeter emission to be distinct, as indicated by observations.

We generally find that both brightness profiles used for fitting disk visibilities are sensitive to the initial guesses used for the parameters. In particular, the most robust methodology for determining the disk outer radius in our tests is to first fit using a power-law surface brightness profile with a sharp cut-off and then use the best fit parameters as the initial guesses for a longer parameter space search using the Nuker brightness profile. 

Our method for modeling the disk visibilities is able to describe real data well. An example of an observed visibility profile and best-fit model is shown in Figure \ref{example_vis} for the disk AS 209 where we find good agreement with the data from \citet{2012ApJ...760L..17P}, even out to large radial coordinates. This fit is characteristic for the objects in our sample. 

\section{Archival Data}\label{data}

We analyze multi-wavelength millimeter observations of 6 disks: AS 209, HD 163296, FT Tau, DR Tau, DoAr 25, and CY Tau. These disks were chosen from the literature as they all have relatively recent resolved millimeter observations at more than two wavelengths (see Table \ref{obs}). These objects also all have published reduced complete visibility profiles readily available for this modeling work. We also update our model analysis of the disk TW Hya. While TW Hya has an abundance of data, the completely reduced interferometric visibilities at the relevant wavelengths are not always provided in the literature. We therefore model this disk using the dust lines as derived in \citet{2017ApJ...840...93P}. These are not the only disks that seem to have decreasing radial extent at longer wavelengths; however, we aim to provide the tools in this work such that the community at large will be able to reproduce our disk models with their own data.

The archival observations of these disks are detailed in Table \ref{obs}. When available we compare our derived surface density profile to the profile derived from detailed modeling of both integrated dust emission and resolved CO emission. When surface density profiles from integrated dust emission are used for comparison we compare to a single surface density profile derived via multi-wavelength millimeter continuum observations. All disks besides DoAr 25 have a single surface density profile derived via multi-wavelength dust observations or resolved CO emission with which to compare. 

Disks with surface density profiles from resolved CO emission are AS 209 \citep{2016ApJ...823L..18H}, HD 163296 \citep{2016ApJ...830...32W}, and TW Hya \citep{2012ApJ...757..129R}. Disks with surface density profiles from multi-wavelength continuum observations are FT Tau \citep{2016AA...588A..53T}, DR Tau \citep{2016AA...588A..53T}, CY Tau \citep{2011AA...529A.105G}, AS 209 \citep{2016AA...588A..53T}, and HD 163296 \citep{2016AA...588A.112G}. For DoAr 25 the only surface density profiles from integrated dust emission available are from individual observed millimeter wavelengths. We therefore primarily consider a surface density profile calculated by fitting a similarity solution, which comes from models of viscously evolving disks, to our derived surface density points (see Equation (\ref{similarity}), parameters shown in Table \ref{parameters}). As shown in Section \ref{modeleddisks}, this surface density profile fit happens to be nearly identical in shape to the dust surface density profile inferred for this disk from observations at 2.8 mm. 

\bigskip

\begin{deluxetable*}{lll} 
\tablecolumns{4}
\tablecaption{Archival Observations \label{obs}}
\tablehead{   
  \colhead{Object} &
  \colhead{Millimeter Dust Observations} &
  \colhead{CO and Other Relevant Observations} 
}
\startdata
AS 209     & \citet{2012ApJ...760L..17P} & \citet{2016ApJ...823L..18H}; \citet{2016AA...588A..53T} \\  
HD 163296 & \citet{2016AA...588A.112G}  & \citet{2016ApJ...830...32W}; \citet{2011ApJ...740...84Q}; \citet{2007AA...469..213I} \\
FT Tau   & \citet{2016AA...588A..53T} & \citet{2014AA...567A.141G} \\
CY Tau     &  \citet{2015ApJ...813...41P} & \citet{2011AA...529A.105G} \\
DR Tau    &  \citet{2016AA...588A..53T} &  \\
DoAr 25    &  \citet{2015ApJ...813...41P} & \citet{2008ApJ...678L.133A} \\
\hline
TW Hya    &  \citet{2012ApJ...744..162A,2016ApJ...820L..40A}; & \citet{2012ApJ...757..129R} \\
  &  \citet{2015ApJ...799..204C}; \citet{2014AA...564A..93M} &  \\
\enddata 
\end{deluxetable*} 

.\\

\begin{deluxetable*}{llll} 
\tablecolumns{5}
\tablecaption{Disk Surface Density and Temperature Profile Parameters \label{parameters}}
\tablehead{   
  \colhead{Object} &
  \colhead{R$_\text{crit}$ [au]} &
  \colhead{$\gamma$} &
  \colhead{Reference}
}
\startdata
AS 209     &  98 &0.91 & (1), integrated dust emission \\  
      & 100 & 1 & (2), CO emission (neglecting ring)\\  
HD 163296  & 119 & 0.88 & (3), integrated dust emission \\
 & 213 & 0.39 & (4), CO emission \\
FT Tau    & 28 & 1.07 & (1), integrated dust emission \\
CY Tau      & 65.6 & 0.28 & (5), integrated dust emission \\
DR Tau     & 20 & 1.07 & (1), integrated dust emission  \\
DoAr 25     & 105 & 0.36 & \\
\hline
TW Hya   & 30 & 1 & (6), CO emission
\enddata 
\tablerefs{(1) \citet{2016AA...588A..53T}, (2) \citet{2016ApJ...823L..18H}, (3) \citet{2016AA...588A.112G}, (4) \citet{2016ApJ...830...32W}, (5) \citet{2011AA...529A.105G}, (6) \citet{2012ApJ...757..129R}}
\end{deluxetable*}

The parameters for the observationally derived surface density profiles used to model these disk are given in Table \ref{parameters}. Every surface density profile that we use in our modeling follows from the self-similar solution to the viscous equations as shown in \citet{1974MNRAS.168..603L} and \citet{1998ApJ...495..385H}. However, the exact functional form of this profile varies in the literature according to the author's preference. For readability, we convert all surface density profiles to our preferred form:  

\begin{equation}\label{similarity}
\Sigma_g(r) = \Sigma_0\left(\frac{r}{r_c}\right)^{-\gamma}\exp\left[-\left(\frac{r}{r_c}\right)^{2-\gamma}\right]
\end{equation} 

\noindent where this similarity solution profile is a shallow power law at small radii and follows an exponential fall off, goverened by the parameter $\gamma$ at radii larger than the critical radius, $r_c$. For the disk DoAr 25, which does not have a similarity solution profile derived from simultaneous modeling of multiwavlength data or CO emission, we use $\chi^2$ minimization to derive a similarity solution profile that fits the derived surface densities at the disk dust lines well. In our modeling we derive new values of $\Sigma_0$, is the surface density profile normalization, which are given in Table \ref{sigma_crit}.

\section{Modeled Disks}\label{modeleddisks}

We first use the method detailed in Section \ref{fitroutine} to determine the location of the disk outer edge at each observed wavelength (the disk dust line) for the six disks that we consider. These radii are listed in Table \ref{outer_radius} where the wavelengths quoted are from the original published archival data. The dust lines for the disk TW Hya are taken directly from \citet{2017ApJ...840...93P}. The stellar luminosity, and stellar mass used in our modeling are also noted. 

\begin{figure*}[tbp]
\epsscale{.8}
\plottwo{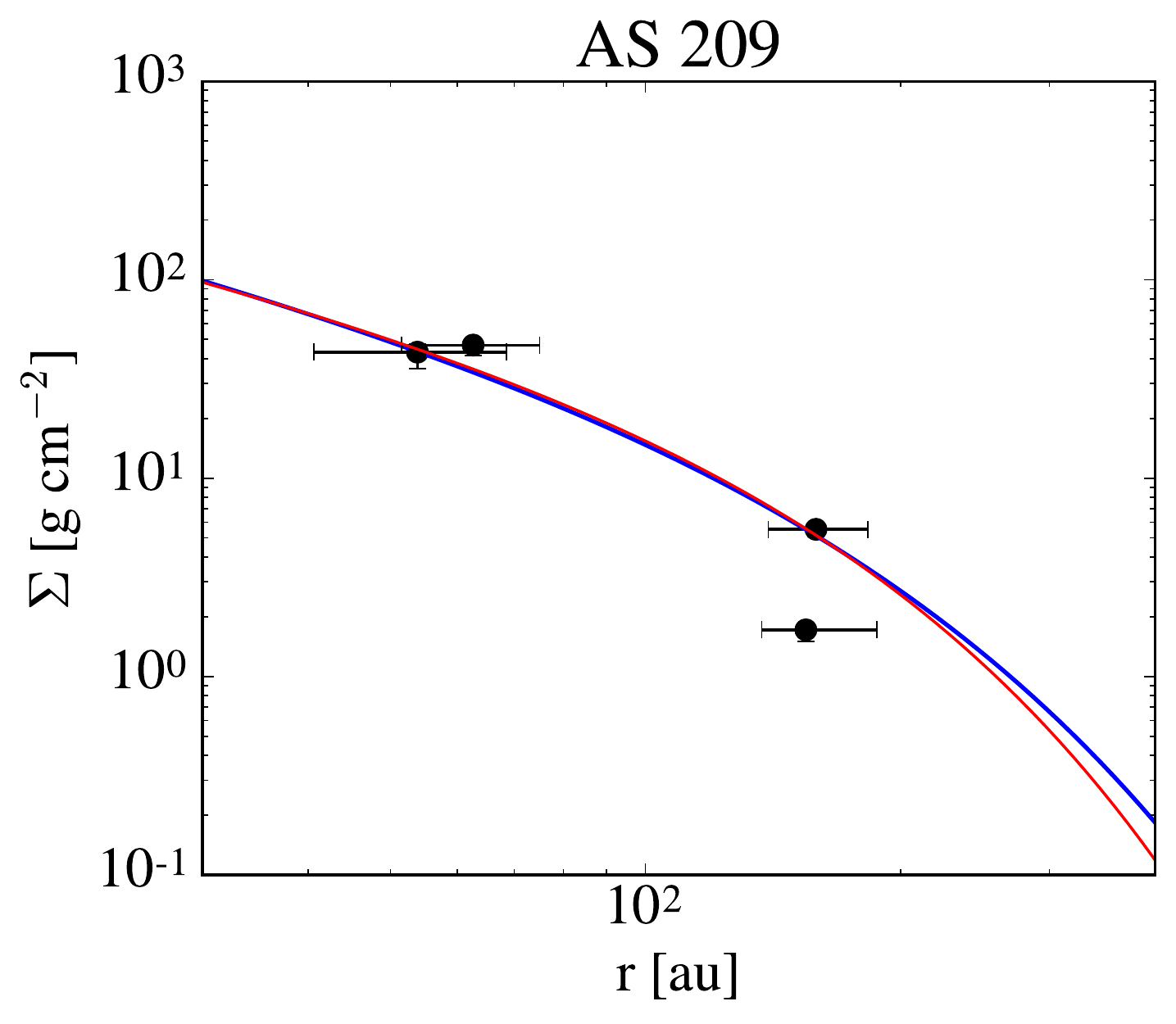}{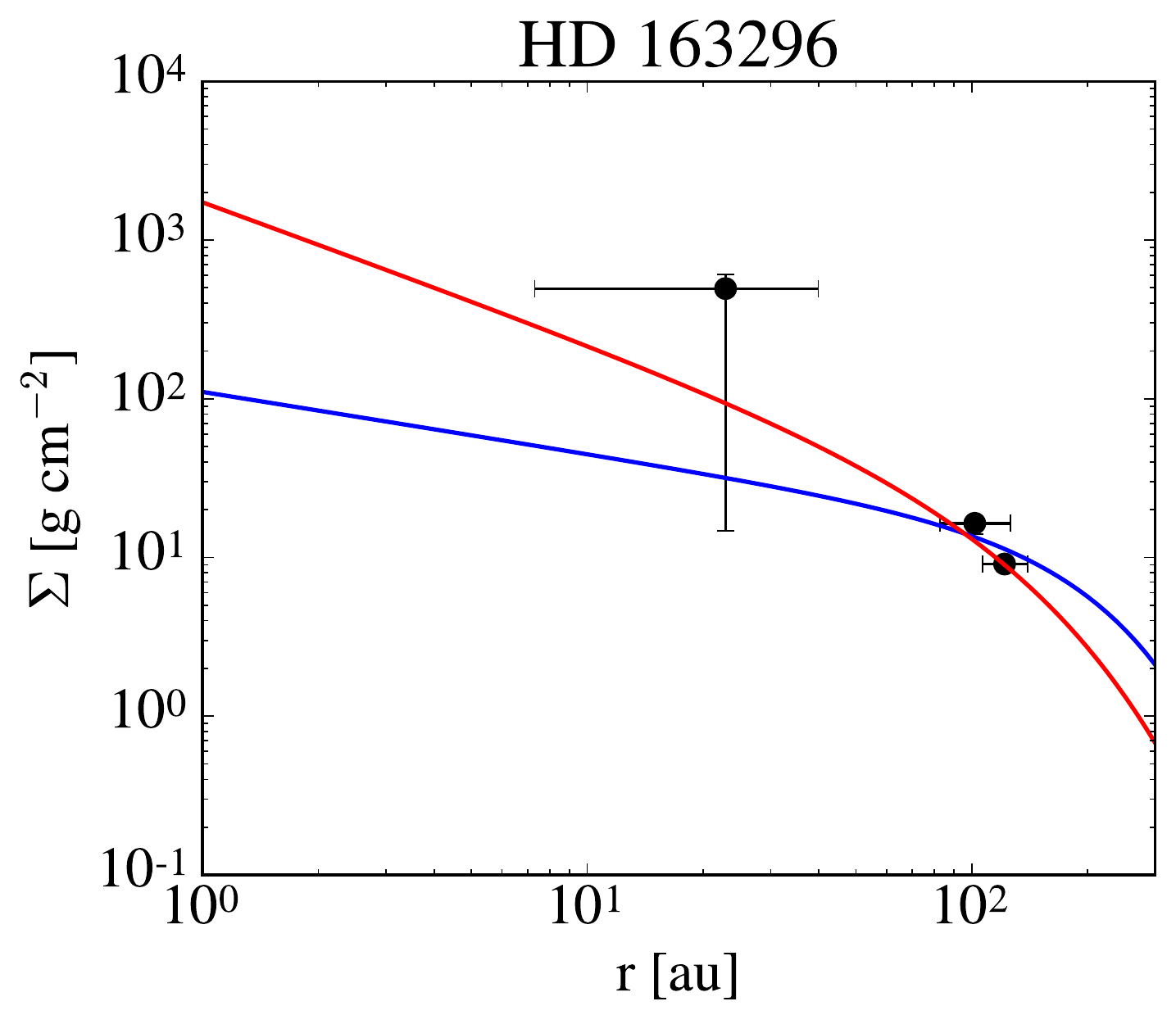}
\plottwo{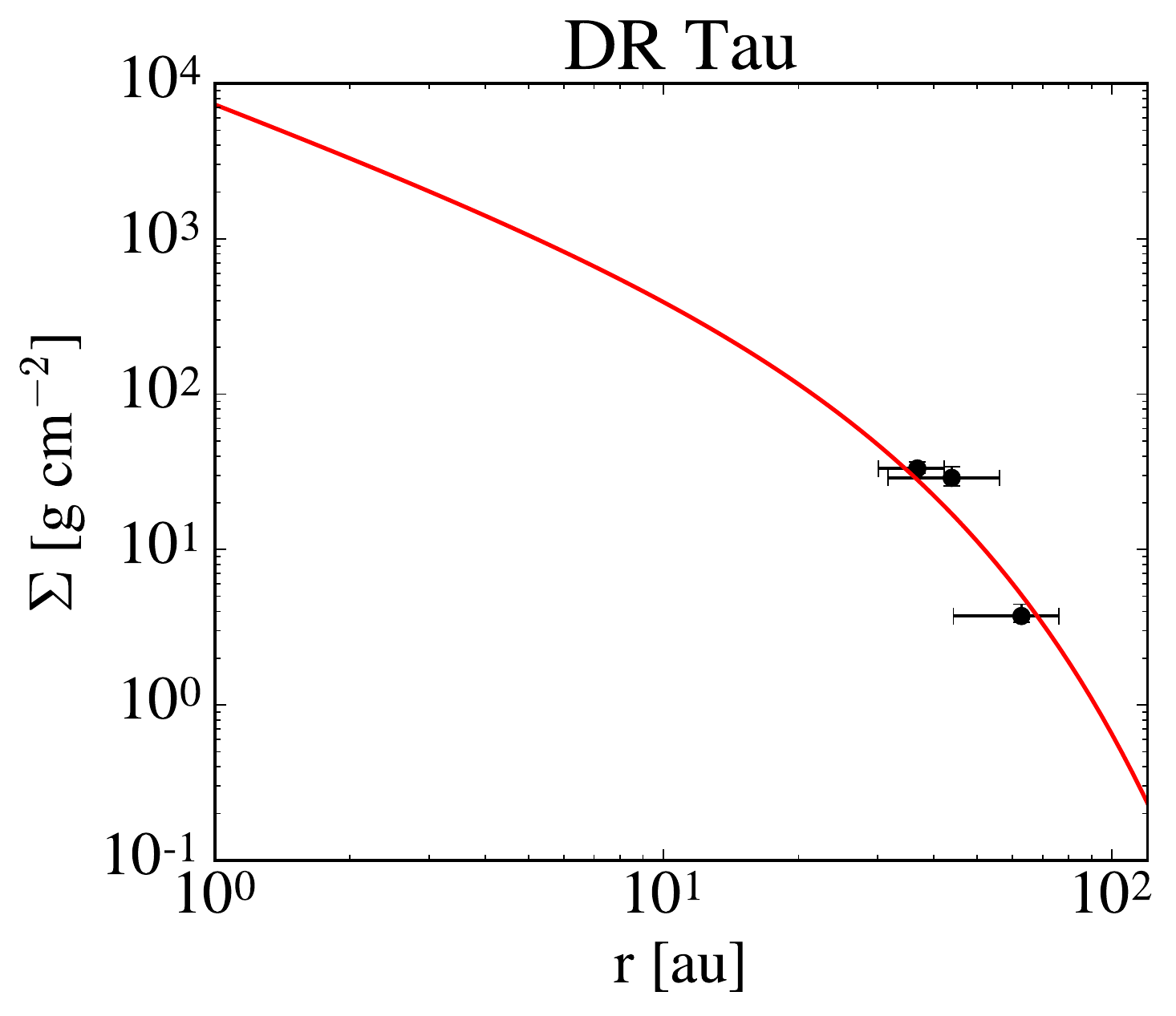}{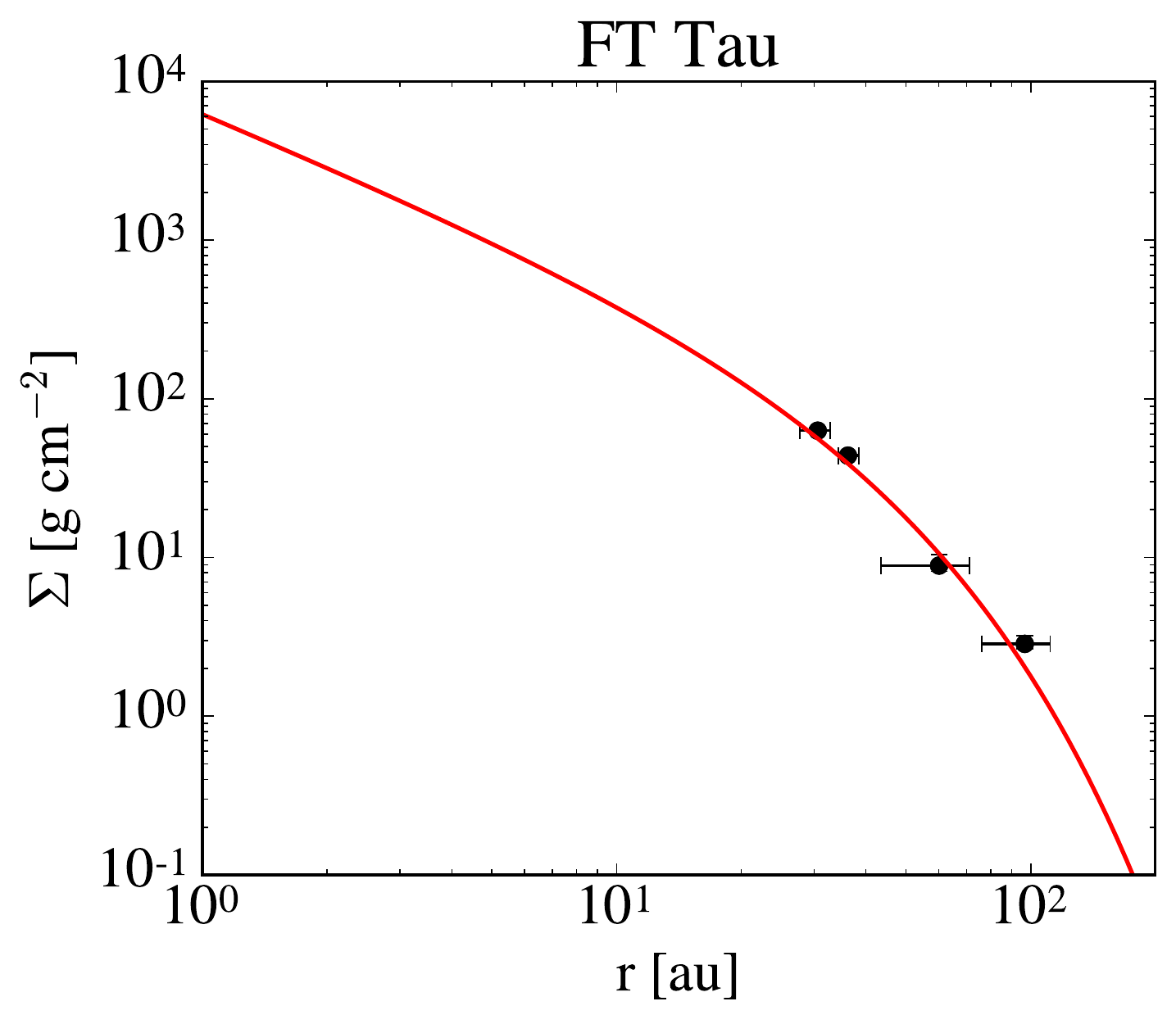}
\plottwo{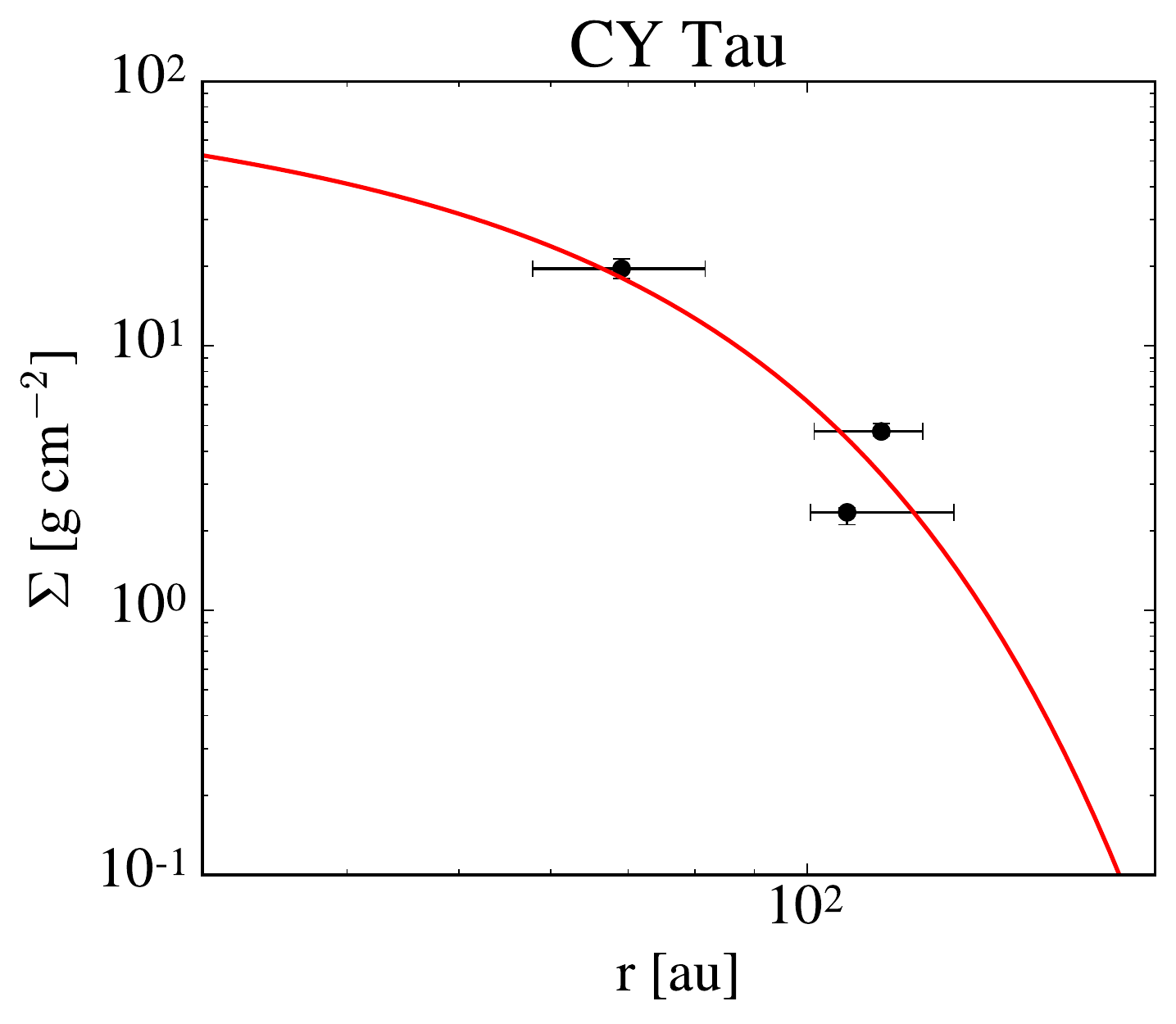}{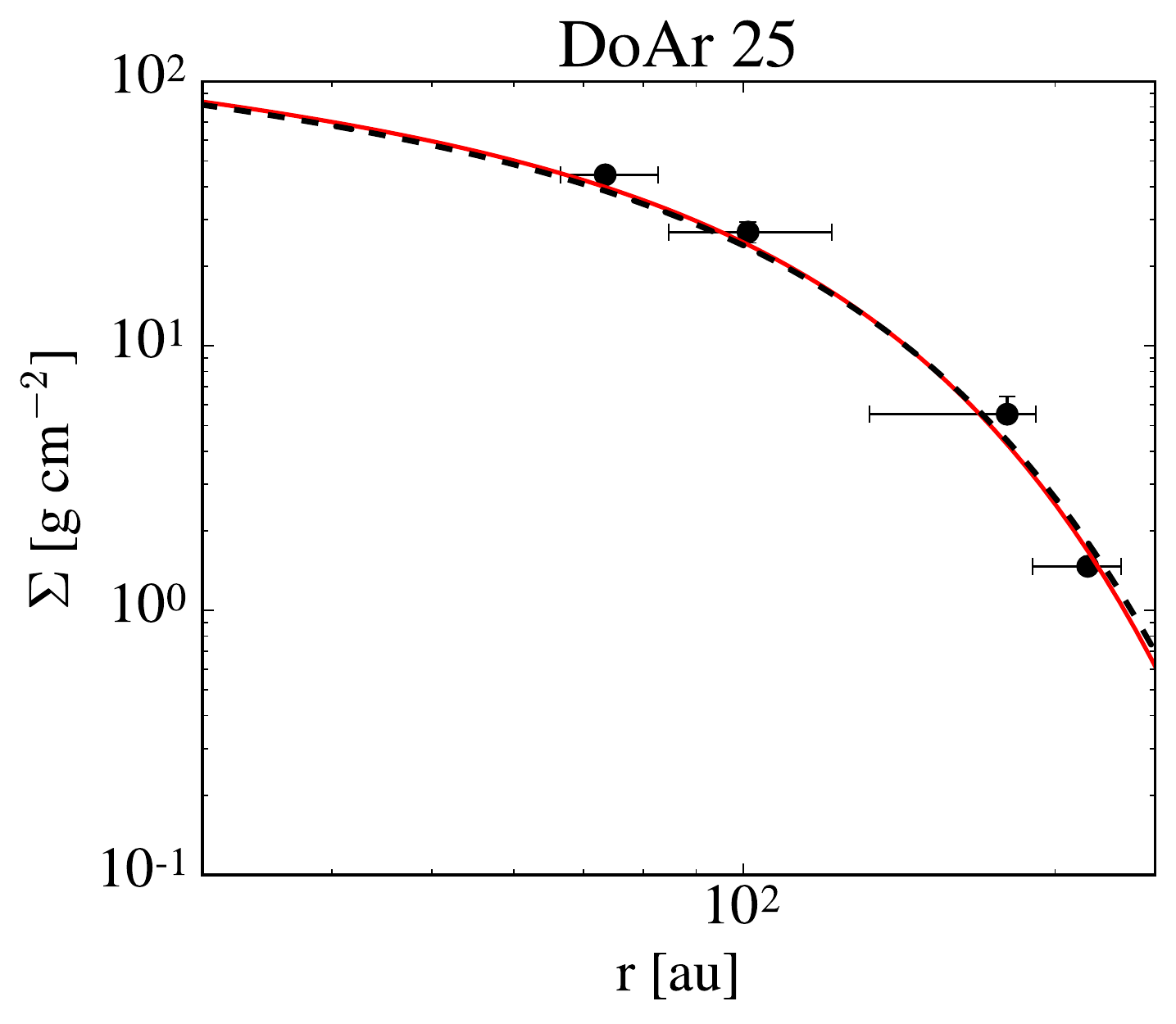}
\epsscale{0.42}
\plotone{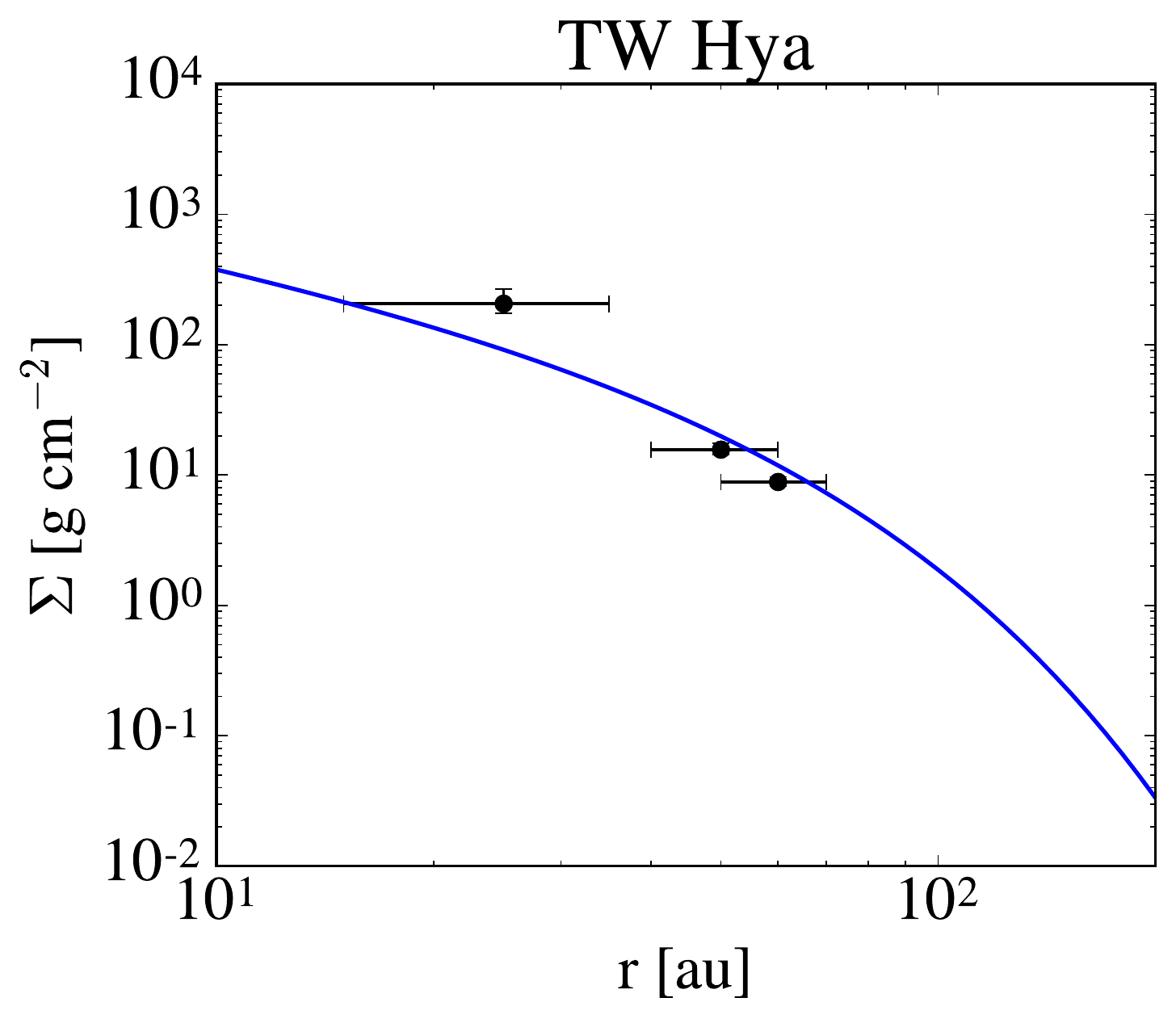}
\caption{Our newly derived surface density values (black points) can be well matched by renormalized surface density profiles derived from either multiwavelength dust observations (red lines) or CO observations (blue lines). These plots are provided for a range larger than is probed by the observations to provide an idea of the general shape and scale of these systems. The plotted radii vary for each disk because the surface density profiles have different shapes and scales. The disk DoAr 25 does not have one single published surface density profile derived from multiwavelength millimeter observations or from CO emission. Instead, the best fit surface density profile for the newly derived surface densities (black, dashed line) is shown as well as the normalized dust surface density profile from observations of the disk at 2.8 mm (red line).}\label{nuker_fits}
\end{figure*}

Using the information in Table \ref{outer_radius} we derive total disk surface densities at the location of disk dust lines as shown in Figure \ref{nuker_fits}. We then renormalize the disk surface density profiles based on either multiwavelength dust emission or resolved CO emission, as described in Table \ref{parameters}, to derive the new disk surface density profile. The newly determined values of $\Sigma_0$ are given in Table \ref{sigma_crit} along with the constant that determines the disk temperature profile (see Equation \ref{disk_temperature}). 

For DoAr 25 we show both the derived fit to the new surface density values and the comparison to the dust surface density profile derived from modeling dust emission at the one wavelength that best matches the newly derived values. In each case we find good agreement in shape between the empirically derived surface density points and the surface density profiles derived from other observational methods. For disks with profiles derived from both CO emission and integrated dust observations, we typically choose the dust surface density profile as the canonical surface density profile in which to determine disk properties. This is because, as discussed below, the mass derived from dust is more consistent with our newly derived masses. However, for the disk TW Hya we primarily consider the renormalized profile derived from CO observations as there is not currently a published profile derived based on the simultaneous fitting of multiwavelength millimeter dust observations for this object. 

\begin{deluxetable*}{lllll}
\tablecolumns{6}
\tablecaption{Dust Lines \label{outer_radius}}
\tablehead{   
  \colhead{Object} &
  \colhead{Stellar Lum.} &
  \colhead{Stellar Mass} &
  \colhead{Observed Wavelength} &
  \colhead{Outer Radius/Dust Line} 
}
\startdata
AS 209\tablenotemark{a} &  1.5 L$_\sun$ & 0.9 M$_\sun$  & 0.85 mm & 154.7$^{+17.5}_{-32.8}$ au \\ 
  & & & 2.8 mm & 159$^{+19.3}_{-24.1}$ au \\   
  & &  & 8 mm & 53.8$^{+13.2}_{-14.7}$ au \\   
  & &  & 10 mm & 62.6$^{+11.1}_{-12.4}$ au \\   
HD 163296\tablenotemark{b}  &  36 L$_\sun$ & 2.3 M$_\sun$ & 0.85 mm  & 121.7$^{+15.2}_{-18}$ au  \\
  & & & 1.3 mm  & 101.8$^{+19.1}_{-24.4}$ au  \\
  & & & 9.8 mm  & 22.9$^{+15.6}_{-17}$ au  \\
FT Tau\tablenotemark{c}   &  0.31 L$_\sun$ & 0.55 M$_\sun$ & 1.3 mm & 96.8$^{+14.7}_{-20.5}$ au \\
  & & & 2.6 mm & 60.1$^{+11}_{-16.6}$ au \\
  & & & 8.0 mm & 36.2$^{+2.3}_{-1.9}$ au \\
  & & & 9.83 mm & 30.6$^{+2.2}_{-2.9}$ au \\
CY Tau\tablenotemark{d}  &  0.4 L$_\sun$ & 0.48 M$_\sun$  & 1.3 mm & 108.3$^{+25.7}_{-7.6}$ au \\
  & &  & 2.8 mm & 115.9$^{+10}_{-14.5}$ au \\
  & &   & 7.14 mm & 69.1$^{+12.5}_{-11.5}$ au \\
DR Tau\tablenotemark{c}  &  1.09 L$_\sun$ & 0.8 M$_\sun$   & 1.3 mm & 62.8$^{+13.2}_{-18.5}$ au \\
  & &  & 7.05 mm & 36.8$^{+5.4}_{-6.7}$ au \\
  & &  & 7.22 mm & 43.9$^{+12.3}_{-12.2}$ au \\
DoAr 25\tablenotemark{f}  &  1.3 L$_\sun$ & 1 M$_\sun$  & 0.88 mm & 215.1$^{+16.7}_{-24.9}$ au \\
  & & & 2.8 mm & 179.8$^{+11.9}_{-47.5}$ au \\
  & & & 8 mm & 101$^{+20.7}_{-16.3}$ au \\
  & &  & 9.8 mm & 73.5$^{+9.2}_{-7}$ au \\
 \hline
TW Hya\tablenotemark{g} &  0.28 L$_\sun$ & 0.8 M$_\sun$ & 0.87 mm  & 60 $\pm$ 10 au \\
  & & & 1.3 mm  & 50 $\pm$ 10 au   \\
  & & & 9 mm  & 25 $\pm$ 10 au \\
\enddata 
\tablerefs{Dust lines are calculated from Archival data as detailed in Table \ref{obs}. Dust lines for TW Hya are from \citet{2017ApJ...840...93P}. Superscript letters denote references for stellar parameters and a discussion and references for stellar ages are given in the text. \tablenotemark{a}\citet{1988cels.book.....H} \tablenotemark{b}\citet{2004AA...416..179N} \tablenotemark{c}\citet{2010AA...512A..15R} \tablenotemark{d}\citet{2007AA...473L..21B} \tablenotemark{f}\citet{2008ApJ...678L.133A} \tablenotemark{g}\citet{2007ApJ...660.1556R,2013Sci...341..630Q}}
\end{deluxetable*}

Integrating the re-normalized surface density profiles allows us to solve for the total disk mass. The total disk masses are given in Table \ref{mass}. The disk masses derived following the method described in this work are given in the first column. Previously published mass estimates using conventional tracer calculations are given in the third and fourth columns. 

Several disks in our sample have age estimates that vary significantly across the literature. We choose a single age for our modeling purposes that is consistent with literature values for each disk as discussed below. The largest range we found in the literature for the stellar age is 7 Myr for TW Hya. We note that this is the most well-studied object in our sample, suggesting that other disks may also have large age uncertainties. We again note that the derived surface density values depend linearly on disk age as shown in Equation (\ref{surf_dens}). The stellar age likely introduces the largest uncertainty in our modeling. In Table \ref{mass}, we provide errors for our mass estimates resulting from the variation of stellar ages quoted in the literature.

\textit{AS 209}

\begin{deluxetable*}{lll} 
\tablecolumns{5}
\tablecaption{Derived Disk Surface Density Profile Normalization and Temperature Constant\label{sigma_crit}}
\tablehead{   
  \colhead{Object} &
  \colhead{T$_0$ [K]} &
  \colhead{$\Sigma_0$ [g cm$^{-2}$]} 
}
\startdata
AS 209    & 131 & 44\tablenotemark{a} \\
    & ``   " & 40\tablenotemark{b} \\
HD 163296 & 284 & 29\tablenotemark{a} \\
 & ``   " & 14\tablenotemark{b}\\
FT Tau & 89 & 183\tablenotemark{a}\\
CY Tau & 98 & 55 \tablenotemark{a}\\
DR Tau & 121 & 315\tablenotemark{a}\\
DoAr 25 & 123 & 68\tablenotemark{a}\\
\hline
TW Hya & 82 & 175\tablenotemark{b} \\
\enddata 
\tablenotetext{a}{For the normalized dust surface density profile.}
\tablenotetext{b}{For the normalized CO surface density profile.}
\end{deluxetable*}

 The most massive disk in our sample is AS 209 with a disk mass that is 27\% as massive as its host star. The newly derived disk mass is a factor $\sim$15 times larger than previous estimates based on dust observations. The new disk mass is a further 115 times larger than the mass derived from CO observations. While both masses are inconsistent with the newly measured mass, the CO derived mass is particularly small. This indicates that the disk AS 209 may be significantly depleted in CO compared to the ISM. AS 209 is a relatively young disk; we choose the commonly quoted age of 1.6 Myr \citep{2009ApJ...700.1502A}, however, age estimates for this system in the literature range as low as 0.5 Myr \citep{2006AA...452..245N,2018AA...610A..24F} which would bring the disk mass estimate down by a factor of $\sim$3.

\textit{DoAr 25}

The disk around DoAr 25 has a mass that is 23\% the mass of its host star and is the second most massive disk in our sample. The newly derived mass is roughly a factor of 8 larger than the mass derived from dust observations. For DoAr 25 we choose an age of 2 Myr as this is consistent with age estimates given for the Ophiuchus star forming region in \citet{2017ApJ...851...83C} though the age estimates in the literature are as high as 4 Myr \citep{2009ApJ...700.1502A}. 

\textit{CY Tau}

The disk around CY Tau has a mass that is 21\% of its host's stellar mass. The newly derived mass is roughly a factor of 6 larger than the mass derived from dust observations.  Age estimates for the disk CY Tau vary from 0.8 Myr to $\sim$ 3 Myr \citep{2007AA...473L..21B,2009ApJ...701..260I,2013ApJ...771..129A,2014A&A...567A.117G}. We choose an age of 1 Myr as this is consistent with the literature \citep[e.g.,][]{2009ApJ...701..260I,2013ApJ...771..129A} and is representative of disk ages in the Taurus star forming region.

 \textit{FT Tau}
 
The disk around FT Tau has a mass that is 18\% of its host's stellar mass. The newly derived mass is roughly a factor of 7 larger than the mass derived from dust observations. For FT Tau we model using the only commonly quoted literature age of 1.6 Myr \citep{2014AA...567A.141G}. 

\textit{TW Hya}

The disk around TW Hya has a mass that is 14\% as massive as its host star. The newly derived mass is roughly a factor of 6 larger than the mass derived from dust observations. The new disk mass is also a factor of 37 larger than the mass derived from CO observations. TW Hya therefore shows moderate depletion of CO at a lower level than for the disk AS 209 but significantly larger than the disk HD 163296. For TW Hya we choose an age of 5 Myr though the literature age estimates range from 3-10 Myr \citep{2006AA...459..511B,2011ApJ...732....8V}. 

\textit{DR Tau}

 \begin{deluxetable*}{ccccccc}
\tablecolumns{7}
\tablecaption{Total Gas Disk Mass \label{mass}}
\tablehead{   
  \colhead{Object} &
  \colhead{Age} &
  \colhead{Dust Line Derived} &
   \colhead{Mass Uncertainty} &
  \colhead{Mass From Integrated} &
  \colhead{Mass From} & 
  \colhead{Mass From} \\
  \colhead{} &
  \colhead{} &
  \colhead{Disk Mass\tablenotemark{a}} &
  \colhead{Based on Disk Age} &
  \colhead{Dust Emission} &
  \colhead{CO Line Emission} &
  \colhead{HD Line Emission} 
}
\startdata
AS 209  &  1.6 Myr   & 0.24 M$_\sun$\tablenotemark{b}  & $- 0.165$ M$_\sun$ & 0.0149 M$_\sun$ & 0.002 M$_\sun$\tablenotemark{a}&  - - - \\  
HD 163296 &  5 Myr & 0.21 M$_\sun$\tablenotemark{b} & $- 0.084$ M$_\sun$ & 0.12 M$_\sun$ & 0.048 M$_\sun$ &  - - -\\
 ``   " &  ``   " & 0.16 M$_\sun$\tablenotemark{c}& $- 0.064$ M$_\sun$ &  ``   " &  ``   " &  - - -\\
FT Tau  &  1.6 Myr  & 0.10 M$_\sun$\tablenotemark{b} & - - - & 0.015 M$_\sun$ &  - - -&  - - -\\
CY Tau   &  1 Myr  & 0.10 M$_\sun$\tablenotemark{b} & $- 0.02$/$+0.2$ M$_\sun$ &0.0165 M$_\sun$ & - - -&  - - -\\
DR Tau   &  1 Myr &  0.09 M$_\sun$\tablenotemark{b}&  $-0.081$/$+0.18$ M$_\sun$ &0.014 M$_\sun$ &  - - -&  - - -\\
DoAr 25  &  2 Myr  &  0.23 M$_\sun$\tablenotemark{b} & $+ 0.23$ M$_\sun$ & 0.03 M$_\sun$ &  - - -&  - - -\\
\hline
TW Hya  &  5 Myr &  0.11 M$_\sun$\tablenotemark{c} & $- 0.044$/$+ 0.11$ M$_\sun$ & 0.018 M$_\sun$ & 0.003 M$\sun$\tablenotemark{a} & $ > 0.05$ M$_\sun$\\
\enddata 
\tablerefs{\citet{2016AA...588A..53T}; \citet{2007AA...469..213I}; \citet{2016ApJ...830...32W}; \citet{2011AA...529A.105G}; \citet{2016ApJ...823L..18H};\\ \citet{2017ApJ...851...83C}; \citet{2014AA...564A..93M}; \citet{2012ApJ...757..129R}; \citet{2013Natur.493..644B}. \\ Mass uncertainty based on disk age is quoted based on age ranges from the literature. A line divides TW Hya from \\the rest of the objects in this sample as the dust lines used in this modeling are taken from \citet{2017ApJ...840...93P}.}
\tablenotetext{a}{Mass derived from integrating the observationally determined surface density profile.}
\tablenotetext{b}{Mass derived from integrating the normalized dust surface density profile.}
\tablenotetext{c}{Mass derived from integrating the normalized CO surface density profile.}
\end{deluxetable*}

The disk around DR Tau has a mass that is 11\% as massive as its host star. The newly derived mass is roughly a factor of 6 larger than the mass derived from dust observations. For the disk DR Tau age estimates range from 0.1 Myr to $\sim$ 3 Myr \citep{2009ApJ...701..260I,2013ApJ...771..129A}. We again choose an age of 1 Myr as this is consistent with many age estimates given in the literature. 

\textit{HD 163296}

 The disk around HD 163296 has a mass that is 9\% of its host's stellar mass and is the third most massive disk in our sample. This disk orbits a Herbig Ae star and is one of the two oldest disks in our sample. HD 163296 has similarity solutions derived using both integrated dust observations and CO emission. We quote newly derived masses considering each of these two profiles in Table \ref{mass}. For the discussion in the text, however, we consider the mass derived via renormalizing the similarity solution profiled derived from integrated dust observations. The newly derived mass estimate is almost a factor of 2 larger than the mass derived from dust observations in \citet{2007AA...469..213I}. We note, however, that integrating the surface density profile derived in \citet{2016AA...588A.112G} from optically thin millimeter dust emission derives a disk mass of $\sim 0.01$ M$_\odot$ (c.f. Figure \ref{zoom_fits}), which is more than an order of magnitude lower than our derived disk masses. The new disk mass is also a factor of 4.5 larger than the mass derived from CO observations. This indicates that HD 163296 may not exhibit as marked a depletion of CO as the other disks in our sample with CO mass estimates. While the disk HD 163296 has age estimates as low as 3 Myr \citep{2017AA...600A..62P}, we choose the common literature age value of 5 Myr for our modeling \citep{1998A&A...330..145V,2009AA...495..901M}. For HD 163296 we also re-derive the dust line locations with the updated Gaia DR2 distance of 100 pc. We find dust lines located at 94$^{+29.7}_{-38.2}$ au, 84$^{+7.1}_{-13.0}$ au, 23.2$^{+9.7}_{-10.7}$ au corresponding to the observed wavelengths of 0.85 mm, 1.3 mm, 9.8 mm respectively. Using these dust lines to derive mass, without updating the inferred stellar parameters, results in a mass estimate of 0.27 M$_\sun$.

 \begin{figure}[tbp]
\plotone{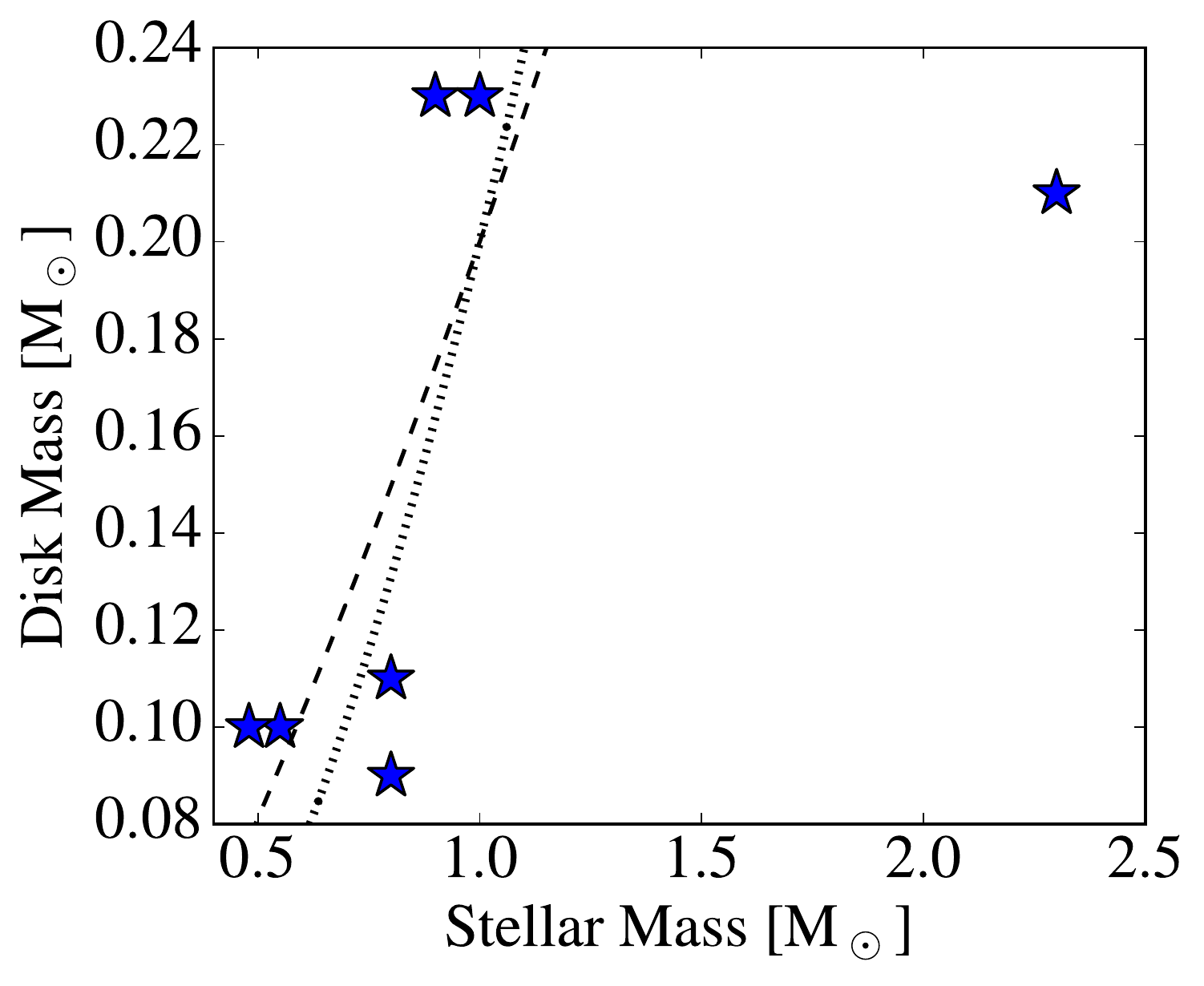}
\caption{The derived disk masses in this sample (blue stars) are consistent with the \citet{2016ApJ...831..125P} steeper than linear scaling relationship between disk dust mass and stellar mass, except for the disk HD 163296 which orbits a massive Herbig Ae star. Our normalized relationship is a factor of 50 larger than the best-fit relationship derived assuming an ISM dust-to-gas ratio in \citet{2016ApJ...831..125P} such that M$_\text{disk}\sim$ 0.2M$_\sun$ (M$_\star$/M$_\sun$)$^{1.3}$ (dashed line) and M$_\text{disk}\sim$ 0.2M$_\sun$ (M$_\star$/M$_\sun$)$^{1.9}$ (dotted line)}
\end{figure}\label{disk_star_mass}

 \begin{figure*}[tbp]
\epsscale{0.8}
\plottwo{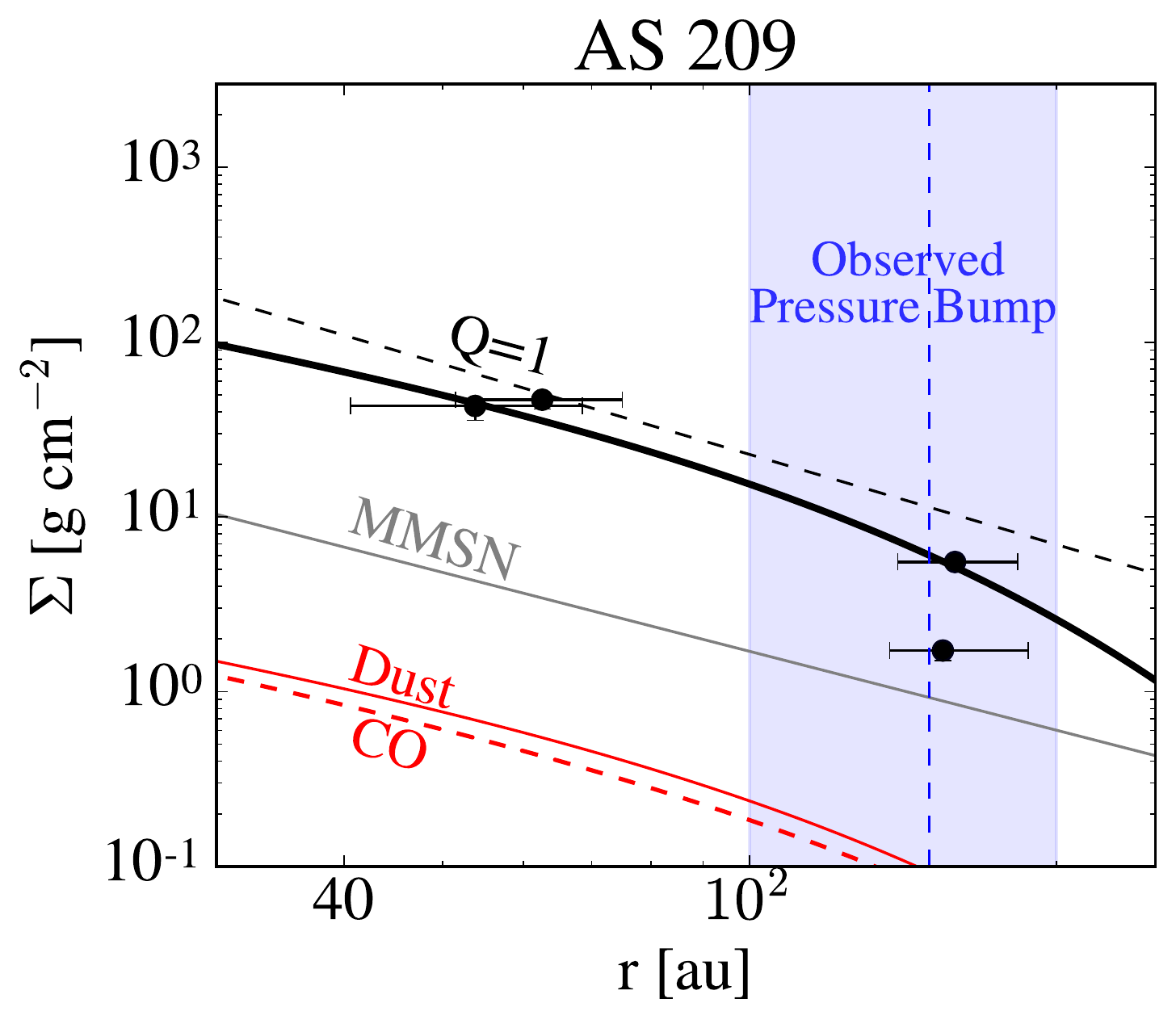}{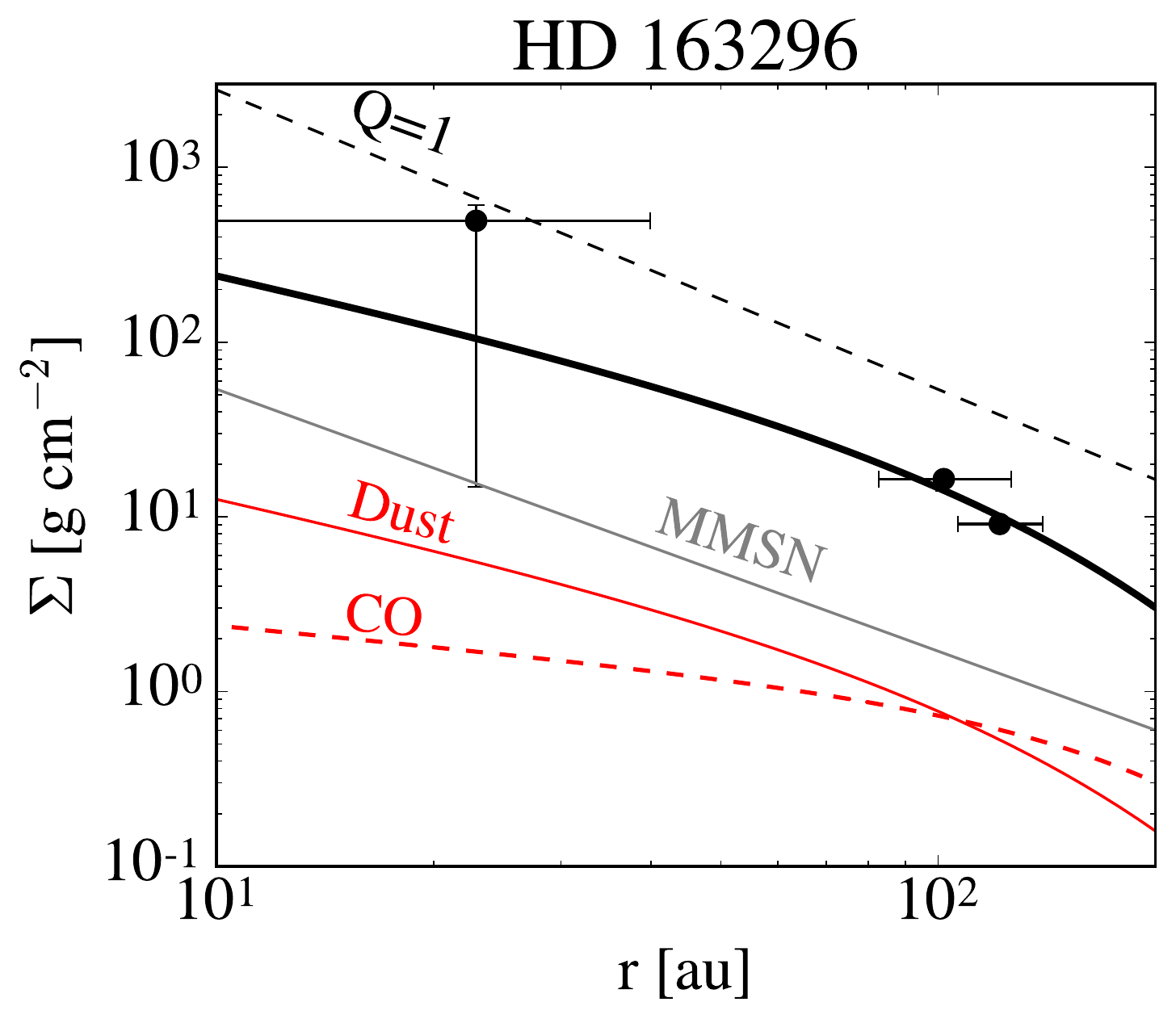}
\plottwo{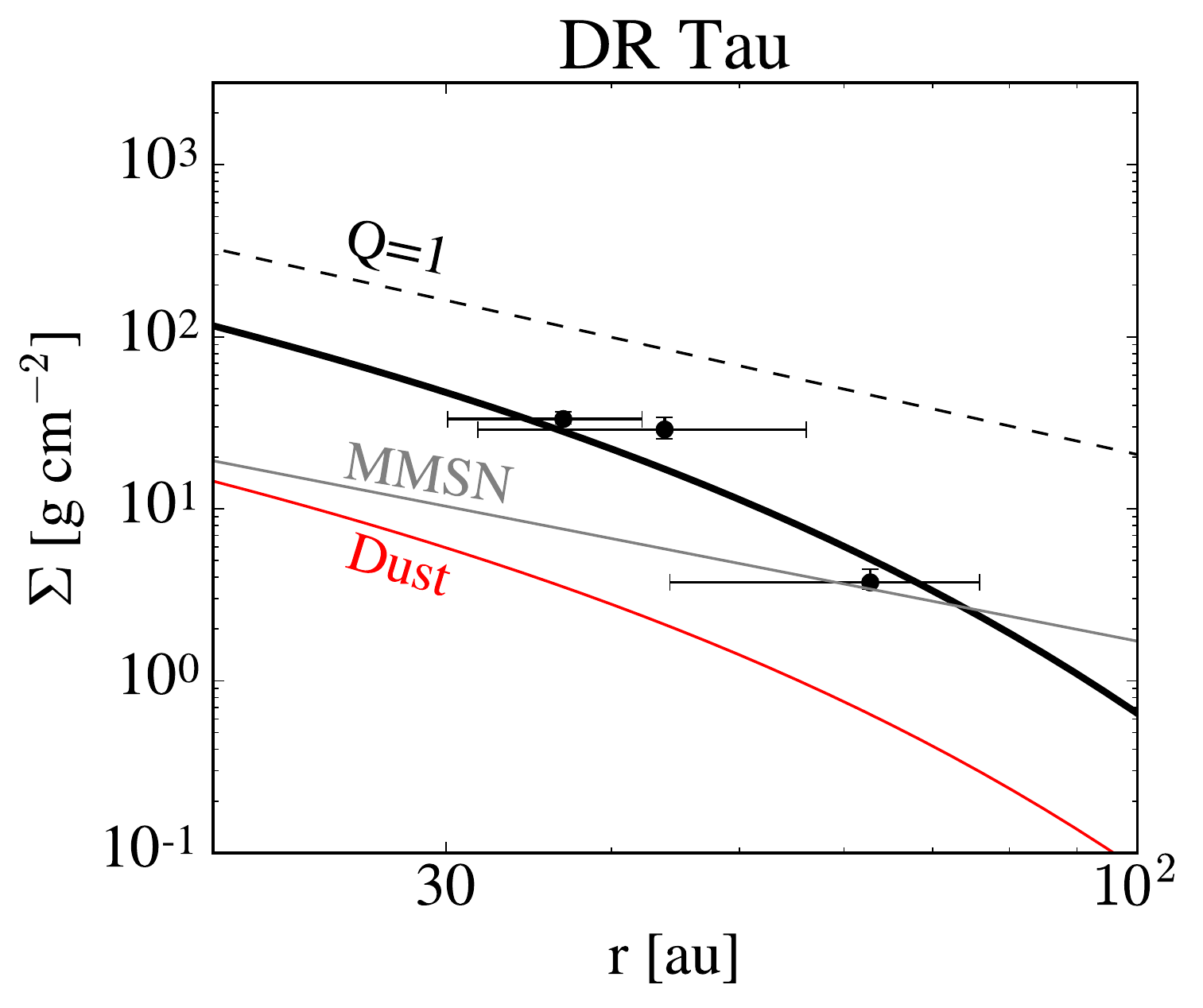}{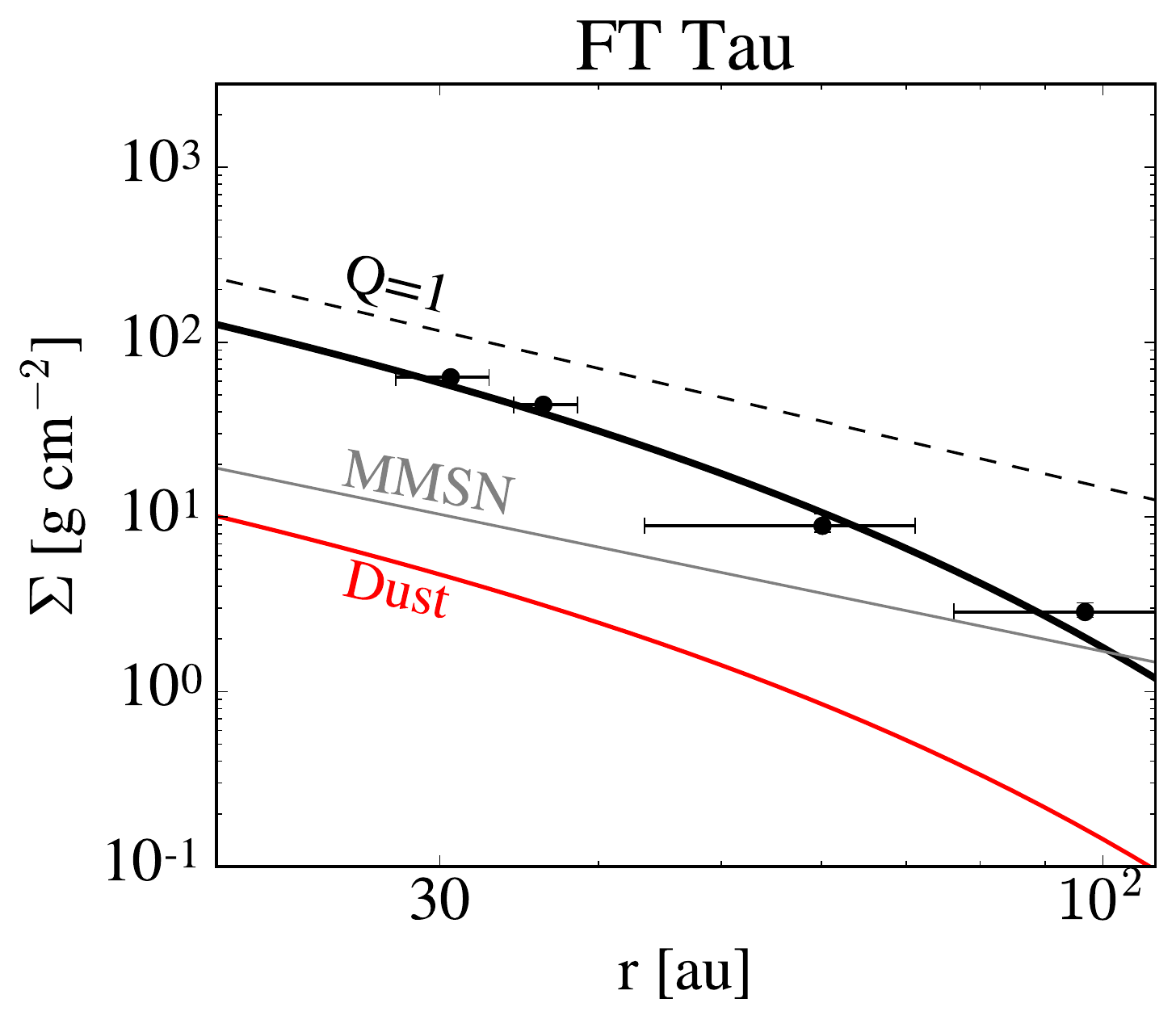}
\plottwo{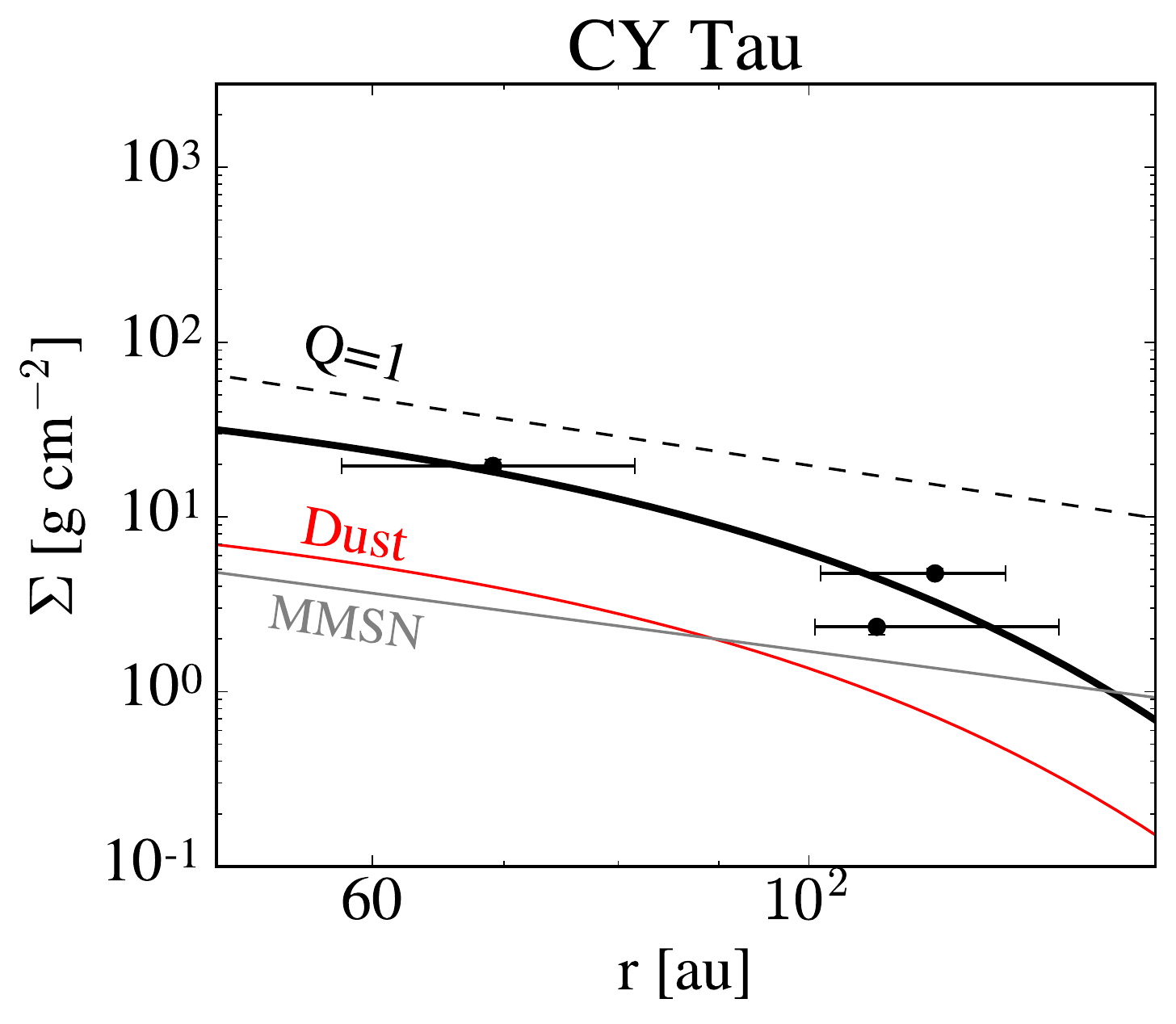}{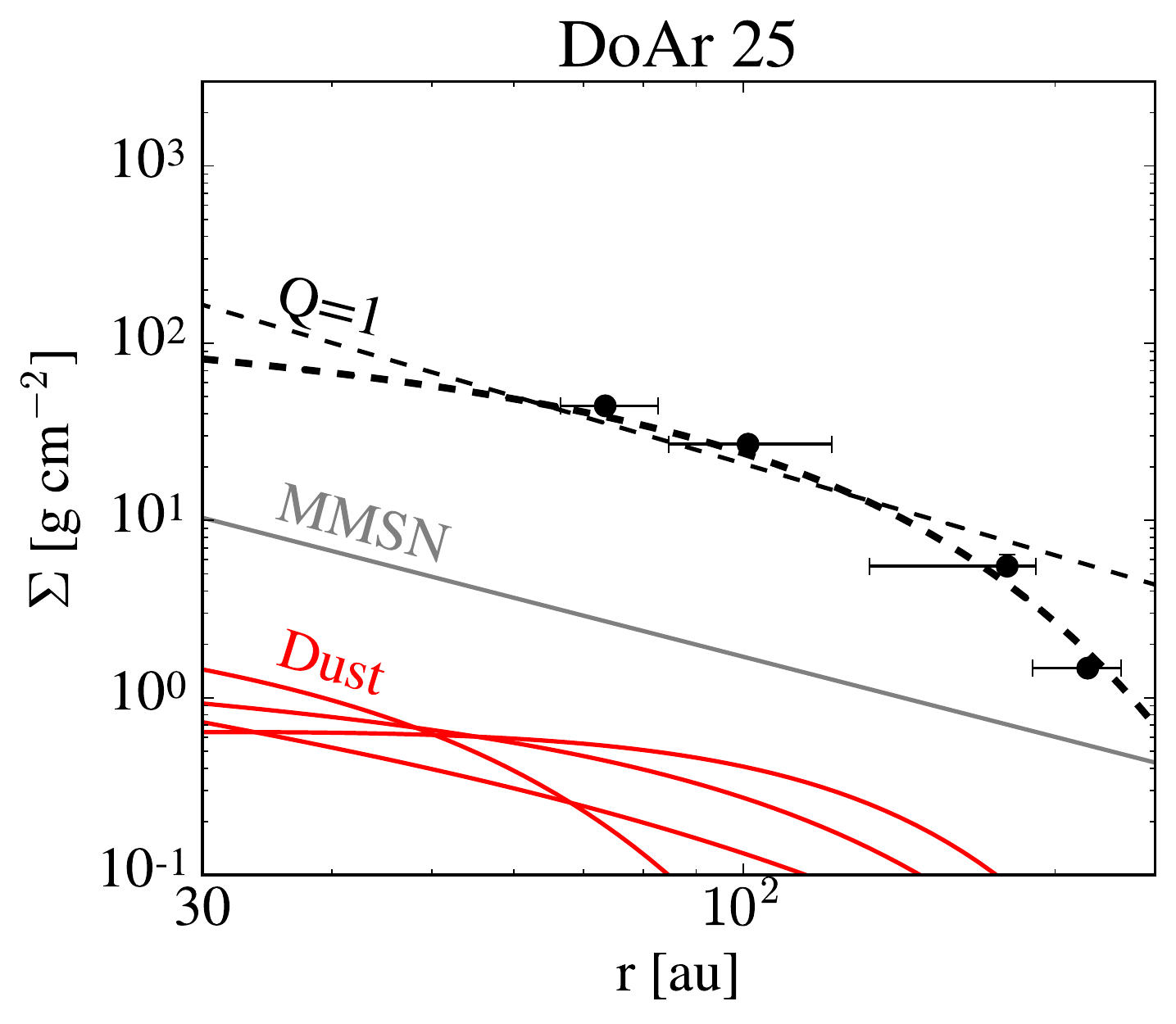}
\epsscale{0.42}
\plotone{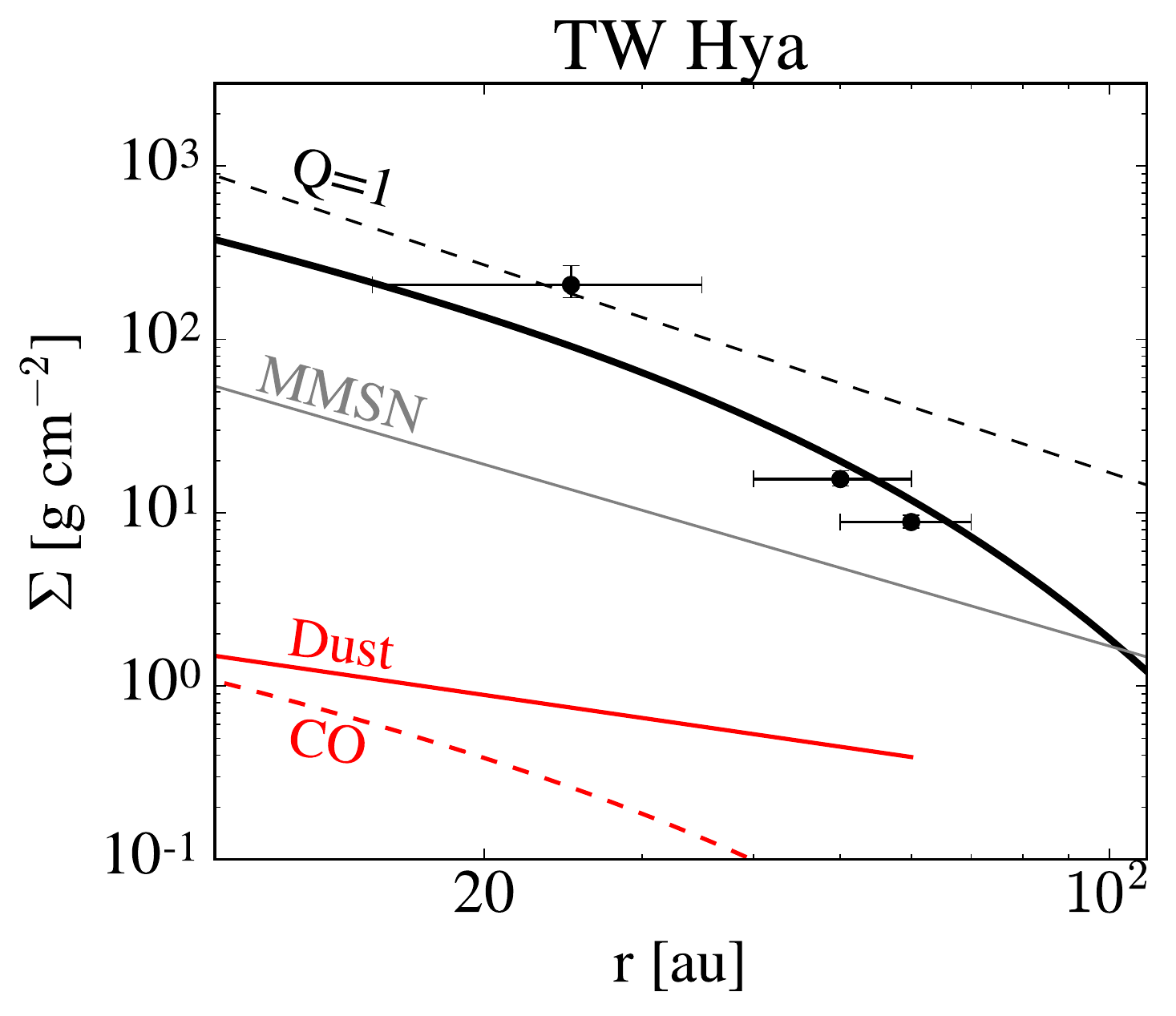}
\caption{The renormalized total surface density profiles (black lines) as shown in Figure \ref{nuker_fits} are plotted for comparison with several other profiles: the total surface density profile derived from integrated dust emission (red lines) or CO emission (red, dashed lines), the minimum mass solar nebula (gray lines) and the gravitational stability limit (dashed, black lines) derived from Toomre-Q stability analysis. We show renormalized profiles derived from integrated dust emission except for the disks TW Hya and DoAr 25 (see text). We choose radius ranges probed by the resolved millimeter continuum observations. AS 209 has a ring of emission observed in both CO and dust observations (blue, dashed line) as discussed in Section \ref{atrap}.}
\end{figure*}\label{zoom_fits} 

\subsection{Trends}
\label{trendz}

Given the derived disk masses found in this work, it is possible that all disks are more massive than was thought previously. All of the disks in this sample have a newly estimated disk mass that is larger than the mass derived from either integrated dust emission or CO observations. Furthermore, the disks in our sample have masses that range from 9 - 27 \% of their host's mass. This may well be a selection effect as our sample is biased towards bright, massive disks that are most readily observed. However, this indicates that the typically assumed dust-to-gas ratio of 10$^{-2}$ is likely incorrect. This conclusion lends weight to the idea that grain growth and drift should alter the dust-to-gas ratio throughout the disk \citep[e.g.,][]{2012A&A...539A.148B, 2012MNRAS.423..389H}. These results are also in agreement with the less-favored result in \citet{2007A&A...469.1169B}, where they found that high disk masses may bring the drift timescales of millimeter grains into agreement with disk lifetimes. Furthermore, the CO-to-H$_2$ ratio also seems to be altered in these disks from the typically assumed ISM value. This also supports the idea that CO chemistry or other physical processes in the disk are more complicated than was initially assumed such that disks can appear to be depleted of gaseous CO. Interestingly, while the factor needed to match integrated dust emission derived masses (2-15) with the newly derived masses is similar across all of the disks in our sample, the factor needed to adjust the CO masses (5 - 115) seems to vary significantly across individual disks. It may therefore be likely that dust is a better tracer of the total mass inventory in disks than CO line emission, although the ratio of dust-to-gas should be carefully chosen. In Section \ref{numericy}, we find an average value of $\sim$10$^{-3}$ for the dust-to-gas ratio of the disks in our sample.

While this is not a statistical sample of disks, in Figure \ref{disk_star_mass} we compare our derived disk masses to their host stellar mass to see if we recover the steeper than linear scaling of disk dust mass with stellar mass from \citet{2016ApJ...831..125P}. The disks in our sample follow a trend with a roughly consistent slope. The \citet{2016ApJ...831..125P} relations were derived using an ALMA survey at 887 $\mu$m for disks orbiting host stars with masses $\sim$ 0.03 to 2 M$_\sun$ in the Chamaeleon star forming region and comparing the mass in dust with the host's stellar mass. There is some expected scatter in this relationship which may be due to differences in disk age and accretion history. If the disk dust mass is directly proportional to the total mass, then we can naively expect this relation to hold for the disks in our sample. This is indeed what we find. A normalized scaling relation of M$_\text{disk}\sim 0.2$M$_\sun$ (M$_\star$/M$_\sun$)$^{1.3}$ or M$_\text{disk}\sim 0.2$M$_\sun$ (M$_\star$/M$_\sun$)$^{1.9}$, a factor of 50 larger than the best-fit relationship derived assuming a dust-to-gas ratio of 10$^{-2}$ from \citet{2016ApJ...831..125P}, match the derived data well. The disk HD 163296 is a significant outlier in this trend, which is not surprising as this scaling relation would predict a disk mass of 0.56-1 M$_\sun$, far exceeding the limit for gravitational stability (see Section \ref{toomreq_time}). The scaling relation for more massive stars, therefore, likely has a different scaling. 

 \begin{figure}[tbp]
\plotone{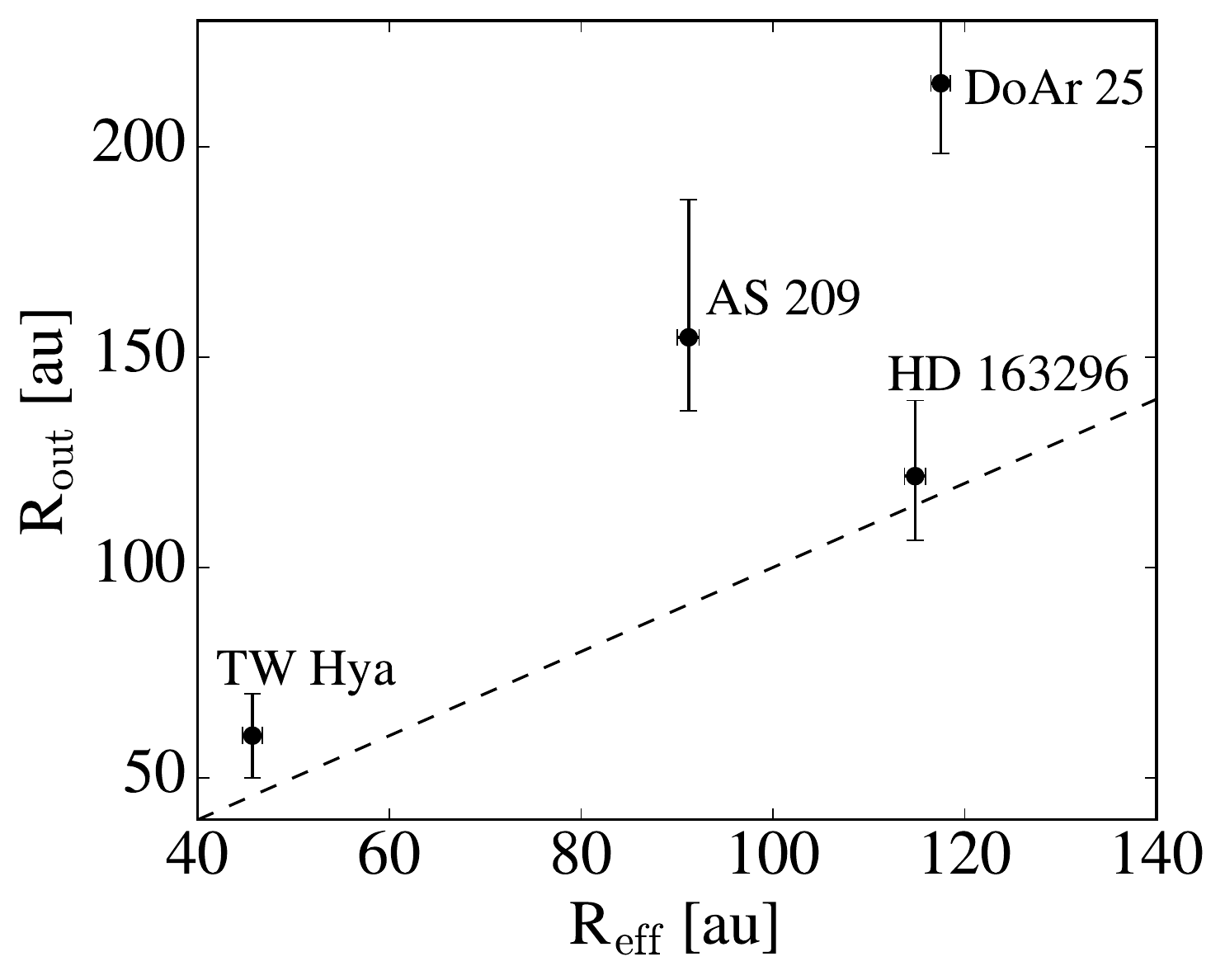}
\caption{The calculated disk outer radius, R$_\text{out}$, which measures the location at which the flux falls off steeply, is larger for every disk in our sample than the effective radius, R$_\text{eff}$, which measures the radius that encompasses 68 \% of the disk flux as calculated in \citet{2017ApJ...845...44T}. The one-to-one line is also shown (dashed).}
\end{figure}\label{size_comp}

Interestingly, there does not seem to be a clear correlation between the stellar age and disk mass. We note that the disk masses in this sample are all the same to within a factor of 3, although there is a wide range in stellar ages. The lack of a clear trend between disk mass and stellar age is likely not apparent in this sample because older disks with lower masses may not be massive enough for current high resolution observations. Our sample does indicate, however, that some disks may not be able to viscously evolve efficiently (see Section \ref{toomreq_time}). 

The newly determined total surface density profiles can also be placed in context of other typically assumed disk profiles such as the minimum mass solar nebula (MMSN). The new profiles for these disks can also be directly compared to their previously derived disk surface density profiles. This comparison is shown in Figure \ref{zoom_fits} for the regions of the outer disk where our modeling work is the most readily tied to empirical evidence. For the MMSN we use the following standard prescription \citep{1977Ap&SS..51..153W,1981PThPS..70...35H}:

\begin{equation}
\Sigma_\text{MMSN} = 1700 \text{ g cm}^{-2}\text{ } r_\text{au}^{-3/2}
\end{equation}

For every disk in our sample the newly derived surface density profile and mass exceeds the MMSN (M$_\text{MMSN} = 0.02 M\odot$) at most radii which itself exceeds the estimate derived from other observational tracers. While the disks in our sample are more massive than the MMSN, the disks DR Tau and FT Tau are comparable at large radii past the critical radius.

The four of the disks in our sample with disk radii measured at $\sim$ 0.9 mm (340 Ghz) were also included in the analysis done by \citet{2017ApJ...845...44T}. The outer radii measured in this work are indeed larger than the effective disk radii from \citet{2017ApJ...845...44T} as expected, although there is not a systematic offset as shown in Figure \ref{size_comp}. However, this is to be expected as the effective disk radius depends on the inner disk index such that disks with more centrally concentrated intensity profiles have smaller effective radii \citep{2017ApJ...845...44T}. The outer radius in contrast is only determined by the radius where disk emission approaches zero. For a discussion on the errors of these measurements see Section \ref{fitroutine}. 

\subsection{Gravitational Stability}\label{toomreq_time}

\begin{figure}[tbp]
\epsscale{1.15}
\plotone{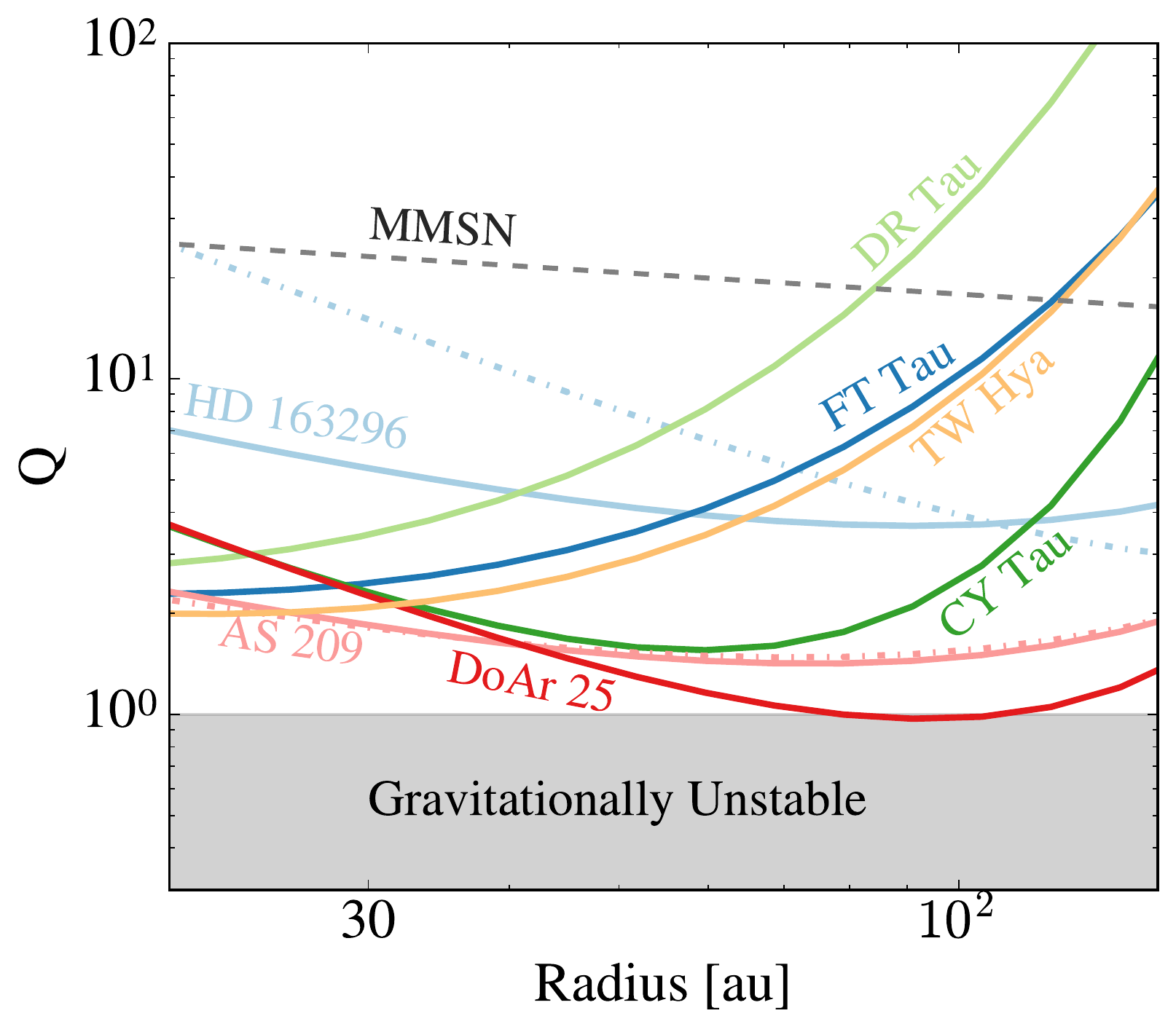}
\caption{All of the disks in our sample are stable against gravitational collapse. The Toomre-Q parameters as a function of radius are shown for the renormalized dust surface density profiles (solid lines) and renormalized CO surface density profiles (dot dashed lines). The disk DoAr 25 is close to exceeding the limit for stability as its Toomre-Q parameter approaches 1 near its critical radius. The Toomre-Q parameter for the MMSN is provided for comparison. }
\end{figure}\label{toomre}

We briefly analyze the stability against gravitational collapse of the newly derived disk surface densities through a Toomre-Q stability analysis. Following the Toomre-Q instability criterion, a disk is unstable to collapse if the local gravity in a region of any arbitrary size overcomes rotational and thermal support, which requires 

\begin{equation}
Q \equiv \frac{c_s\Omega}{\pi G \Sigma}  \lesssim 1 \;\;,
\end{equation}

\noindent where $c_s$ is the sound speed, $\Omega$ is the orbital frequency, and $\Sigma$ is the disk surface density \citep{1964ApJ...139.1217T}. 

The surface density that corresponds to a Toomre-Q parameter of unity (roughly the first point in which a profile becomes unstable to collapse) is shown for each disk in Figure \ref{zoom_fits}. The Toomre-Q parameter varies throughout the disks and reaches a minimum at a particular semi major axis. The value of the Toomre-Q parameter for the full disk sample as a function of radius is shown in Figure \ref{toomre}. All of the disks in our sample, except for the disk DoAr 25, respect the Toomre-Q stability criterion as expected by their smooth morphologies. The disk DoAr 25 has a derived surface density profile that just reaches this limit at roughly 100 au. This disk was previously thought to be massive based on classical dust emission observations which also indicate that this object may be approaching gravitational instability \citep{2009ApJ...700.1502A}, although high resolution imaging at 1.3 mm shows three bright rings located at 86, 111, and 137au instead of spiral arms \citep{2018arXiv181204193H}. Furthermore, while our newly derived estimates of disk surface density for the other 6 disks in our sample do respect the gravitational stability limit, they all reach Q values less than 10. 

We conclude that at least among the brightest sample of disks in the sky, low Toomre-Q values are not uncommon. This idea is supported by recent ALMA observations in which 4 disks have so far been shown to have spiral arm structures that suggest instability to collapse \citep{2018arXiv181204193H,2016Sci...353.1519P}.  

We note that it is seemingly easier theoretically to produce massive disks that approach gravitational instability as disks may form near the limit of stability \citep[e.g.,][]{2011ARAA..49...67W} and it is non-trivial to trigger viscous evolution in disks via the magneto-rotational instability \citep[e.g.,][]{2007NatPh...3..604C,2009ApJ...701..737B,2015ApJ...808...87M,2015ApJ...799..204C}. Understanding why some disks have been able to accrete efficiently while others have not may shine a light on how non-ideal magnetohydrodynamics (MHD) operates in protoplanetary disks. 

\subsection{Derived Dust Surface Densities and Numerical Validation}\label{numericy}

\begin{figure}[tbp]
\epsscale{1.15}
\plotone{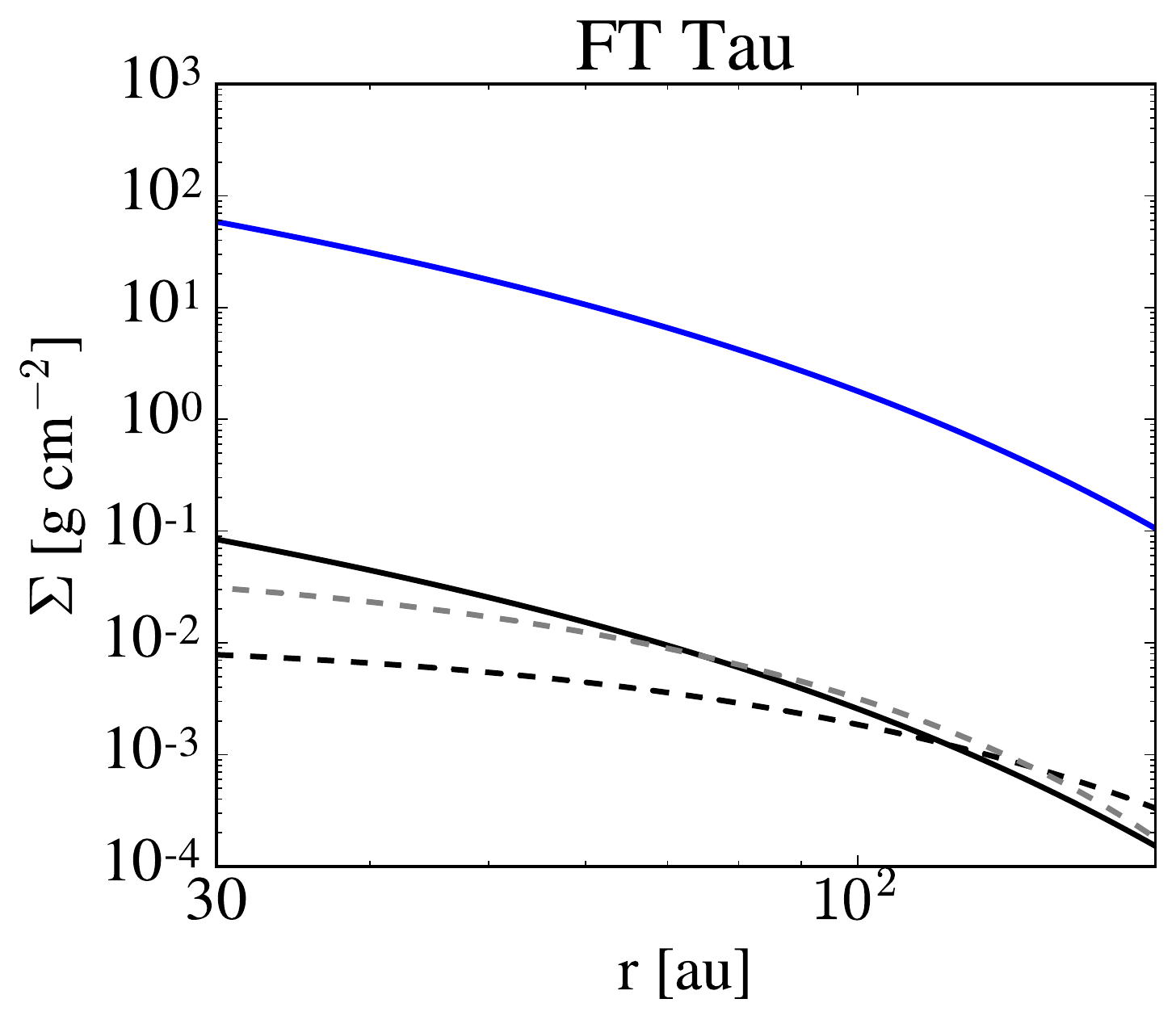}
\caption{Our model for dust surface density for the disk FT Tau (black dashed line) is roughly consistent with the observationally derived dust surface density profile \citep[black solid line,][]{2016AA...588A..53T}. The newly derived dust surface density profile is also fairly consistent with the dust surface density derived when we input our derived disk parameters into the \citet{2012A&A...539A.148B} dust evolution code (gray dashed line). The newly derived gas surface density profile using our model (blue line) is also shown.}
\end{figure}\label{sd_rough}

Following the method described in Section \ref{dustsd}, we now derive the dust surface density for our modeled disks through a consideration of a particle coagulation. We derive the dust-to-gas ratio and hence the disk surface density using Equation (\ref{dust-to-gas}) and multiply this by our total gas surface density profile to derive a dust surface density profile. An example of our derived dust surface densities is shown in Figure \ref{sd_rough} for the disk FT Tau which is representative of the other disks in our sample. Using our order-of-magnitude derivation we find rough agreement with observations. We find this level of agreement particularly encouraging as there are many unaccounted for sources of error in both dust observations (i.e. the dust grain opacity) and our model (see Section \ref{uncertain}).

Every disk in our sample has a derived dust-to-gas ratio of approximately 10$^{-3}$ in the outer disk with the exception of the disks TW Hya and HD 163296 which have an average dust-to-gas ratio of 10$^{-4}$ in the outer disk, in good agreement with the dust surface density profiles derived from integrated dust emission.

\begin{figure}[tbp]
\epsscale{1.2}
\plotone{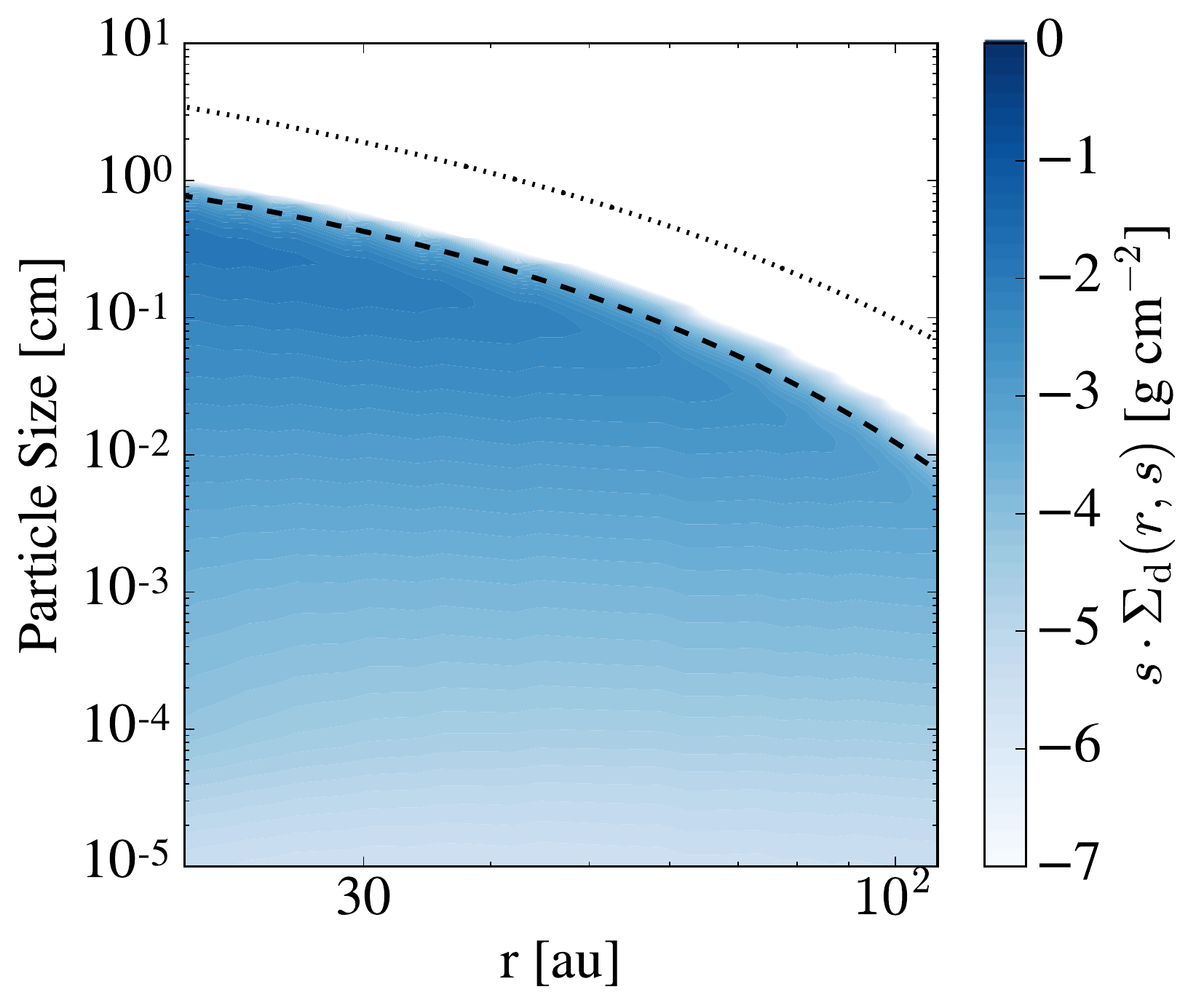}
\caption{There is good agreement between the location of the millimeter particles found numerically when we input our derived disk parameters into the \citet{2012A&A...539A.148B,2015ApJ...813L..14B} dust evolution code and the location of the particles in the observations. Shown are the location of particles at 1.6 Myr for the disk FT Tau as modeled using this numerical code. The contour lines represent the surface density of reconstructed particle size distribution as a function of radius. The dashed line represents the drift limit (the largest particles present) and the dotted line is the fragmentation limit. This simulation had an initial dust-to-gas ratio of 10$^{-2}$, indicating that this disk formed with more solid material than is available at present.}
\end{figure}\label{part_til}

While we derive lower present day dust-to-gas ratios than the typically assumed ISM value of 10$^{-2}$, the implication is that at earlier times the dust mass was much higher. The larger disk masses in both gas and dust (at earlier times) may help resolve problems in planet formation theory as applied to extrasolar system formation as presented in \citet{2018A&A...618L...3M}. 

While we do not use a dust opacity model to calculate a dust-to-gas ratio, we can use our results to roughly derive an opacity model. The derived dust opacity in our modeling is only different than typically assumed values for each disk due to the difference in dust-to-gas ratio. As our calculated dust surface density profiles are in rough agreement with those inferred from integrated dust emission, the total dust opacity is in agreement with the dust opacity assumed in the literature (see Table \ref{obs}) when relating dust emission to total mass for each object with a modifying constant due to the decreased dust-to-gas ratio. The assumed literature dust opacities vary for the objects in our sample from $\sim$0.01 - 8 cm$^2$ g$^{-1}$ at $\sim$1.3 mm \citep{2007AA...469..213I,2011AA...529A.105G,2014AA...564A..93M,2016AA...588A..53T,2017ApJ...851...83C}. The average literature opacity value for objects in our  sample at 1.3 mm, removing the high and low outliers, is $\sim$$0.5$ cm$^2$ g$^{-1}$ ($\sim$$1.5$ cm$^2$ g$^{-1}$ with outliers), which is lower than the $\sim$3 cm$^2$ g$^{-1}$ at the same wavelength adopted in the DHSARP survey \citep{2018ApJ...869L..45B}.

The total dust opacity used to derive dust masses from disk fluxes in the sub-mm could therefore be described simply as:

\begin{equation}
k^{'}_{\lambda} = f_d k_\lambda,
\end{equation}

\noindent where $k^{'}_{\lambda}$ is the total dust opacity at a given wavelength in units of cm$^{2}$ g$^{-1}$, $f_d$ is the dust-to-gas ratio, which varies from $\sim$$10^{-4}-10^{-3}$ for objects in our sample, and $k_\lambda$ is the previously assumed dust absorption opacity at a given wavelength also in units of cm$^{2}$ g$^{-1}$ which can differ from disk to disk. Other than the change in the assumed dust-to-gas ratio, our inferred dust opacities are in rough agreement with those commonly assumed in the literature for the different objects in our sample (c.f. Figure \ref{sd_rough}).

We further verify our disk model using the publicly available dust evolution code from \citep{2012A&A...539A.148B,2015ApJ...813L..14B}. The model from \citet{2012A&A...539A.148B,2015ApJ...813L..14B} evolves the disk from early times and reconstructs a full particle size distribution. They find that this semi-analytic model matches well with more detailed numerical modeling in several tested regimes of interest. We input our derived disk parameters into the \citep{2012A&A...539A.148B} dust evolution code and find that we are able to reproduce the particle locations as a function of radius. This is shown in Figure \ref{part_til} for the disk FT Tau where we assume $\alpha = 10^{-3}$, no viscous gas evolution, and an initial dust-to-gas ratio of 10$^{-2}$ and run the model to the current age of the system, 1.6 Myr. In this code we also derive a low dust-to-gas ratio that varies as a function of radius with an average value of $\sim10^{-3}$ for the radii of interest in agreement with the results from our model. We therefore find our new disk model to be both numerically reproducible and in good agreement with observational work as shown in Figure \ref{sd_rough}. 

In summary, an increase in the total gaseous surface density allows for larger particles to remain coupled to the gas for longer such that their drift is slowed. The increased gas mass derived in our modeling, therefore, readily explains the observed locations of differently sized dust particles.

\section{Discussion}\label{discuss}
\subsection{Disk Substructure}\label{atrap}
Recent observations of disks in the millimeter using ALMA have revealed the richness of disk substructure in the form of rings, gaps, spiral arms, vortices, and more \citep[e.g.,][]{2015ApJ...808L...3A,2016Sci...353.1519P,2018arXiv181204193H,2017ApJ...844...99L,2041-8205-869-2-L41}. In particular, the prevalence of ring structures suggest that these features may be fundamental to the majority of protoplanetary disks \citep[e.g.,][]{2018arXiv181204193H,2017ApJ...844...99L}. Several theoretical models exist to explain the generation of disk rings such as: planet-disk interactions \citep[e.g.,][]{1986ApJ...307..395L,2012ApJ...755....6Z}, or disk specific mechanisms such as large scale instabilities causing pressure bumps \citep{2016MNRAS.457L..54L}, grain growth around ice lines \citep[e.g.,][]{2041-8205-806-1-L7,2017A&A...602A..21S}, and many more. While the cause of these rings is an active area of research, these models tend to create ring-like features through the presence of dust traps that can slow drift and cause particle pileups at particular radial locations. We provide a brief discussion of how dust traps may influence our method of determining total disk surface density and leave detailed modeling to future work. 

\begin{figure}[tbp]
\epsscale{1.2}
\plotone{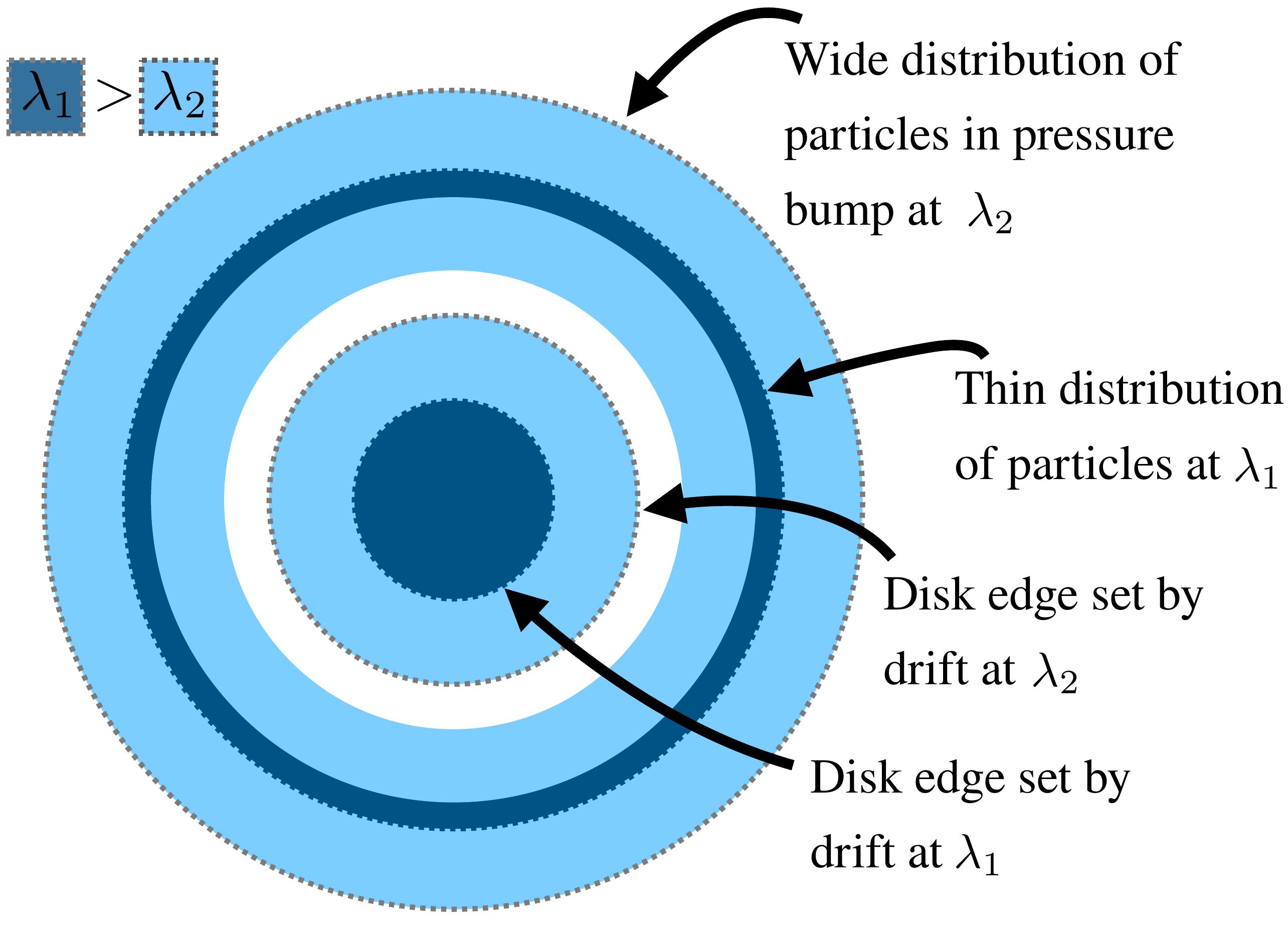}
\caption{Cartoon of how the presence of a pressure trap in a disk will alter the signature of particle drift. Particles of size $2\pi/\lambda_1$, where $\lambda_1$ is the longer observational wavelength, will be strongly affected by the pressure trap and should exhibit a narrow ring of emission when viewed at $\lambda_\text{obs} = \lambda_1$. Interior to the pressure trap there will be a large gap as these particles drift relatively quickly. Particles of size $2\pi/\lambda_2$, where $\lambda_2$ is the shorter observational wavelength, will be less strongly effected by the pressure trap and should generate a wider ring of emission when viewed at $\lambda_\text{obs} = \lambda_2$. Interior to the pressure trap there will be a smaller gap as these smaller particles are slower drifters.}
\end{figure}\label{pressure_trap}

\subsubsection{Dust Lines and Efficient Dust Traps}

If dust traps are indeed prevalent in disks they will have clear signatures depending on the efficiency of the trapping. If the dust trap is efficient in trapping particles then we should see a distinct increase in emission at the same dust trap radial location at each millimeter wavelength. The radial width of the dust trap may vary at different wavelengths, however, as larger particles that are more influenced by drift are trapped more strongly in the pressure bump than smaller particles (see Figure \ref{pressure_trap}). If an efficient dust trap is present exterior to the maximum location that a particle could be present due to drift then it is possible that there will be an uptick in disk emission at the radial location of the dust trap. The dust line would therefore be set by dust trapping instead of particle drift. In this case, resolved disk images would show a ring of emission present across millimeter wavelengths with an interior gap in emission. At longer wavelengths that probe larger grains we would expect this gap to appear larger as the large grains interior to the dust trap would drift quickly. Correspondingly, at shorter wavelengths we would expect to see a smaller gap. This is summarized in Figure \ref{pressure_trap}. In the case that multiwavelength high resolution images of disks show these signatures of a pressure bump, the disk edge set by drift can thus be disentangled from the empirical disk dust line which we have defined as the disk outer edge.

Alternatively if the dust trap is interior to the disk dust line but also strongly efficient then it could be possible that the drop off in emission is sufficiently sharp interior to the radial location set by drift such that modeling disk visibilities places the dust line location interior to the drift location. Given our modeling sensitivities (see Figure \ref{diff_tested_disks}) this outcome is not likely. Furthermore, in this scenario we would also expect an efficient dust trap to be present at the same location across different millimeter wavelengths. 

\subsubsection{Evidence of Dust Traps in Our Sample}

In our multiwavelength modeling of dust lines, only two objects show strong evidence of efficient dust trapping for various particle sizes: AS 209 and CY Tau. Three of the other disks in our sample, HD 163296, FT Tau and DoAr 25, have recent high resolution ALMA observations that indicate ringed substructure in observations taken at 1.3 mm \citep{2017ApJ...844...99L,2018arXiv181204193H,2041-8205-869-2-L49}. However, as we do not derive the same dust line at different observed wavelengths, these substructures likely correspond to less efficient particle trapping. As these disks do not have published reduced visibility data suitable for our derivation of disk dust lines, we compare these disk outer radii based on the analysis of the new ALMA observations to our derived radii at the same or similar wavelengths. For HD 163296, our derived disk outer radius at 1.3 mm is consistent with the radius of the outermost disk ring as presented in \citet{2018arXiv181204047I}. We note that for this disk, the ring features seen in dust emission are not present at the same contrast in similarly resolved CO emission \citep{2016PhRvL.117y1101I,2018arXiv181204047I}, indicating a weaker pressure bump than is viewed in AS 209 (see below). While we do not model DoAr 25 at 1.3 mm in this work, the derived location of the outermost ring in \citet{2018arXiv181204193H} is consistent with our disk dust line at 2.8 mm. For FT Tau, recent modeling work finds that 90\% of the disk flux is contained within 42 au \citet{2017ApJ...844...99L}. This is somewhat different from the outer radius that we derive for this disk at 1.3 mm. However, the radius that encompasses 90\% of the disk flux is different from the dust line measurement presented in this work as we are interested in the radius at which disk emission goes to zero. This distinction is described in more detail in our discussion of the outer radius as defined in \citet{2017ApJ...845...44T} (see Section \ref{fitroutine}). 

One of the key features of efficient dust trapping, as shown in Figure \ref{pressure_trap}, is the presence of disk dust lines at the same location at different observed wavelengths. Our analysis of the disk AS 209 is consistent with such efficient trapping. Observations in CO find an increase in emission at roughly 150 AU \citep[see Figure \ref{zoom_fits},][]{2016ApJ...823L..18H,2018ApJ...869L..48G} and a ring of emission in millimeter dust observations at roughly the same radius \citep[e.g.,][]{2018AA...610A..24F}. We correspondingly find two dust lines located at the same radial distance and recent observations of AS 209 using ALMA at 1.3 mm find an outermost ring in emission consistent with the dust lines derived at 0.85 and 2.8 mm \citep{2018arXiv181204193H,2018ApJ...869L..48G}. This may indicate the presence of a pressure bump creating a dust trap at 150 AU which may obscure the location of the outer edge caused by drift for these particle sizes. In the case of AS 209 the larger particles with dust lines interior to 150 au have seemingly drifted interior to this point and must not be efficiently trapped at 150 au. This may constrain the way in which dust traps affect particles of different sizes. Alternatively, the dust trap may have formed after these particles drifted inwards to their present location which could constrain the timescale over which the dust trap formed in the disk. Either way, the presence of dust lines interior to such a trap allows for accurate scaling of previously derived surface density profiles such that the method for deriving surface density profiles presented in this paper is still useful.

The disk CY Tau is also a candidate for having an outer ring due to an efficient particle trap. The dust lines at the shorter wavelengths (1.3 mm and 2.8 mm) for this disk are both located at roughly the same location. Again, the presence of dust lines interior to this particle trap help constrain our disk modeling work. 

\subsubsection{Dust Lines and Inefficient Dust Traps}

For dust traps that do not efficiently trap grains, we may expect to see two different behaviors. For disks with closely spaced inefficient pressure traps, we would expect that they will slow particle drift and generate a multiplicative factor in the drift velocity that translates to a multiplicative factor for the disk surface density. This multiplicative factor would cause the derived disk surface density to decrease. If the drift efficiency is only moderately slowed, the effect on this modeling should be within the observational error. We comment as an aside that the derived dust line locations may look like a step function in this case because the particles will spend most of their time in the pressure traps. If the inefficient pressure traps are not close together then we would expect that, while particles may spend a longer time in the pressure trap, the disk outer edge caused by drift will be apparent in high resolution imaging interior to this location as described above. 

\subsection{Porous Aggregate Particles}
Recent laboratory work and numerical simulations have indicated that particles in disks may form as porous aggregates with filling factors as low as 10$^{-4}$ corresponding to particles with very low densities \citep[e.g.,][]{2012ApJ...752..106O,2013A&A...554A...4K,2013A&A...557L...4K}. Using effective medium theory (EMT), \citet{2014A&A...568A..42K} show that the absorption mass opacity of aggregate particles can be characterized by the product of the particle radius ($s$) and the filling factor ($f$). This is because the absorption mass opacity depends directly on the imaginary refractive index, which is proportional to the filling factor, and the size parameter, which is proportional to the particle size.  At certain wavelengths the absorption mass opacity of compact grains show distinct interference patterns that are not present in the absorption mass opacity of aggregates with the same value of $sf$. However, for the different values of $sf$ shown in \citet{2014A&A...568A..42K} that correspond to our observed wavelengths (see their Figure 3), the absorption opacity for compact grains differs by less than an order of magnitude for aggregates with the same characteristic parameter given our assumption that $\lambda_\text{obs} = 2\pi s_\text{obs}$. The observed particle size is therefore given by $s_\text{obs} \approx sf$. Here $s_\text{obs}$ refers to to the particle size we would expect to dominate the emission and to drift if the actual particles are of a the same effective size. 

Therefore the true particle size, which sets the aerodynamic properties of the grains is given by $s = s_\text{obs}/f$. The particle density is correspondingly different from the typically assumed internal density of compact grains such that $\rho_\text{agg} = \rho_s f$. The equation for deriving surface density can therefore be rewritten for aggregate particles:

\begin{equation}
\Sigma_g(r) \approx \frac{t_\text{disk}v_0 \rho_\text{agg} s}{r} \approx \frac{t_\text{disk}v_0 \rho_s f s_\text{obs}}{f r} \approx \frac{t_\text{disk}v_0 \rho_s s_\text{obs}}{r}
\end{equation}

\noindent which reduces such that it is equivalent to Equation (\ref{surf_dens}). 

This model for deriving gaseous disk surface densities is therefore robust for both compact and aggregate dust grains as it is roughly independent of the particle filling factor. 

While our derived gas disk surface densities do not depend on grain porosity, the coagulation process of porous grains may differ from compact grains. For porous particles with larger cross sections, particle growth may well be more efficient if their sticking efficiency and fragmentation threshold are otherwise similar to compact grains. Thus, if aggregates are an order of magnitude more efficient at particle growth then this would change our derived value of the dust-to-gas ratio and therefore our derived dust surface density profile.

\subsection{Implications for Other Disks}
\begin{figure}[tbp]
\epsscale{1.}
\plotone{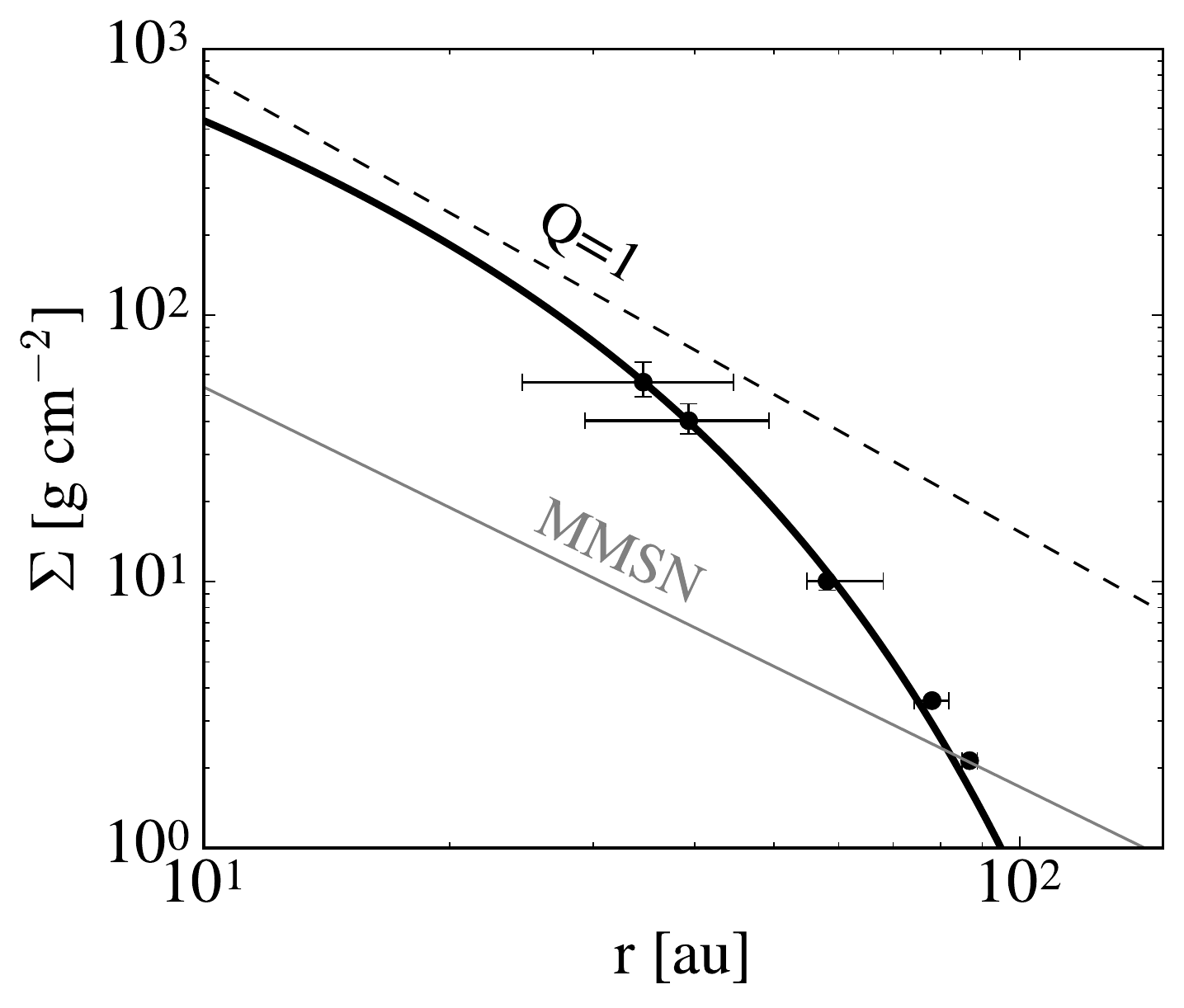}
\caption{The disk UZ Tau E may also be well described following this method. We derived surface density points using the radii from \citet{2018ApJ...861...64T} in which they consider a disk radius that encompasses 68\% of the total flux (black points). Because these radii are not the disk dust lines, this result is approximate.  Nevertheless, a similarity solution fit to the derived surface density points (black line) is a good fit and is stable against collapse. The limit of gravitational stability (dashed, black line) and the surface density profile for the MMSN (gray line) are shown for reference. }
\end{figure}\label{uz_tau}

There are several other disks in the literature with resolved multiwavength observations that may be well suited to this type of modeling work. However, as many of these disks have only been observed at two observational wavelengths and many others were not published with their complete reduced visibility profiles they are not included in this initial work. 

We briefly model the disk UZ Tau E based off of the analysis from \citet{2018ApJ...861...64T} as shown in Figure \ref{uz_tau}. In their analysis they determine an effective disk radius ($R_\text{eff}$) that corresponds to a fixed fraction (68 \%) of the total luminosity as we discuss previously (see Section \ref{trendz}). As this is not the true dust line (outer radius) in the sense used in our modeling, this model for UZ Tau E is a rough approximation. We note, however, that this seems to be a good candidate disk for this modeling work in the future as the derived surface density points are stable against collapse and seem to follow a profile that is believable for the surface density of a protoplanetary disk. We us a $\chi^2$ minimization to fit the derived surface density points as a similarity solution profile (see Equation \ref{similarity}) and derive the following parameters: $r_\text{crit}$ = 23 au, $\gamma$ = 0.9, $\Sigma_0$ = 369 g cm$^{-2}$. 

This work is not the only evidence for disks being massive as discussed in Section \ref{intro}. Four disks recently observed in high resolution using ALMA show evidence of spiral arms \citep{2016Sci...353.1519P,2018arXiv181204193H}, indicating gravitational instability. Furthermore, in the survey by \citet{2016ApJ...831..125P}, several disks in their sample are within a factor of 3 of being gravitationally unstable. If the dust-to-gas ratio for these objects is instead $10^{-3}$, then a significant fraction of their sample will be approaching the limit of gravitational instability. If future observations confirm the prevalence of disks that show features of gravitational instability, and massive masses in dust, it further indicates that many protoplanetary disks are likely more massive than previously assumed.

\subsection{Further Observational Verification of the Dust Line Model}
We briefly discuss two observational diagnostics as described in \citet{2017ApJ...840...93P} that may provide independent verification of the disk surface density model described in this work. For further details of these diagnostics and observational tools see Section 5 of \citet{2017ApJ...840...93P}.

The first of these diagnostics is whether or not the surface density profile derived from a disk ice line matches the surface density profile derived from disk dust lines. As the location of a species's ice line depends on disk surface density, the surface density can be derived from ice line locations if the radial location of the ice line in the disk midplane is well-constrained and if drift is important in influencing the dynamics of grains at this location or the species's abundance is well constrained.

The second of these diagnostics is whether or not the dust and ice lines scale oppositely with disk surface density. We expect that this will happen because disks with dust lines at larger radial scales should have ice lines located at smaller radii for a given disk temperature structure and molecular abundance. This diagnostic may be approachable with a large sample of well-observed disks. 

\section{Summary and Conclusions}\label{summarize}
We apply a novel method of determining the surface density of protoplanetary disks to a set of 7 diverse objects that does not rely on a tracer-to-H$_2$ ratio or an assumed dust opacity model. We use an MCMC method to model spatially resolved images of disks at multiple wavelengths and infer the location of the disk outer edge (i.e. a disk dust line). This measurement is then related to the maximal radial locations in which particles of size $2\pi/\lambda_\text{obs}$ are observed. Then, through a consideration of the aerodynamic properties of these grains, the total gaseous disk surface density is derived at specific radial locations. These derived surface density values are then used as benchmarks to scale previously modeled surface density profiles derived either from combined multiwavelength dust or CO emission observations. This method may be particularly robust as it does not rely on an assumed dust-to-gas or CO-to-H$_2$ ratio to derive total gas surface densities. This new method is appropriate for disks that have evolved ages ($t_\text{disk} \gtrapprox 1$ Myr) and have different radial extents at different observed wavelengths. For the 7 bright protoplanetary disks in our sample, we derive total gas disk masses and compare these masses to previous values determined from CO and dust emission. We further derive disk dust-to-gas ratios and dust surface density profiles.

Our new disk masses for objects in our sample are 9-27\% as massive as their stellar hosts and have minimum Toomre-Q values below 10 even for the disks in our sample that are relatively old. Our sample is biased towards the brightest and most massive disks in the sky as they are the most readily observed. However, understanding why some disks may be able to efficiently viscously evolve away from the limit of gravitational instability with time while others do not may shed light on the mechanisms that govern magnetohydrodynamics in protoplanetary disks. 

Most of our newly derived masses are larger than the total mass obtained by dust observations by a factor of $\sim$6-8. The disk HD 163296 has a new mass that is roughly consistent with the previous dust emission mass measurement and AS 209 has a new mass that is a factor of 15 larger than measured from integrated dust emission. The three disks with resolved CO observations have new constraints on disk mass that exceed the mass derived from CO emission alone. The amount of observed depletion of CO varies significantly for the three disks in our sample and ranges from a factor of 3-115. This supports the popular idea that CO may be depleted or missing in protoplanetary disks. Though more massive, our new total gas masses scale more consistently with masses inferred from integrated dust emission than from CO emission, indicating that dust is a more robust tracer of total gas mass. 

We further consider the growth of the observed particles to infer the disk dust-to-gas ratio and thus the disk surface density profile. Our model dust surface density profiles match the surface density profiles derived via millimeter observations well. The dust surface densities and locations of the particles in this model can also be reproduced with semi-analytic simulations when our new disk parameters are used as input initial conditions. 

The derived dust-to-gas ratio is typically $\sim$$10^{-3}$ for the disks in our sample in the outer disk. It is perhaps more appropriate to use this value when calculating total disk mass from integrated dust observations that probe the outer disk. The exceptions in this sample are the disks HD 163296 and TW Hya which have a dust-to-gas ratio of 10$^{-4}$. These low dust-to-gas ratios suggest that there was significantly more dust mass available earlier in the lifetime of these disks before particles begin to drift rapidly. The larger disk dust mass at earlier times may help resolve problems in the application of planet formation theory to extrasolar planetary systems.

Some of the disks in our sample appear to have dust lines set by dust traps as their outer radius is at roughly the same location at multiple wavelengths. In particular, two disks in our sample show evidence of an outer ring that may efficiently trap particles that are relatively small in size. For one disk this ring is also present in CO observations. We provide a method of qualitatively determining the location of the disk outer edge set by drift in the case where there is an efficient particle trap present in the disk. We further show that this method of determining surface density is roughly independent of particle porosity. This method could be applied to many disks that are currently observed and may be observed in the future with ALMA at multiple millimeter wavelengths. To continue to validate (or invalidate) this method we stress the importance of having a large sample of objects in which to perform this analysis.

\section{Acknowledgements}
We would like to thank Meredith MacGregor for providing us with her code to produce FITS image files from model surface brightness profiles. We also thank Uma Gorti, Xi Zhang, and Eugene Chiang for their useful comments and insightful discussion. We thank the anonymous referee for constructive comments that have helped improve this manuscript. This material is based upon work supported by the National Science Foundation Graduate Research Fellowship under Grant DGE1339067. R.M.C. acknowledges support from NSF grant number AST-1555385. L.P. acknowledges support from CONICYT project Basal AFB-170002 and from FONDECYT Iniciaci\'on project \#11181068. H.E.S. gratefully acknowledges support from the National Aeronautics and Space Administration under grant No. $17~\rm{NAI}18\_~2-0029$ issued through the NExSS Program. 

\bibliography{refs}
\bibliographystyle{aasjournal}

\appendix

\section{Order of Magnitude Derivations of Particle Relative Velocities}
We provide a framework for understanding particle relevant velocities used in growth calculations in an order of magnitude sense. We consider the case when the relative velocity between two particles is determined by turbulence. This appendix is provided as a tool to increase intuition regarding particle growth in an order of magnitude sense. We did not find such a derivation in previously published work and this was greatly helpful in our understanding of this work and our derivation in Appendix \ref{earlygrow}. 

For a full discussion of the complete expression that we use to determine particle relative velocities in all regimes, see Equation 16 from \citet{2007A&A...466..413O}.  

\subsection{Kolmogorov Cascade}\label{appgrow}
The Kolmogorov cascade describes how energy is transferred from large to small scales in a turbulent fluid.  In a fluid that is in steady state we can use dimensional arguments to derive the following scalings: 
\begin{equation}
E_k \propto k^{-5/3} 
\end{equation}
\begin{equation}
v_k \propto k^{-1/3}
\end{equation}
\noindent where $k$ is the wavenumber with units of length$^{-1}$. These scalings follow directly from balancing the energy in and out of a particular eddy scale. 

In the following discussion we take particle 1 to have a stopping time ($t_1$) and hence Stokes number, $St_1$ (see Equation \ref{stopping}) that is larger than that for particle 2 ($t_2$), $St_2$. 

\subsection{Tightly Coupled Regime -- $t_1,\;t_2 < t_{\eta}$}
For tightly coupled particles with stopping times less than the turnover time of the smallest scale eddies ($t_\eta = Re^{-1/2}t_L$ where $t_L$ is the turnover time of the largest eddies which we take to be the local orbital period and Re is the Reynold's number, defined as the ratio between the turbulent and molecular kinematic viscosities. In ``$\alpha$ notation" this is: Re$=\alpha c_s H/\nu$ where $c_s$ is the sound speed, $H$ is the scale height, and $\nu$ is the kinematic viscosity of the gas), the particle will be able to reach an equilibrium with eddies of every scale. This regime is valid if both particles have stopping times less than the turnover time of the smallest eddy such that $t_1,t_2 < t_{\eta}$. Therefore, when a particle enters any eddy it forgets its initial motion and aligns itself with the motion of the gas that comprises that eddy. In this regime, a particle's velocity relative to the gas can be described as the particle's settling terminal velocity. 

These particles are continually accelerated by drag forces such that the particle's velocity relative to the gas can be thought of as the velocity where the acceleration from the eddies balances the acceleration from drag. An eddy of scale $k$ can accelerate a particle up to a velocity $v_k \propto k^{-1/3}$ on an eddy turnover timescale $t_k \propto (kv_k)^{-1} \propto k^{-2/3}$. This is the minimum acceleration needed to reach a velocity $v_k$ in a turnover time and is relevant because a particle can only couple to an eddy that can accelerate it to the eddy's velocity in less than or equal to a turnover time. Thus we have
\begin{align}
a_k \propto v_k/t_k \propto k^{1/3}
\end{align}
which is the acceleration that an eddy of scale $k$ provides in a turnover time. Thus the acceleration is dominated by the smallest scale eddies (eddies with the largest $k$). 

Equating the drag force with the acceleration from the smallest scale eddy gives:
\begin{align}
\frac{F_d}{m} &\sim a_{\eta}\\
\frac{v_{pg}}{t_s} &\sim \frac{v_\eta}{t_\eta}\\
v_{pg} &\sim v_\eta\frac{t_s}{t_\eta}
\end{align}

\noindent where $F_d$ is the drag force, $a_\eta$ is the acceleration from the smallest eddy, $v_{pg}$ is the relative velocity between the particle and the gas, $v_\eta$ is the velocity of the smallest eddy and $t_s$ is the stopping time of the particle which is equal to the stokes number of the particle divided by $\Omega$. 

Thus the relative velocity between two particles with stopping times $t_1$ and $t_2$ (assuming without loss of generality that $t_1 > t_2$) is:
\begin{align}
v_{12} \sim \frac{v_\eta}{t_\eta}\left(t_1 - t_2 \right)
\end{align}

We can put this equation into more familiar terms by using the following expressions: $v_L^2\sim v_{gas}^2 \sim v_\eta^2Re^{1/2} \sim v_\eta^2 t_L/t_\eta$ and $t_L \sim \Omega^{-1}$

\begin{equation}
v_{12}^2 \sim \frac{v_{gas}^2}{t_Lt_\eta} \left(t_1 - t_2 \right)^2 \sim v_{gas}^2\frac{\Omega}{t_\eta} \left(t_1 - t_2 \right)^2 \sim  v_{gas}^2\frac{\Omega^{-1}}{t_\eta} \left(St_1 - St_2 \right)^2
\end{equation}

\noindent where $St$ is the Stokes number of a particle given by

\begin{equation}\label{stopping}
\begin{array}{@{} r @{} c @{} l @{} }
&St &{}\approx \Omega t_s = \displaystyle
\begin{cases}
\Omega\rho_s s/\rho c_s & s< 9\lambda/4, \text{  Epstein drag},\\
\Omega4\rho_s s^2/9 \rho c_s \lambda &  s> 9\lambda/4, \text{Re} \lesssim 1 \text{  Stokes drag}
\end{cases}
\end{array}
\end{equation}

\noindent \citep[summarized in][]{2010AREPS..38..493C}. Here $t_s$ is the particle's stopping time, $\rho$ is the gas midplane density, $\rho_s = 2$ g cm$^{-3}$ is the density of a solid particle, $s$ is the particle size, and $\lambda = \mu/\rho \sigma_\text{coll}$ is the gas mean free path where $ \sigma_\text{coll} = 10^{-15}$ cm$^2$. Now we can derive our full expression: 

\begin{align}
v_{12}^2 = v_{gas}^2 \frac{t_L}{t_\eta}\left(St_1 - St_2 \right)^2
\end{align}
which is Equation (27) of \citet{2007A&A...466..413O}.

In the tightly coupled regime both the small and large particles are relevant in determining the relative velocity between the two particles. 

\subsection{Intermediately Coupled Regime -- $t_\eta\le t_1\le t_L$ or St$_1 < 1$}
In this regime the larger particle ($t_1$) becomes decoupled from some but not all eddies. A particle is coupled to all eddies with turnover times longer than the particle's stopping time.\footnote{Note that in the above discussion we have also assumed that the time for a particle to cross over an eddy due to laminar drift, $t_{cross}$, is long. This is because, for $t_s \ll t_L$, we expect $v_{rel}(k)$, the relative velocity between the particle and the eddy with scale $k$, to be small, i.e. $\eta v_k St_1/l < t_{cross}$, such that the particle will not drift over an eddy. See \citet{2007Icar..192..588Y} for further discussion.} Smaller eddies have shorter turnover times and smaller velocities. Both particles are well-coupled to large scale eddies; the velocities of the particles are correlated and their relative velocities are low. On the scale at which one particle decouples, the relative velocity is of order the total eddy velocity. We are thus interested in the eddy scale for which the eddy turnover time is $t_1$ because that is the decoupled eddy with the largest velocity. We refer to the eddy turnover time at this scale as $t^*$.

The eddy length scale is $l=1/k$ and $v_k\sim l/t$ is the eddy velocity. We can derive the following scalings: $t_k = 1/(kv_k) \propto l^{2/3}$, $v_k\propto l^{1/3}$. This gives us $v_k\propto t_k^{1/2}$. Decoupling eddies are eddies such that $t<t^*$, the eddy fluctuation time is smaller than the particle's stopping time and the particle is not well-coupled to the eddy. We can therefore say that the relative velocities should roughly be: $v_k \propto \sqrt{t^*}$. We now have $t_* \sim t_s$ \citep[c.f. Equation (3) of][]{2007A&A...466..413O}. As shown in \citet{2007A&A...466..413O}, this is indeed the case -- for small particles a good approximation for $t^*$ is $t^* = y_a^*t_s$ where $y_a^*$ is roughly 1.6. This gives $v_k \propto \sqrt{1.6t_s}$.

Thus, relative velocities in this regime should be proportional to the square root of the Stokes number of the particle. From examination of Equation (28) of \citet{2007A&A...466..413O} this is indeed the case. A more detailed calculation yields the following expression:

\begin{align}
v_{12}^2 = v_{gas}^2\left[2y_a-(1+\epsilon)+\frac{2}{1+\epsilon}\left(\frac{1}{1+y_a}+\frac{\epsilon^3}{y_a+\epsilon}\right)\right]St_1
\end{align}

\noindent where $\epsilon = St_1/St_2$. When particles grow from collisions with like-sized grains, their relative velocity can be approximated as $v_{12} \sim \sqrt{2v_\text{gas}^2St_1}$.

In the intermediately coupled regime the large particle dominates the relative velocity between the two particles.

\subsection{Heavy Particle Regime -- $St \gg 1$}\label{appgrow1}
A well known expression for the RMS velocity (relative to inertial space) of a particle with $St \gg 1$ is
\begin{align} \label{eq:v_rms_large}
v_p = \frac{v_{gas}}{\sqrt{1 + St}}
\end{align}
This is derived in \citet{2007Icar..192..588Y} and \citet{2018ApJ...861...74R}. In this regime particles receive many uncorrelated ``kicks" from the largest scale eddies over a single stopping time, causing the particle to random walk in velocity. These random walk kicks are balanced by settling. 
In general, we can write the RMS particle-particle relative velocity $\braket{\delta v_{12}^2}$ as
\begin{align} \label{eq:v_pp_general}
\braket{\delta v_{12}^2} = \braket{\delta v_{1}^2} + \braket{\delta v_{2}^2} - 2 \braket{\delta v_1 \delta v_2}
\end{align}
Smaller particles can couple strongly to the same eddy, which will cause correlations in their velocity and lead to a non-zero value of $\braket{\delta v_1 \delta v_2}$. For $St_1 \gg 1$ however, the large particle does not couple strongly to any eddy size, so we expect no correlation between the two particles' velocities, i.e. $\braket{\delta v_1 \delta v_2} = 0$. In that case using Equation \eqref{eq:v_rms_large} in Equation \eqref{eq:v_pp_general} gives

\begin{align}
\braket{\delta v_{12}^2} &= \braket{\delta v_{1}^2} + \braket{\delta v_{2}^2}\\
&=v_{gas}^2 \left( \frac{1}{1+St_1} + \frac{1}{1+St_2} \right)
\end{align}
which is Equation (29) of \citet{2007A&A...466..413O}.

We note here that the smaller particle dominates the relative velocity between the two particles. 

\section{Early Stage Particle Growth}\label{earlygrow}

\begin{figure}[tbp]
\epsscale{0.5}
\plotone{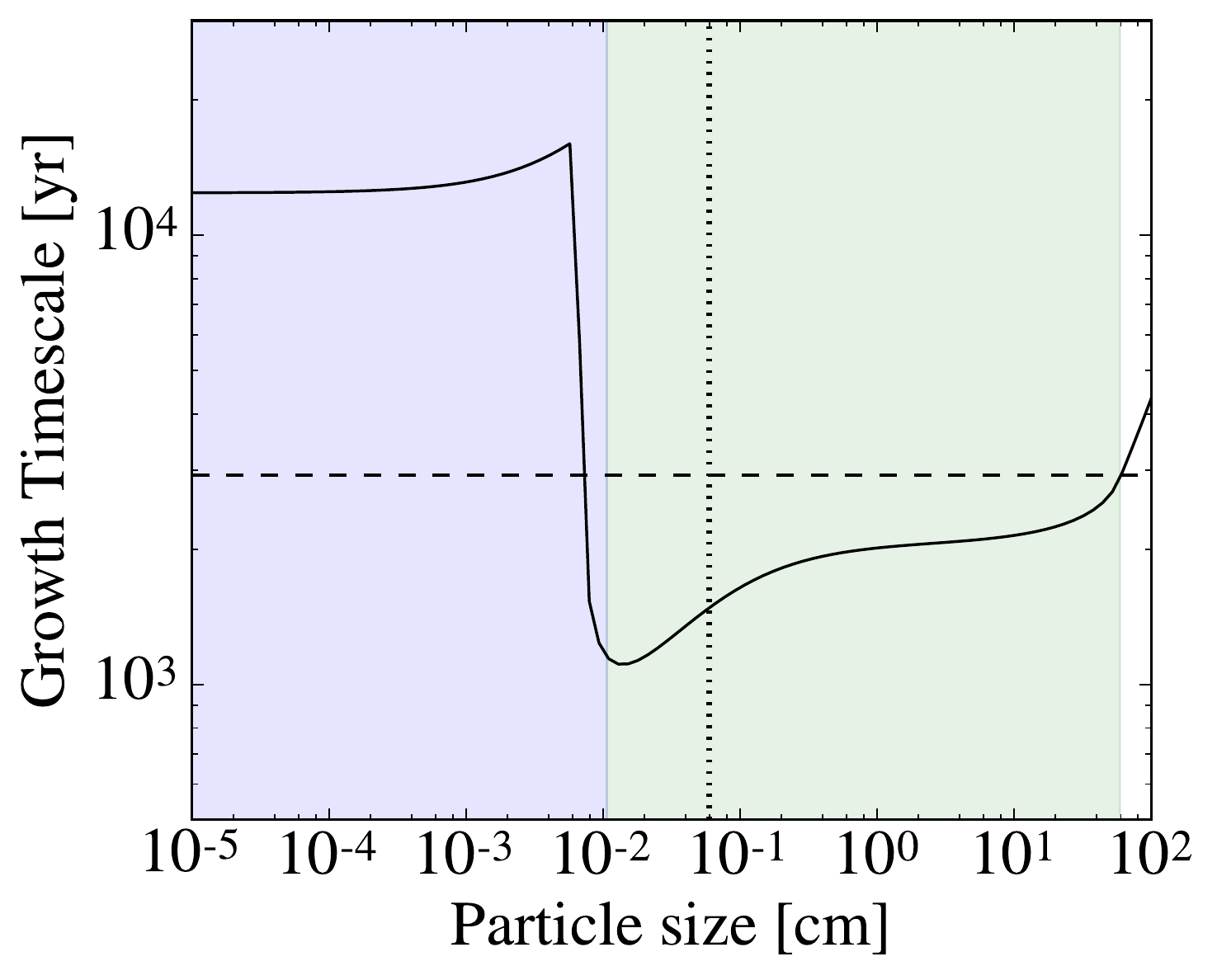}
\caption{The growth timescale as a function of particle size for a fiducial disk with a particle size distribution having most of the mass in the largest sizes. The horizontal dashed line is the growth timescale calculated by \citet{2012A&A...539A.148B} ($1/(\Omega f_d)$ see Equation \ref{egrow_app}), which is intended to approximate the growth timescale for intermediately sized grains. The dotted vertical line is the point at which the particle's stopping time is equal to $\alpha$; to the right of this line the dust scale height has settled to a value smaller than the gas scale height. The blue shaded region represents the small particle growth regime while the green shaded region represents the intermediately sized particle growth regime. }
\end{figure}\label{regimes_app}

At early times particles must grow by several orders of magnitude in size from small submicron grains inherited from the ISM to large grains affected by particle drift or fragmentation. To approximate the time that a disk is in this regime we consider growth by particle collisions such that $\tau_\text{grow} = m/\dot{m}$ where $\dot{m}$ is given in Equation (\ref{grow_rate}). 

For this broad range of particle sizes there are two relevant particle relative velocity regimes: the tightly coupled and intermediately coupled regimes (see Section \ref{appgrow}). In both regimes, particle growth is dominated by collisions with similarly sized grains when considering a Dohnanyi (or similar) size distribution. This is because for such a size distribution most of the mass is in the largest sizes which thus dominate the overall growth rate in spite of their slower relative velocities. 

Very small tightly coupled particles have relative velocities, and thus growth rates, that depend on $\alpha$ though they are roughly independent of particle size (see Figure \ref{regimes_app}). The growth of the tightly coupled particles is generally slower than the growth of intermediately coupled particles for values of $\alpha \lesssim 10^{-1}$.

Once particles grow to a large enough size that they begin to decouple from the gas, their growth rate increases. In the intermediate regime, the relative velocity between similarly sized particles can be approximated as $\Delta v \sim \sqrt{\alpha \text{St}} c_s$ (see Section \ref{appgrow}). In the Epstein drag regime, the Stokes number is given by $St = \Omega\rho_s s/\rho_g c_s$. The scale height for these particles can be approximated as $H_d = H\sqrt{\alpha/St}$. For these particles the growth timescale can therefore be approximated as $\tau_\text{grow} \sim 1/(\Omega f_d)$ \citep{2012A&A...539A.148B} which is independent of both size and $\alpha$.  Following \citet{2012A&A...539A.148B}, because the growth timescale is roughly independent of particle size, the timescale that it takes to grow several orders of magnitude acts like a Coulomb logarithm and can be roughly approximated as

\begin{equation}\label{egrow_app}
t_\text{early grow} = \tau_\text{grow} \ln \left(\frac{a_\text{max}}{a_0}\right) \sim 1/(\Omega f_d) \ln \left(\frac{a_\text{max}}{a_0}\right).
\end{equation}

\noindent where $a_\text{max}$ is the maximum particle size at a given location and $a_0$ is the size of the smallest particles inherited from the ISM.

We repeat this derivation replacing $\Delta v$ with the full particle relative velocity expression from \citet{2007A&A...466..413O} (see their Equation 16) and $H_d = H\sqrt{\alpha/(\alpha+St)}$ \citep{2012ApJ...747..115O} such that particles with St $\lesssim \alpha$ have a scale height equal to the gas scale height. We also assume $St_1 = 0.9 St_2$, roughly the size difference that produces the maximum growth rate in a Dohnanyi size distribution as particles closer in size have relative velocities that approach zero. The growth timescale for particles over a range of sizes is shown in Figure \ref{regimes_app} for an $\alpha$ of 10$^{-3}$ assuming a dust-to-gas ratio of 10$^{-2}$.  

The particle growth timescale is roughly independent of size for both very small and intermediately sized particles. However, the absolute scale between these regimes differs. The average growth timescale considering the growth of very small grains differs from the timescale given in Equation (\ref{egrow_app}) by a factor that is a function of $\alpha$ as shown in Figure \ref{coefficient}. For $\alpha$ values roughly less than or equal to 10$^{-2}$, the early growth coefficient is inversely correlated with $\alpha$ such that the early particle growth rate can be well described as

\begin{equation}\label{new_earlygrow}
t_\text{early grow} = \tau_\text{grow} \ln \left(\frac{a_\text{max}}{a_0}\right) \sim \frac{0.033\alpha^{-0.63}}{\Omega f_d} \ln \left(\frac{a_\text{max}}{a_0}\right).
\end{equation}

\begin{figure}[tbp]
\epsscale{0.5}
\plotone{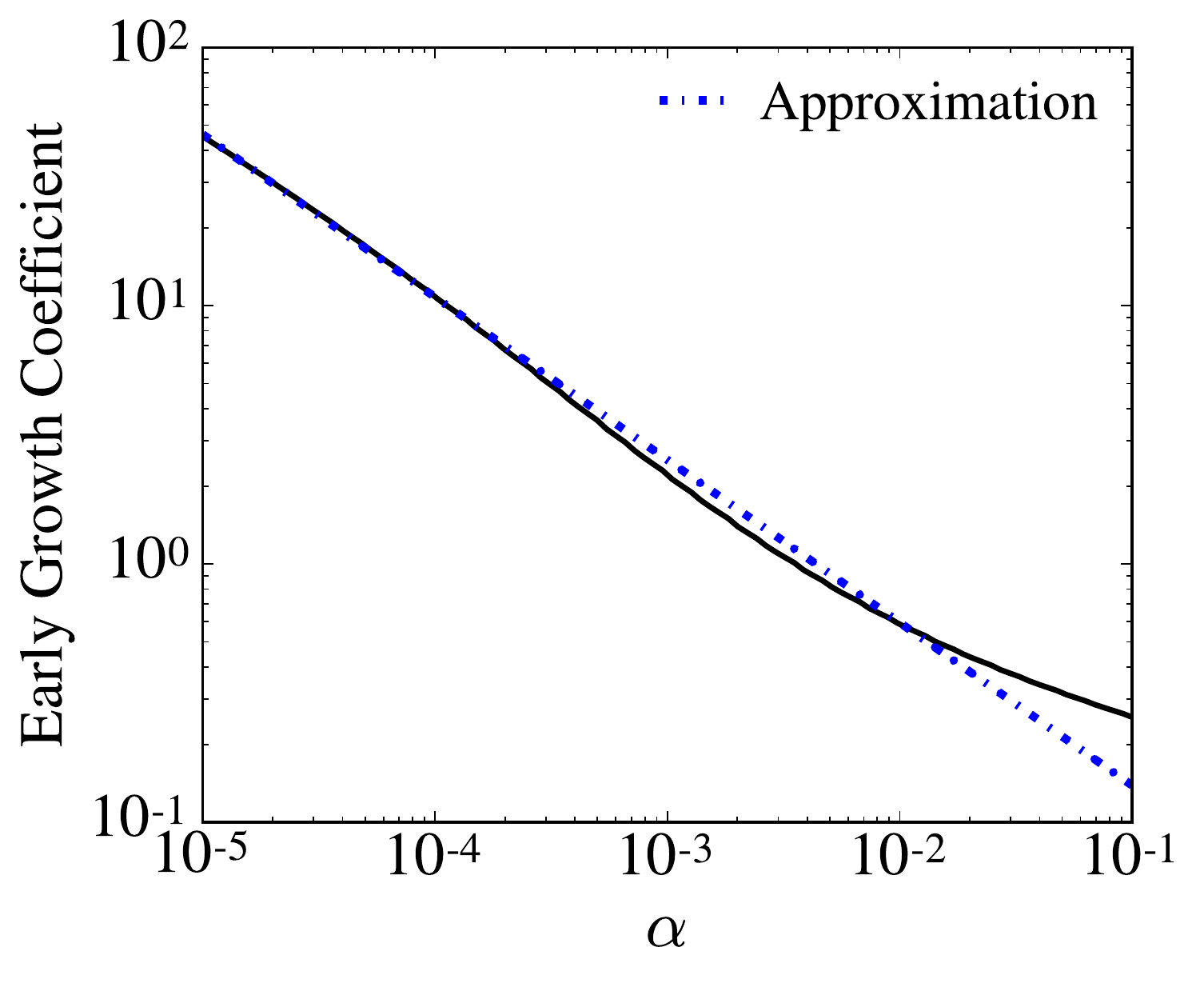}
\caption{The coefficient that determines the magnitude of the early growth timescale (black line) as a function of alpha. An approximation (blue, dashed line) of this relationship is added as a coefficient in Equation (\ref{new_earlygrow}). This approximation is appropriate for $\alpha \lesssim 10^{-2}$.}
\end{figure}\label{coefficient}

\noindent In this work, we assume an $\alpha$ of 10$^{-3}$. For this value of $\alpha$ it is appropriate to increase the early growth timescale in Equation (\ref{slowdatgrow}) by a factor of 2. We find that the early growth timescale does not matter to our results unless $\alpha \lesssim 10^{-7}$.

\end{document}